\documentclass[a4paper,fleqn,usenatbib]{mnras}

\usepackage{mathptmx}

\usepackage[T1]{fontenc}

\usepackage[dvipdf]{graphicx}	
\usepackage{amsmath}	
\usepackage{amssymb}	
\usepackage{pdflscape}
\usepackage{color}

\title[Dust formation and mass loss around intermediate-mass AGB stars in the early Universe]{Dust formation and mass loss around intermediate-mass AGB stars with initial metallicity $Z_{\rm ini} \le 10^{-4}$ in the early Universe I: Effect of surface opacity on the stellar evolution and dust-driven wind}
\author[Shohei Tashibu, Yuki Yasuda and Takashi Kozasa]
       {Shohei Tashibu,$^1$\thanks{E-mail: e163053c@yokohama-cu.ac.jp}
         \thanks{present address: School of Medicine, Yokohama City University,
         Yokohama 236-0027, Japan}
  Yuki Yasuda,$^2$
  Takashi Kozasa,$^{2,3}$ \\
  $^1$Department of Astronomy, The University of Tokyo, 7-3-1 Hongo,
  Bunkyo-ku, Tokyo 113-0033, Japan\\
  $^2$Division of Earth and Planetary Sciences, Faculty of Science,
  Hokkaido University, Sapporo 060-0810, Japan\\
  $^3$Department of Cosmosciences, Graduate School of Science,
  Hokkaido University, Sapporo 060-0810, Japan}

\date{Accepted XXX. Received YYY; in original form ZZZ}

\pubyear{2016}

\begin{document}
\label{firstpage}
\pagerange{\pageref{firstpage}--\pageref{lastpage}}
\maketitle

\begin{abstract}
Dust formation and resulting mass loss
  around Asymptotic Giant Branch (AGB) stars with
  initial metallicity in the range of
    $0 \leq Z_{\rm ini} \leq 10^{-4}$ and initial mass
    $2\leq M_{\rm ini}/M_{\sun} \leq 5$ 
  are  explored by the hydrodynamical calculations of 
  dust-driven wind (DDW) along the AGB evolutionary tracks.
   We employ the MESA code to simulate the evolution of stars, assuming
  an empirical mass-loss rate in
  the post-main sequence phase, and considering the three types of low-temperature
  opacities (scaled-solar, CO-enhanced, and 
  CNO-enhanced opacities) to elucidate the effect on the stellar evolution and the DDW. 
  We find that the treatment of low-temperature opacity strongly affects the dust formation and 
  resulting DDW; in 
  the carbon-rich AGB phase, the maximum $\dot{M}$ of  
   $M_{\rm ini} \geq$ 3 $M_{\sun}$  star with the CO-enhanced opacity  is at least one order of 
  magnitude smaller than that with the CNO-enhanced opacity. 
   A wide range of stellar parameters being covered,  
  a necessary  condition for driving efficient DDW with $\dot{M} \ge 10^{-6}$ $M_{\sun}$ yr$^{-1}$ 
   is expressed as the effective temperature   
  $T_{\rm eff} \la 3850$ K and
    $\log(\delta_{\rm C}L/\kappa_{\rm R} M)
    \ga 10.43\log T_{\rm eff}-32.33 $
  with the carbon excess $\delta_{\rm C}$ defined as
  $\epsilon_{\rm C} - \epsilon_{\rm O}$ and  the Rosseland mean 
  opacity $\kappa_{\rm R}$ in units of cm$^2$g$^{-1}$ in the
  surface layer,  and the stellar mass (luminosity) $M$ $(L)$
  in solar units.
  The derived fitting formulae of gas and dust mass-loss rates in terms of input stellar parameters 
  could be useful for investigating the dust yield from AGB stars in
  the early Universe being consistent with the stellar evolution calculations.
\end{abstract}




\section{Introduction}

While the major source of interstellar dust in the early
universe at redshift $z \ga 5$ is believed to be core-collapsed supernovae 
(e.g., Todini \& Ferrara 2001; Nozawa et al. 2003), the possibility that asymptotic giant branch (AGB) stars are an important source of dust
has been suggested and investigated. Dwek et al. (2007) claimed that 
core-collapse supernovae (CCSNe) cannot reproduce the dust mass of about $4 {\bf \times} 10^8$ $M_{\sun}$ in
the high-redshift ($z$ = 6.4) quasar J1148+5251, unless the produced dust mass
is much more than that evaluated from the observations of CCSNe in nearby
galaxies, or the dust destruction
efficiency is much less than that inferred from theoretical calculations. 
Valiante et al. (2009, 2011) have shown that AGB stars can
  contribute to dust enrichment even at redshift $z < 8-10$, based on the dust
  yield of AGB stars with initial metallicity $Z_{\rm ini} =10^{-3}$
  calculated by Zhukovska, Gail \& Trieloff (2008).

So far, the investigations on dust formation around AGB
stars have suggested that AGB stars with initial metallicity $Z_{\rm ini} \la 10^{-3},$   
cannot be assigned as the source of  Si-bearing dust such as silicate, since the abundance of silicon being 
scaled by the initial metallicity is 
 so small to prevent the formation of Si-bearing dust in the winds (Di Criscienzo et al.  2013).  
Thus, only carbon dust is expected to form around AGB stars with initial metallicity  
$Z_{\rm ini} \lesssim 10^{-3}$, owing to the progressive enrichment
 of carbon of the surface regions, favoured by repeated Third Dredge-Up (TDU) events. 

The upper limit on the initial mass of the star to have 
production of carbon dust during the AGB phases decreases with 
decreasing $Z_{\rm ini}$. This is because the core mass of the star is higher
the lower is the metallicity and when the core mass is above a given 
threshold ($\sim$0.8 $M_{\sun}$) the stars experience hot bottom burning  
(HBB, Renzini \& Voli 1981), with the destruction of the surface carbon.
Di Criscienzo et al. (2013) have inferred 
that only low-mass stars of initial mass $M_{\rm ini} \lesssim 1$ $M_{\sun}$ with
$Z_{\rm ini} \leq 10^{-4}$ can produce carbon dust significantly, and that AGB 
stars cannot be considered as important dust manufacturers at 
$Z_{\rm ini} < 10^{-4}$: this conclusion was based on their calculation for 
$Z_{\rm ini} = 3.0 \times 10^{-4}$ and on stellar evolution calculations with 
$Z_{\rm ini} \la  2\times 10^{-5}$ by Campbell \& Lattanzio (2008).
On the other hand,  Constantino et al. (2014) confirmed 
that the mass threshold for HBB is different between models with and without
composition-dependent low-temperature opacity. Also, 
even if stars experience HBB, the stars could become carbon-rich after the 
cease of HBB, depending not only on the treatment of
 low-temperature opacity but also 
on the initial mass as well as the mass-loss rate during the evolution 
(e.g., Ventura \& Marigo, 2010;  Nanni et al. 2013).
Thus, the pros and cons of formation of carbon dust in the AGB with
$Z_{\rm ini} \leq 10^{-4}$ have yet to be
explored by 
  investigating how the treatment of low-temperature opacity affects the stellar
  evolution and dust formation.

Formation of dust and the resulting mass loss around AGB stars 
  are not only determined by the abundances of
dust-forming elements in the surface layer, but also
  sensitively depend on the effective temperature (see Gail \& Sedlmayr 2013
and references therein). In this context, the most relevant
input on the
stellar evolution and the dust formation is the low-temperature
opacity. During this last decade, it has been emphasized that the
low-temperature opacity varying with the change of surface
elemental composition due to the TDU and HBB during the thermally-pulsing AGB (TP-AGB)
phase strongly affects the 
evolution of stars with $Z_{\rm ini} \geq 10^{-4}$ 
(e.g., Marigo 2002; Cristallo et al. 2007;
Ventura \& Marigo 2010; Constantino et al. 2014;
Fishlock, Karakas \& Stancliffe 2014). In particular,
these authors have demonstrated that the composition-dependent
low-temperature opacity makes the effective temperature drastically decrease 
in carbon-rich (C-rich) stars,  in comparison with the scaled-solar opacity.  
Thus, it can be expected that the treatment of low-temperature opacity directly influences the
formation of dust and the resulting dust-driven wind (DDW).

Dust formation around AGB stars is a complicated process associated with
the dynamical as well as the thermal behaviours of gas above the
photosphere; dust condenses in the high-density gas induced 
by the shock originating from the stellar pulsation, then the mass loss is driven through
the radiation pressure force acting on newly 
formed dust (the so-called pulsation-enhanced DDW, 
e.g. Fleischer, Gauger \& Sedlmayr 1992;
Winters et al. 2000). Thus, dust formation has to be treated with the 
consequent gas outflow from
AGB stars self-consistently, considering the periodic change of the
stellar properties and the corresponding wind structure
simultaneously. However, most of previous studies on the dust yields of
low-metallicity AGB stars (e.g., Ventura et al. 2012a, 2012b, 2014;
Di Criscienzo et al. 2013; Nanni et al. 2013) have followed the scheme
developed by Ferrarotti \& Gail (2006), without
solving the formation processes of dust grains and the resulting
density structure of outflowing gas 
self-consistently; the dust yield has been evaluated from the calculations of
dust growth in a stationary wind, given the number density of dust seed
particles and the mass-loss rate. Thus, the derived properties of newly formed
dust, such as the amount and the size, may suffer 
ambiguities inherent in the treatment.
Although the self-consistent hydrodynamical calculation of DDW for
C-rich AGB stars with subsolar metallicities has been carried out
(Wachter et al. 2008), to our knowledge, so far no attempt has been done for  
 AGB stars with $Z_{\rm ini} \leq 10^{-4}$ in the early Universe. 

In order to explore whether AGB stars can produce and supply carbon dust in
the early universe, first we simulate the evolution of stars
whose initial mass ranges from 2 to 5 $M_{\sun}$ with initial metallicity
$Z_{\rm ini} \leq 10^{-4}$. In the simulations, three types of low-temperature opacities 
(scaled-solar, CO-enhanced, and CNO-enhanced opacities)  
are  considered to clarify how the
  treatment of low-temperature opacity influences the stellar parameters
  related to dust formation during the AGB-phase.
  Then, applying the calculated stellar parameters along
  the evolutionary track of TP-AGB to the hydrodynamical model of
  pulsation-enhanced DDW, we investigate the dependence of 
  the properties of dust and DDW produced in  the TP-AGB 
  phase on the treatment of low-temperature (surface) opacity as well as 
  on the initial mass and metallicity. In addition, we evaluate
  a necessary
    condition for realizing DDW, and
  derive the fitting formulae
    for gas and dust mass-loss rates caused by DDW in terms of
  the input 
stellar parameters.

This paper is organized as follows. In Section 2, we briefly address the
stellar evolution model, focusing on the tools implemented in the Modules
for Experiments in Stellar Astrophysics (MESA) code
(Paxton et al. 2011, 2013), and introduce the hydrodynamical model of
pulsation-enhanced DDW used in this study. 
Section 3 provides the results of stellar evolution
  calculations and shows how the stellar parameters 
  (e.g., effective temperature and elemental composition in surface layer)
    controlling the dust formation and the mass loss during theTP-AGB phase are 
    affected by the treatment of
  low-temperature opacity. 
 Then, the dependences of the dust formation and  resulting mass loss during the
  C-rich AGB phase on the low-temperature opacity as well as on the initial
  mass and metallicity are presented in Section 4. In Section 5,   
  a necessary condition for producing the efficient 
  DDW with
  mass-loss rate $\dot{M} \geq 10^{-6}$ $M_{\sun}$ yr$^{-1}$ and the formulae of
  gas and dust mass-loss rates are 
  derived, and the implication on the  evolution of AGB stars and
  dust formation in the early Universe discussed.  The summary is presented 
  in Section 6. The input stellar parameters of the hydrodynamical calculations 
  as well as the derived properties of DDW with $\dot{M} \ge 10^{-7}$ $M_{\sun}$ yr$^{-1}$ are tabulated 
  for the models with the CO-enhanced and CNO-enhanced opacities in Appendix A.

\section{The models}
The formation of carbon dust and the resulting mass loss
around AGB stars  with
$Z_{\rm ini} \leq 10^{-4}$ after the C-rich star stage is reached are investigated through two separate steps;
first, given the initial mass and metallicity, the stellar evolution is simulated
from the pre-main-sequence to the end of the AGB phase. Second, the stellar parameters at roughly every $0.05$ $M_{\sun}$
along the evolutionary track on TP-AGB are applied to the hydrodynamical
model of pulsation-enhanced DDW to evaluate the formation of
carbon dust and resulting DDW in the C-rich AGB phase. Here we
briefly describe the models used in these two steps.

\subsection{Stellar evolution}

We employ the MESA code for the calculation of stellar evolution models, evolved from the pre-main sequence up to the end of AGB phase of stars with 
initial masses $M_{\rm ini}\,=\,2,\,3,\,4$ and 5 $M_{\sun}$ and metallicities 
$Z_{\rm ini}\,=\,0,\,10^{-7},\,10^{-6},\,10^{-5}$ and $10^{-4}$. In addition to the treatment of convection, 
we describe the low-temperature opacities and the mass-loss formula implemented in the MESA code for the purpose 
of the present study in the following subsections.

\subsubsection{Convection} 
In the calculations,
the standard mixing length theory (MLT; Cox \& Giuli 1968) is applied to treat
convection as a diffusive process within convective regions, 
 defined according to the
Schwarzschild criterion, $\nabla_{\rm ad} < \nabla_{\rm rad}$ , where $\nabla_{\rm ad}$ and $\nabla_{\rm rad}$ are the adiabatic and the radiative temperature gradient, respectively. In the convective region involving the nuclear burning, MESA solves the coupled structure, burning and mixing equations as detailed in Paxton et al. (2011, 2013). The mixing length parameter 
$\alpha_{\rm MLT}$ is set to be 2.0 as a standard value to reproduce the evolution of the Sun. 
Overshooting expresses the physical concept of an exponentially decaying velocity field  beyond the 
convective boundary, and the overshoot mixing is treated as a time-dependent, diffusive process with the overshoot mixing diffusion coefficient defined as $D_{\rm OV} = D_{\rm conv,0} \exp(-2z/fH_{\rm P})$, where $D_{\rm conv,0}$ is the MLT diffusion 
coefficient at the boundary, $z$ is the distance from the boundary, $f$ is a free parameter called overshooting parameter, and $H_{\rm P}$ is the pressure scale height (Herwig 2000). For overshooting parameters, we adopt $f$ = 0.014 at all convective boundaries, except for the bottom of the He-shell flash region at which $f$ is set to be 0.008 throughout the evolution after the first thermal pulse (TP), referring to Paxton et al. (2011); note that we adpot $f=0.014$
at the bottom of convective envelope instead of $f=0.126$ since we consider that the formation of $^{13}$C pocket is not relevant to the purpose of this paper.  We note that the MLT scheme leads to a less efficient HBB than the full spectrum of turbulence (FST) scheme (Canuto \& Mazzitelli 1991), as discussed by Ventura \& D'Antona (2005).
Thus, if the FST scheme was applied,  less carbon dust would be formed since carbon burning by HBB would be much stronger.

\vspace{-0.3cm}
\subsubsection{Low-temperature opacity}
In order to
clarify the role played by low-temperature opacities not only for what regards stellar evolution but also on the formation of carbon dust and resulting mass loss 
during AGB phase, we implement to the MESA code three types of 
low-temperature opacities;
(1) the scaled-solar opacity, with the elemental composition of metals 
scaled by the solar composition 
  (Grevesse \& Noels 1993) according to the
initial metallicity;
(2) the CO-enhanced opacity, in which the opacity is calculated according
to the enhancement of C and O abundances with respect to the scaled-solar values;
(3) the CNO-enhanced opacity also includes the variation of N with respect to the
scaled-solar value, besides the variations of C and O abundances. Since the CN
molecule dominates the Rosseland mean opacity at
low-temperature $\log T \la 3.6$
(Marigo \& Aringer 2009), the CNO-enhanced opacity is, among the three possibilities, the most suitable one to describe the evolution of AGB stars.

The opacity tables are constructed by using the \AE SOPUS tool
(Marigo \& Aringer 2009) and are incorporated into the MESA code; the tables
consist of five grids of metallicity ($Z = 10^{-12}, 10^{-7}, 10^{-6}, 10^{-5}$ and $10^{-4}$), three grids of
the mass fraction of
hydrogen ($X_{\rm H} =$ 0.50, 0.65 and 0.80), and 16 grids of the increment of the mass fraction ($0 \leq dX_i \leq 9.72 \times 10^{-2}$) for
element $i$ ($i =$ C, N, and O). The grids of temperature $T$ (in units of Kelvin) and of the parameter
$R$ $=\rho/(T/10^6)^3$\,  with  gas density $\rho$ in c.g.s units, cover the ranges of $3.20 \leq \log T \leq 4.50$ and
$-8.0 \leq \log R \leq 1.0$, respectively. Note that we adopt
the opacity tables for $Z = 10^{-12}$ as a representative of $Z = 0$. 
For high-temperature opacity ($\log T \geq 4.0$), we adopt the OPAL type 2
opacities (Iglesias \& Rogers 1993, 1996) implemented in the MESA code,
accounting for the enhancement of carbon and oxygen during the evolution.
The opacity in the range of $3.8 \leq \log T \leq 4.0$ is calculated by linearly interpolating between 
the low- and high-temperature opacities  tables at a fixed  $\log R$.

\subsubsection{Mass loss}

Mass loss during the evolution of low- and intermediate-mass stars plays a
decisive role in determining the efficiency of TDU and HBB 
(e.g., Weiss \& Ferguson 2009). In previous studies
empirical and/or theoretical mass-loss formulae have been applied;
for example, Weiss \& Ferguson (2009) applied 
the Reimers formula (Reimers 1975) on the Red Giant
Branch (RGB) and the AGB with pulsation periods $P \leq 400$ days,  and then 
the formulae proposed by van Loon et al. (2005) for oxygen-rich (O-rich) AGB
and Wachter et al. (2002) for C-rich AGB; the formulae by 
Vassiliadis \& Wood (1993) and Bl\"ocker (1995)  were adopted for the whole AGB phase 
in Ferrarotti \& Gail (2006) and Ventura et al. (2012a, 2012b), respectively. 
Although the DDW
mechanism has been believed to be plausible for C-rich AGB stars in the
galaxies with solar and subsolar metallicities at present time, we have
no convincing knowledge on whether the intermediate-mass stars  
in the early universe, with $Z_{\rm ini} \leq 10^{-4}$ can
form carbon dust so efficiently as to drive the
mass loss on the AGB.  

 The aim of this paper is to reveal whether and in what conditions 
     the formation of carbon dust and the consequent DDW onset on 
     the AGB, as the first step to explore the 
     role of AGB stars as the source 
    of dust in the early universe. 
    Thus, we apply an  empirical mass-loss formula in the post-main sequence 
    phase for simplicity. In this paper we adopt the formula by Schr\"oder \& Cuntz  
    (2005, hereinafter SC05) since the formula reasonably reproduces the mass-loss rate 
    on the RGB; although the recent 
    population synthesis model of AGB-stars in metal-poor galaxies prefers a modified SC05 
    as the mass-loss formula on the AGB before the onset of DDW, the application on 
    the RGB is 
    questionable (Rosenfield et al. 2014). The formula by SC05 is given by
\begin{eqnarray}
  \dot{M}_{\rm SC05} =
  \eta \frac{L R}{M}\left( \frac{T_{\rm eff}}{4000 K} \right)^{3.5}
  \left(1+ \frac{1}{4300g}\right) ,
\end{eqnarray}
where the mass-loss rate is in units of $M_{\sun}$ yr$^{-1}$, the effective
temperature $T_{\rm eff}$ is in units of Kelvin ,
  and the stellar mass $M$, luminosity $L$ and
  surface gravity $g$ are in solar units.  In the calculations, we adopt
  the fitting parameter $\eta = 8 \times 10^{-14}$;
  the value of $\eta$ is adjusted by fitting
to the observed mass-loss rates of red giant stars in globular clusters with
different metallicities (SC05), and with which the formula well  
reproduce the observed mass-loss
rates of the galactic giants and supergiants (Schr\"oder \& Cuntz 2007). 

\subsection{Dust-driven wind}

Here, we describe the hydrodynamical model of pulsation-enhanced
DDW by Yasuda \& Kozasa (2012) employed in this paper. 
The hydrodynamical model treats self-consistently the nucleation and
growth processes of carbon dust in the gas lifted up by the
pulsation shock and the consequent DDW. The
model adopts the scheme for the formation of carbon grains proposed
by Gauger, Gail \& Sedlmayr (1990) and includes
the decay process of dust by heating due to backward
radiation. Given the stellar parameters at the photosphere (see below) at a 
given epoch during the AGB phase, the model allows evaluating the physical 
quantities related to dust formation: the time evolution of mass-loss rate, 
gas velocity, condensation efficiency of carbon dust (defined as the ratio of 
carbon locked into dust to carbon available for dust formation) as well as the 
amount and the grain size distribution of dust particles in the wind, together 
with their time-averaged values at the outer boundary of the hydrodynamical 
model.

The stellar parameters necessary for the hydrodynamical model
calculation are the current stellar mass $M$, luminosity $L$, effective
temperature $T_{\rm eff}$, the abundances of H, He, C, N and O, the  
Rosseland mean opacity $\kappa_{\rm R}$
in the photosphere, the period $P_0$ and the velocity
  amplitude $\Delta u_{\rm p}$ of pulsation.
The temporal evolution of the stellar parameters, except for the
period and the amplitude of pulsation, can be obtained from
the stellar evolution calculation. For the pulsation period, we apply
the formula for the fundamental radial mode of pulsation by
Ostlie \& Cox (1986), which is given by
\begin{eqnarray}
\log P_0 = - 1.92 - 0.73\log M + 1.86\log R,
\end{eqnarray}
where $P_{\rm 0}$ is in units of day, and the stellar mass $M$ and stellar
radius $R$ are in solar units.
As for the velocity amplitude of pulsation, little is known as 
mentioned in Gail \& Sedlmayr (2013), though Wood (1986) have estimated to be 
a few km s$^{-1}$, by adjusting the  variable pressure at the inner boundary 
to  a certain photospheric density, deduced from observations.
Thus, in this paper, $\Delta u_{\rm p}$  
is set to be 2 km s$^{-1}$ as a reference value.
We address the reader to Yasuda \& Kozasa (2012) for the details on 
the numerical
schemes for the formation process of carbon grains and the dust-driven wind.

In the calculations, we use the optical constants of astronomical graphite
(Draine 1985) for carbon dust. The physical quantities characterizing the dust 
formation and resulting DDW presented in the following
sections are specified by the values averaged over
the final 60 pulsation cycles at the outer boundary 
placed at 25 times the initial stellar radius of the hydrodynamical model.
Note that we avoid the calculations at current stellar  
  mass in the short time intervals of TPs during the
  evolution, since
  the convergence problems often occur just after the onset of TP in the
  stellar evolution calculations,  
  as presented in the following subsection.
  
  \subsection{Convergence problem}
During the stellar evolution calculation, we often come across 
convergence problems
just after the onset of TP when the effective temperature
decreases down to $\sim 3200$ K. The failed convergency is not caused by the dominance
of radiation pressure in the convective envelope, 
i.e. the small ratio of gas
pressure to total pressure, $\beta$, but may be associated with the opacity, 
as argued by Karakas \& Lattanzio (2007). While the investigation of the
precise cause of the convergence problems is postponed to a future work, in the
present calculations we avoid convergence difficulties
as follows: we set a minimum temperature ($T_{\rm min,op}$)
from just before the onset of  TP to the beginning of
  TDU, the low-temperature opacity is replaced with that calculated by using
  $T_{\rm  min, op}$ in a region where the local temperature is lower than
  $T_{\rm min,op}$.
  
  This method might affect the TDU
efficiency parameter $\lambda$, defined as the ratio between the mass dredged up
 after a thermal pulse, $\Delta M_{\rm DUP}$, and the increment of core mass during the preceding interpulse phase, 
$\Delta M_{\rm c}$ (e.g., Herwig 2005); in the calculations of model stars with
the CNO-enhanced opacity, for example,
the largest difference of $\lambda$ between
successive TDUs with and without the convergence problem is 0.24 for
$M_{\rm ini} =$ 5 $M_{\sun}$, with $Z_{\rm ini} = 10^{-4}$. Such a degree of
difference can be seen between successive TDUs with no convergence problem.
By setting the different values of $\log T_{\rm min,op}$
for the successive TDUs at $M=$ 1.69 $M_{\sun}$ and 1.60 $M_{\sun}$ during the
evolution of $M_{\rm ini}=$ 3 $M_{\sun}$ with $Z_{\rm ini}=10^{-7}$, 
  the largest difference of $\lambda$ reaches 0.404.  However, 
  only two models 
  ($M_{\rm ini}=$ 3 $M_{\sun}$ with 
  $Z_{\rm ini}=10^{-7}$ and $M_{\rm ini}=$ 4 $M_{\sun}$ with $Z_{\rm ini}=0$) 
    suffer such a larger increment at a TDU among
the last few TDUs before the evolution calculation finally stops. 
In addition, the effective temperature quickly changes in a short time 
interval of TP. Thus, we consider that the prescription for dealing with
the convergence problem could not cause a serious problem against the aim of
this paper.

\section{Evolution of low-metallicity AGB stars}
Table 1 summarizes the stellar models and the calculated
quantities characterizing their TP-AGB phase: the initial metallicity
$Z_{\rm ini}$; the type of low-temperature opacity;
the final stellar (core) mass $M_{\rm tot,f}$ ($M_{\rm c,f}$); the total number of TP;
the total number of TDU; the threshold stellar
mass $M_{{\rm C/O}>1}$, defined as the mass of the star at the time after which
C/O-ratio in the surface layer keeps exceeding unity; the maximum temperature
at the bottom of convective envelope during the TP-AGB phase  
$T_{\rm bce,max}$; the minimum effective temperature during the interpulse
phases $T_{\rm eff, min}$;  the final carbon excess (defined as
$\delta_{\rm C} \equiv \epsilon_{\rm C}-\epsilon_{\rm O}$ with
$\epsilon_{\rm C}$ ($\epsilon_{\rm O}$) the abundance of C (O) by number
relative to H) as well as the final mass fractions of C, N, and O in the surface
layer. Note that "final" does not always mean the end of AGB phase. The models
that fail to evolve to the end of TP-AGB phase due to the convergence problems
(see below) are denoted by the superscript $f$. Also, the models undergoing HBB (week HBB) 
  during the evolution are
  specified by the superscript H (WH) 
  attached to the value of $T_{\rm bce,max}$.
\begin{table*}
\begin{center}
\caption{Models and characteristic quantities during AGB phase: 
    the initial metallicity $Z_{\rm ini}$, 
    the type of low-temperature opacity; $\kappa_{\rm solar}$, $\kappa_{\rm CO}$, 
    and $\kappa_{\rm CNO}$ denote the scaled-solar, the 
    CO-enhanced, and the CNO-enhanced opacities, respectively, 
    the final stellar mass $M_{\rm tot,f}$ in solar units, 
    the final core mass $M_{\rm c,f}$ in solar units with the superscript
    $f$ for the models that fail to evolve to the end of AGB phase,  
    the total number of TP $N_{\rm TP}$, the total number of TDU $N_{\rm TDU}$, 
    the stellar mass below which C/O$>$1 $M_{\rm C/O>1}$ in solar units,
    the maximum temperature at the bottom of convective envelope during
    interpulse phases $T_{\rm bce,max}$ in units of 10$^6$ K with the
    superscript H (WH) for the models that experience HBB (week HBB),  
    the minimum effective temperature during interpulse phases
    $T_{\rm eff,min}$ in units of K, the final carbon excess
    $\delta_{\rm C,f} \times 10^4$, and the mass fractions of C, N, and O in
    the surface layer $X_{\rm f}$(C), $X_{\rm f}$(N), and $X_{\rm f}$(O),
    respectively.}
\label{tab2}
   \begin{tabular}{cllcccclccccc}\hline \hline
      $Z_{\rm ini}$ & $\kappa$ & $M_{\rm tot,f}$ & $M_{\rm c,f}$ & $N_{\rm TP}$ & $N_{\rm TDU}$ & $M_{\rm C/O>1}$ & $T_{\rm bce,max}$ & $T_{\rm eff,min}$ & 
      $\delta_{\rm C,f}\times10^4$ & $X_{\rm f}$(C) & $X_{\rm f}$(N) & $X_{\rm f}$(O)\\ \hline
\multicolumn{2}{c}{$M_{\rm ini} = 2.0 M_{\sun}$}&&&&&&&&&&& \\ \hline
      $10^{-4}$ & $\kappa_{\rm solar}$ & 0.695 & 0.695 & 20 & 19 & 1.88 & 10.0 & 3927 & 49.4 & 4.20${\bf \times}10^{-2}$ & 1.68${\bf \times}10^{-4}$ & 4.74${\bf \times}10^{-3}$ \\
      $10^{-4}$ & $\kappa_{\rm CO}$ & 0.692 & 0.692 & 17 & 16 & 1.88 & 7.24 & 3254 & 22.0 & 2.02${\bf \times}10^{-2}$ & 5.22${\bf \times}10^{-5}$ & 2.29${\bf \times}10^{-3}$ \\
      $10^{-4}$ & $\kappa_{\rm CNO}$ & 0.690$^f$ & 0.690 & 17 & 16 & 1.88 & 6.92 & 3136 & 25.5 & 2.32${\bf \times}10^{-2}$ & 5.97${\bf \times}10^{-5}$ & 2.63${\bf \times}10^{-3}$ \\ 
      $10^{-5}$ & $\kappa_{\rm solar}$ & 0.698 & 0.698 & 17 & 16 & 1.84 & 10.0 & 3974 & 35.8 & 3.12${\bf \times}10^{-2}$ & 3.47${\bf \times}10^{-4}$ & 3.28${\bf \times}10^{-3}$ \\
      $10^{-5}$ & $\kappa_{\rm CO}$ & 0.695 & 0.695 & 15 & 14 & 1.84 & 7.60 & 3439 & 24.1 & 2.16${\bf \times}10^{-2}$ & 3.15${\bf \times}10^{-4}$ & 2.21${\bf \times}10^{-3}$ \\
      $10^{-5}$ & $\kappa_{\rm CNO}$ & 0.689$^f$ & 0.689 & 13 & 12 & 1.84 & 6.04 & 2939 & 22.0 & 2.00${\bf \times}10^{-2}$ & 2.94${\bf \times}10^{-4}$ & 2.09${\bf \times}10^{-3}$ \\
      $10^{-6}$ & $\kappa_{\rm solar}$ & 0.707 & 0.707 & 17 & 17 & 1.83 & 10.0 & 3980 & 25.7 & 2.25${\bf \times}10^{-2}$ & 3.59${\bf \times}10^{-4}$ & 2.27${\bf \times}10^{-3}$  \\
      $10^{-6}$ & $\kappa_{\rm CO}$ & 0.703 & 0.703 & 16 & 16 & 1.83 & 8.07 & 3575 & 20.2 & 1.79${\bf \times}10^{-2}$ & 3.96${\bf \times}10^{-4}$ & 3.02${\bf \times}10^{-3}$ \\
      $10^{-6}$ & $\kappa_{\rm CNO}$ & 0.699$^f$ & 0.699 & 14 & 14 & 1.83 & 6.17 & 3059 & 14.5 & 1.31${\bf \times}10^{-2}$ & 3.26${\bf \times}10^{-4}$ & 1.33${\bf \times}10^{-3}$ \\
      $10^{-7}$ & $\kappa_{\rm solar}$ & 0.712 & 0.712 & 20 & 18 & 1.81 & 10.4 & 3984 &  23.5 & 2.04${\bf \times}10^{-2}$ & 2.08${\bf \times}10^{-4}$ & 2.11${\bf \times}10^{-3}$ \\
      $10^{-7}$ & $\kappa_{\rm CO}$ & 0.711 & 0.711 & 17 & 16 & 1.81 & 8.45 & 3627 & 18.8 & 1.65${\bf \times}10^{-2}$ & 2.11${\bf \times}10^{-4}$ & 1.72${\bf \times}10^{-3}$ \\
      $10^{-7}$ & $\kappa_{\rm CNO}$ & 0.707$^f$ & 0.707 & 17 & 14 & 1.81 & 6.79 & 3153 & 15.0 & 1.34${\bf \times}10^{-2}$ & 1.98${\bf \times}10^{-4}$ & 1.42${\bf \times}10^{-3}$ \\
      $0$ & $\kappa_{\rm solar}$ & 0.727 & 0.727 & 26 & 16 & 1.76 & 10.7 & 3990 & 21.4 & 1.79${\bf \times}10^{-2}$ & 2.70${\bf \times}10^{-4}$ & 1.72${\bf \times}10^{-3}$ \\
      $0$ & $\kappa_{\rm CO}$ & 0.726 & 0.726 & 25 & 14 & 1.76 & 9.02 & 3695 & 14.5 & 1.24${\bf \times}10^{-2}$ & 2.23${\bf \times}10^{-4}$ & 1.18${\bf \times}10^{-3}$ \\
      $0$ & $\kappa_{\rm CNO}$ & 0.723 & 0.723 & 23 & 13 & 1.76 & 7.43 & 3272 & 12.6 & 1.08${\bf \times}10^{-2}$ & 2.21${\bf \times}10^{-4}$ & 1.03${\bf \times}10^{-3}$\\ \hline
\multicolumn{2}{c}{$M_{\rm ini} = 3.0 M_{\sun}$}&&&&&&&&&&& \\ \hline
      $10^{-4}$ & $\kappa_{\rm solar}$ & 0.823 & 0.823 & 29 & 27 & 1.64 & 74.8$^{\rm H}$ & 3912 & 16.1 & 1.56${\bf \times}10^{-2}$ & 1.84${\bf \times}10^{-2}$ & 4.05${\bf \times}10^{-3}$\\
      $10^{-4}$ & $\kappa_{\rm CO}$ & 0.828 & 0.827 & 29 & 27 & 1.67 & 74.3$^{\rm H}$ & 3521 & 13.7 & 1.35${\bf \times}10^{-2}$ & 1.75${\bf \times}10^{-2}$ & 3.50${\bf \times}10^{-3}$\\
      $10^{-4}$ & $\kappa_{\rm CNO}$ & 1.121$^f$ & 0.828 & 28 & 26 & 2.75 & 56.9$^{\rm WH}$ & 2693 & 21.6 & 1.95${\bf \times}10^{-2}$ & 1.49${\bf \times}10^{-3}$ & 2.20${\bf \times}10^{-3}$\\
      $10^{-5}$ & $\kappa_{\rm solar}$ & 0.820 & 0.819 & 28 & 26 & 1.56 & 73.6$^{\rm H}$ & 3960 & 19.9 & 1.87${\bf \times}10^{-2}$ & 1.90${\bf \times}10^{-2}$ & 4.44${\bf \times}10^{-3}$\\
      $10^{-5}$ & $\kappa_{\rm CO}$ & 0.868$^f$ & 0.818 & 29 & 26 & 1.58 & 73.2$^{\rm H}$ & 3660 & 13.1 & 1.31${\bf \times}10^{-2}$ & 1.82${\bf \times}10^{-2}$ & 3.65${\bf \times}10^{-3}$\\
      $10^{-5}$ & $\kappa_{\rm CNO}$ & 1.122$^f$ & 0.822 & 26 & 24 & 2.68 & 56.0$^{\rm WH}$ & 2698 & 22.0 & 1.97${\bf \times}10^{-2}$ & 9.89${\bf \times}10^{-4}$ & 2.09${\bf \times}10^{-3}$\\
      $10^{-6}$ & $\kappa_{\rm solar}$ & 0.815 & 0.815 & 28 & 26 & 1.66 & 72.2$^{\rm H}$ & 3969 & 19.3 & 1.78${\bf \times}10^{-2}$ & 1.87${\bf \times}10^{-2}$ & 4.15${\bf \times}10^{-3}$\\
      $10^{-6}$ & $\kappa_{\rm CO}$ & 0.817 & 0.817 & 28 & 26 & 1.70 & 71.7$^{\rm H}$ & 3685 & 17.5 & 1.62${\bf \times}10^{-2}$ & 1.77${\bf \times}10^{-2}$ & 3.61${\bf \times}10^{-3}$\\
      $10^{-6}$ & $\kappa_{\rm CNO}$ & 1.510$^f$ & 0.820 & 24 & 22 & 2.60 & 53.9$^{\rm WH}$ & 2873 & 18.2 & 1.61${\bf \times}10^{-2}$ & 5.62${\bf \times}10^{-4}$ & 1.47${\bf \times}10^{-3}$\\
      $10^{-7}$ & $\kappa_{\rm solar}$ & 0.816 & 0.816 & 28 & 27 & 1.66 & 71.5$^{\rm H}$ & 3980 & 22.4 & 1.93${\bf \times}10^{-2}$ & 1.90${\bf \times}10^{-2}$ & 3.86${\bf \times}10^{-3}$\\
      $10^{-7}$ & $\kappa_{\rm CO}$ & 0.820 & 0.819 & 28 & 27 & 1.79 & 71.1$^{\rm H}$ & 3726 & 18.4 & 1.61${\bf \times}10^{-2}$ & 1.80${\bf \times}10^{-2}$ & 3.21${\bf \times}10^{-3}$\\
    $10^{-7}$ & $\kappa_{\rm CNO}$ & 1.072$^f$ & 0.828 & 27 & 26 & 2.57 & 47.8$^{\rm WH}$ & 2883 & 22.3 & 1.88${\bf \times}10^{-2}$ & 3.55${\bf \times}10^{-4}$ & 1.77${\bf \times}10^{-3}$\\
      $0$ & $\kappa_{\rm solar}$ & 0.805 & 0.805 & 40 & 37 & 2.55 & 61.0$^{\rm WH}$ & 4011 & 16.2 & 1.30${\bf \times}10^{-2}$ & 1.49${\bf \times}10^{-2}$ & 2.35${\bf \times}10^{-3}$\\
      $0$ & $\kappa_{\rm CO}$ & 0.807 & 0.807 & 40 & 37 & 2.55 & 50.3$^{\rm WH}$ & 3543 & 27.0 & 2.04${\bf \times}10^{-2}$ & 7.72${\bf \times}10^{-4}$ & 1.88${\bf \times}10^{-3}$\\
      $0$ & $\kappa_{\rm CNO}$ & 1.195$^f$ & 0.798 & 32 & 31 & 2.55 & 31.9 & 2997 & 21.2 & 1.62${\bf \times}10^{-2}$ & 3.75${\bf \times}10^{-4}$ & 1.55${\bf \times}10^{-3}$\\ \hline
\multicolumn{2}{c}{$M_{\rm ini} = 4.0 M_{\sun}$}&&&&&&&&&&& \\ \hline
      $10^{-4}$ & $\kappa_{\rm solar}$ & 0.914$^f$ & 0.880 & 36 & 34 & 1.50 & 87.7$^{\rm H}$ & 3928 & 13.5 & 1.17${\bf \times}10^{-2}$ & 1.71${\bf \times}10^{-2}$ & 2.73${\bf \times}10^{-3}$\\
      $10^{-4}$ & $\kappa_{\rm CO}$ & 0.968$^f$ & 0.882 & 36 & 34 & 1.57 & 87.7$^{\rm H}$ & 3659 & 10.4 & 9.31${\bf \times}10^{-3}$ & 1.67${\bf \times}10^{-2}$ & 2.38${\bf \times}10^{-3}$\\
      $10^{-4}$ & $\kappa_{\rm CNO}$ & 1.237$^f$ & 0.882 & 36 & 35 & 1.90 & 87.0$^{\rm H}$ & 2868 & 7.62 & 7.04${\bf \times}10^{-3}$ & 1.41${\bf \times}10^{-2}$ & 1.83${\bf \times}10^{-3}$\\
      $10^{-5}$ & $\kappa_{\rm solar}$ & 0.878 & 0.878 & 36 & 33 & 1.56 & 87.0$^{\rm H}$ & 3968 & 15.4 & 1.30${\bf \times}10^{-2}$ & 1.68${\bf \times}10^{-2}$ & 2.66${\bf \times}10^{-3}$\\
      $10^{-5}$ & $\kappa_{\rm CO}$ & 0.918$^f$ & 0.878 & 36 & 33 & 1.64 & 86.9$^{\rm H}$ & 3768 & 11.9 & 1.04${\bf \times}10^{-2}$ & 1.65${\bf \times}10^{-2}$ & 2.31${\bf \times}10^{-3}$\\
      $10^{-5}$ & $\kappa_{\rm CNO}$ & 1.270$^f$ & 0.878 & 35 & 33 & 1.92 & 86.2$^{\rm H}$ & 2915 & 6.75 & 6.31${\bf \times}10^{-3}$ & 1.42${\bf \times}10^{-2}$ & 1.70${\bf \times}10^{-3}$\\
      $10^{-6}$ & $\kappa_{\rm solar}$ & 0.862 & 0.861 & 36 & 33 & 1.59 & 84.3$^{\rm H}$ & 3975 & 16.4 & 1.37${\bf \times}10^{-2}$ & 1.77${\bf \times}10^{-2}$ & 2.68${\bf \times}10^{-3}$\\
      $10^{-6}$ & $\kappa_{\rm CO}$ & 0.862 & 0.862 & 36 & 33 & 1.62 & 84.2$^{\rm H}$ & 3774 & 15.0 & 1.26${\bf \times}10^{-2}$ & 1.72${\bf \times}10^{-2}$ & 2.46${\bf \times}10^{-3}$\\
      $10^{-6}$ & $\kappa_{\rm CNO}$ & 1.236$^f$ & 0.859 & 35 & 33 & 1.94 & 83.3$^{\rm H}$ & 2882 & 7.42 & 6.89${\bf \times}10^{-3}$ & 1.53${\bf \times}10^{-2}$ & 1.81${\bf \times}10^{-3}$\\
      $10^{-7}$ & $\kappa_{\rm solar}$ & 0.858 & 0.857 & 37 & 32 & 1.57 & 83.5$^{\rm H}$ & 3982 & 16.4 & 1.35${\bf \times}10^{-2}$ & 1.80${\bf \times}10^{-2}$ & 2.68${\bf \times}10^{-3}$\\
      $10^{-7}$ & $\kappa_{\rm CO}$ & 0.893$^f$ & 0.857 & 37 & 32 & 1.62 & 83.4$^{\rm H}$ & 3785 & 13.0 & 1.09${\bf \times}10^{-2}$ & 1.73${\bf \times}10^{-2}$ & 2.24${\bf \times}10^{-3}$\\
      $10^{-7}$ & $\kappa_{\rm CNO}$ & 1.196$^f$ & 0.855 & 35 & 33 & 1.98 & 82.5$^{\rm H}$ & 2860 & 8.50 & 7.64${\bf \times}10^{-3}$ & 1.52${\bf \times}10^{-2}$ & 1.92${\bf \times}10^{-3}$\\
      $0$ & $\kappa_{\rm solar}$ & 0.832 & 0.832 & 44 & 41 & 1.64 & 79.4$^{\rm H}$ & 4019 & 9.16 & 7.77${\bf \times}10^{-3}$ & 2.26${\bf \times}10^{-2}$ & 2.52${\bf \times}10^{-3}$\\
      $0$ & $\kappa_{\rm CO}$ & 0.833 & 0.833 & 45 & 41 & 1.64 & 79.2$^{\rm H}$ & 3834 & 8.79 & 7.46${\bf \times}10^{-3}$ & 2.23${\bf \times}10^{-2}$ & 2.40${\bf \times}10^{-3}$\\
      $0$ & $\kappa_{\rm CNO}$ & 1.193$^f$ & 0.835 & 44 & 41 & 2.12 & 77.9$^{\rm H}$ & 2867 & 8.43 & 7.04${\bf \times}10^{-3}$ & 1.80${\bf \times}10^{-2}$ & 1.96${\bf \times}10^{-3}$\\ \hline
\end{tabular}
\end{center}
\end{table*}
\begin{table*}
\begin{flushleft}
  \textbf{Table 1 continued.} Note that the models of
  $M_{\rm ini}=5$ $M_{\sun}$ with $Z_{\rm ini}=0$ and $10^{-7}$ are excluded
  since the star does not form carbon dust
  without undergoing TDU on TP-AGB.
\end{flushleft}
\begin{center}
   \begin{tabular}{cllcccclccccc}\hline \hline
      $Z_{\rm ini}$ & $\kappa$ & $M_{\rm tot,f}$ & $M_{\rm c,f}$ & $N_{\rm TP}$ & $N_{\rm TDU}$ & $M_{\rm C/O>1}$ & $T_{\rm bce,max}$ & $T_{\rm eff,min}$ & 
      $\delta_{\rm C,f} \times10^4$ & $X_{\rm f}$(C) & $X_{\rm f}$(N) & $X_{\rm f}$(O)\\ \hline
\multicolumn{2}{c}{$M_{\rm ini} = 5.0 M_{\sun}$}&&&&&&&&&&& \\ \hline
      $10^{-4}$ & $\kappa_{\rm solar}$ & 1.133$^f$ & 0.969 & 44 & 41 & 1.53 & 101$^{\rm H}$ & 3984 & 6.59 & 6.55${\bf \times}10^{-3}$ & 1.59${\bf \times}10^{-2}$ & 2.58${\bf \times}10^{-3}$\\
      $10^{-4}$ & $\kappa_{\rm CO}$ & 1.066$^f$ & 0.971 & 45 & 36 & 1.59 & 100$^{\rm H}$ & 3777 & 7.88 & 7.48${\bf \times}10^{-3}$ & 1.53${\bf \times}10^{-2}$ & 2.64${\bf \times}10^{-3}$\\
      $10^{-4}$ & $\kappa_{\rm CNO}$ & 1.329$^f$ & 0.968 & 44 & 41 & 1.73 & 100$^{\rm H}$ & 3096 & 4.49 & 4.96${\bf \times}10^{-3}$ & 1.43${\bf \times}10^{-2}$ & 2.35${\bf \times}10^{-3}$\\
      $10^{-5}$ & $\kappa_{\rm solar}$ & 1.054$^f$ & 0.947 & 42 & 37 & 1.51 & 98.2$^{\rm H}$ & 3985 & 9.15 & 8.31${\bf \times}10^{-3}$ & 1.63${\bf \times}10^{-2}$ & 2.64${\bf \times}10^{-3}$\\
      $10^{-5}$ & $\kappa_{\rm CO}$ & 1.019$^f$ & 0.947 & 42 & 38 & 1.56 & 97.9$^{\rm H}$ & 3837 & 10.0 & 8.97${\bf \times}10^{-3}$ & 1.62${\bf \times}10^{-2}$ & 2.76${\bf \times}10^{-3}$\\
      $10^{-5}$ & $\kappa_{\rm CNO}$ & 1.242$^f$ & 0.945 & 42 & 38 & 1.77 & 97.5$^{\rm H}$ & 3073 & 5.65 & 5.77${\bf \times}10^{-3}$ & 1.50${\bf \times}10^{-2}$ & 2.40${\bf \times}10^{-3}$\\
      $10^{-6}$ & $\kappa_{\rm solar}$ & 1.040$^f$ & 0.941 & 44 & 35 & 1.51 & 98.2$^{\rm H}$ & 3985 & 9.60 & 8.63${\bf \times}10^{-3}$ & 1.62${\bf \times}10^{-2}$ & 2.60${\bf \times}10^{-3}$\\
      $10^{-6}$ & $\kappa_{\rm CO}$ & 0.994$^f$ & 0.942 & 45 & 36 & 1.56 & 98.3$^{\rm H}$ & 3849 & 11.0 & 9.61${\bf \times}10^{-3}$ & 1.59${\bf \times}10^{-2}$ & 2.67${\bf \times}10^{-3}$\\
      $10^{-6}$ & $\kappa_{\rm CNO}$ & 1.317$^f$ & 0.941 & 44 & 35 & 1.77 & 98.0$^{\rm H}$ & 3076 & 4.97 & 5.25${\bf \times}10^{-3}$ & 1.48${\bf \times}10^{-2}$ & 2.22${\bf \times}10^{-3}$\\ \hline
\end{tabular}
\end{center}
\end{table*}
\begin{figure*}
\begin{tabular}{cc}
 \begin{minipage}[b]{0.5\linewidth}
  \centering
  \includegraphics[keepaspectratio, scale=0.5]{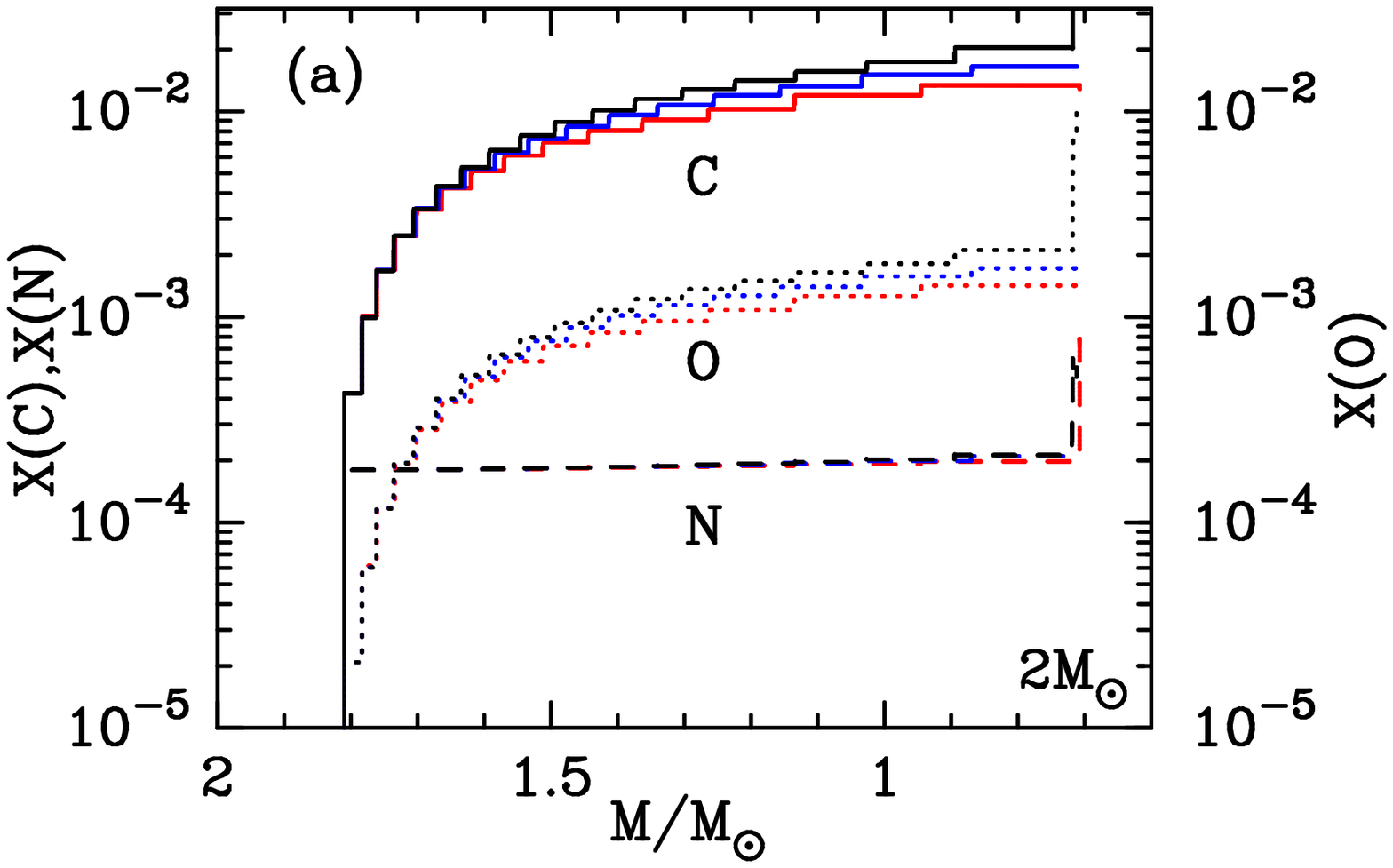}
 \end{minipage}
 \begin{minipage}[b]{0.5\linewidth}
  \centering
  \includegraphics[keepaspectratio, scale=0.5]{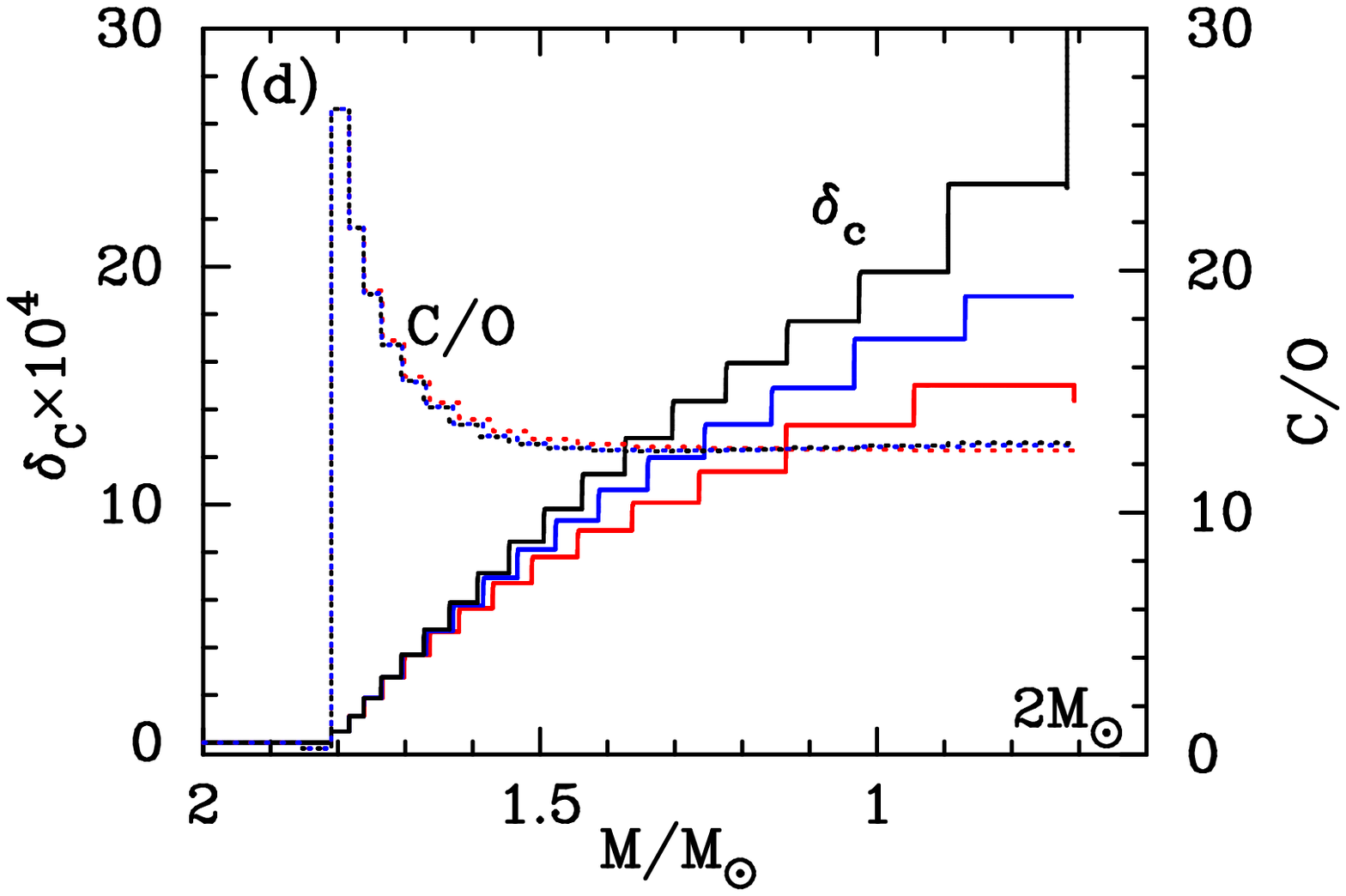}
 \end{minipage}
\end{tabular}
\begin{tabular}{cc}
 \begin{minipage}[b]{0.5\linewidth}
  \centering
  \includegraphics[keepaspectratio, scale=0.5]{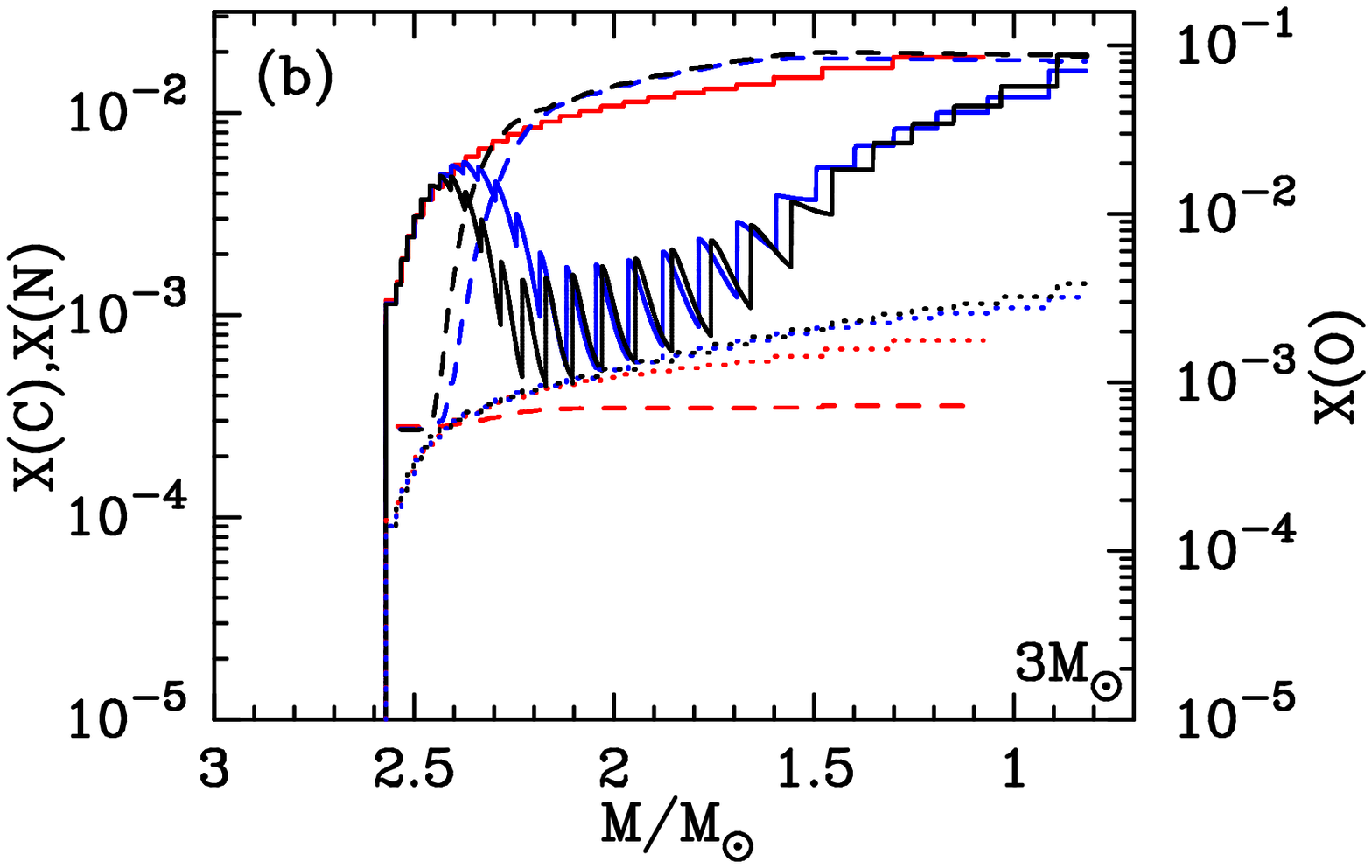}
 \end{minipage}
 \begin{minipage}[b]{0.5\linewidth}
  \centering
  \includegraphics[keepaspectratio, scale=0.5]{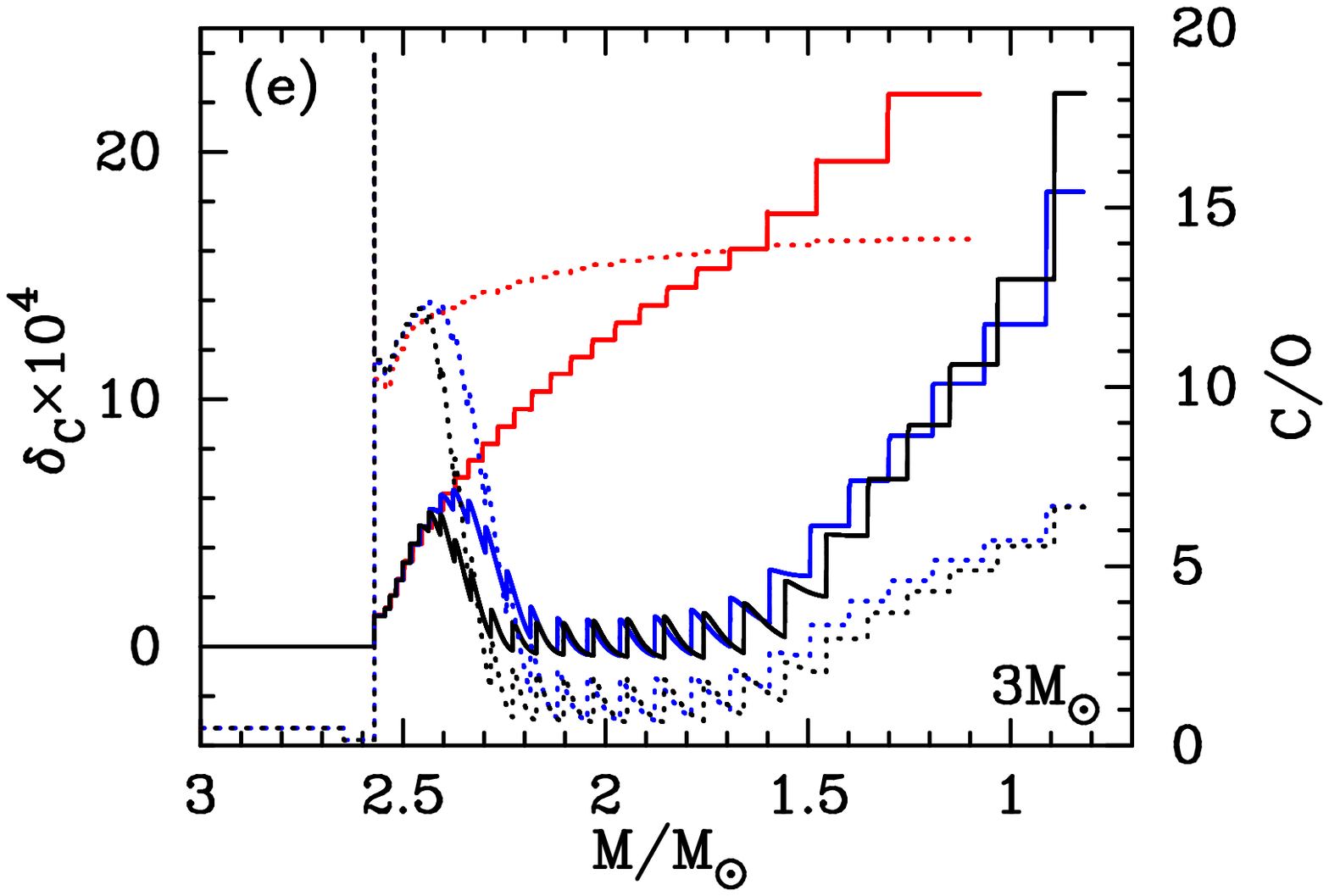}
 \end{minipage}
\end{tabular}
\begin{tabular}{cc}
 \begin{minipage}[b]{0.5\linewidth}
  \centering
  \includegraphics[keepaspectratio, scale=0.5]{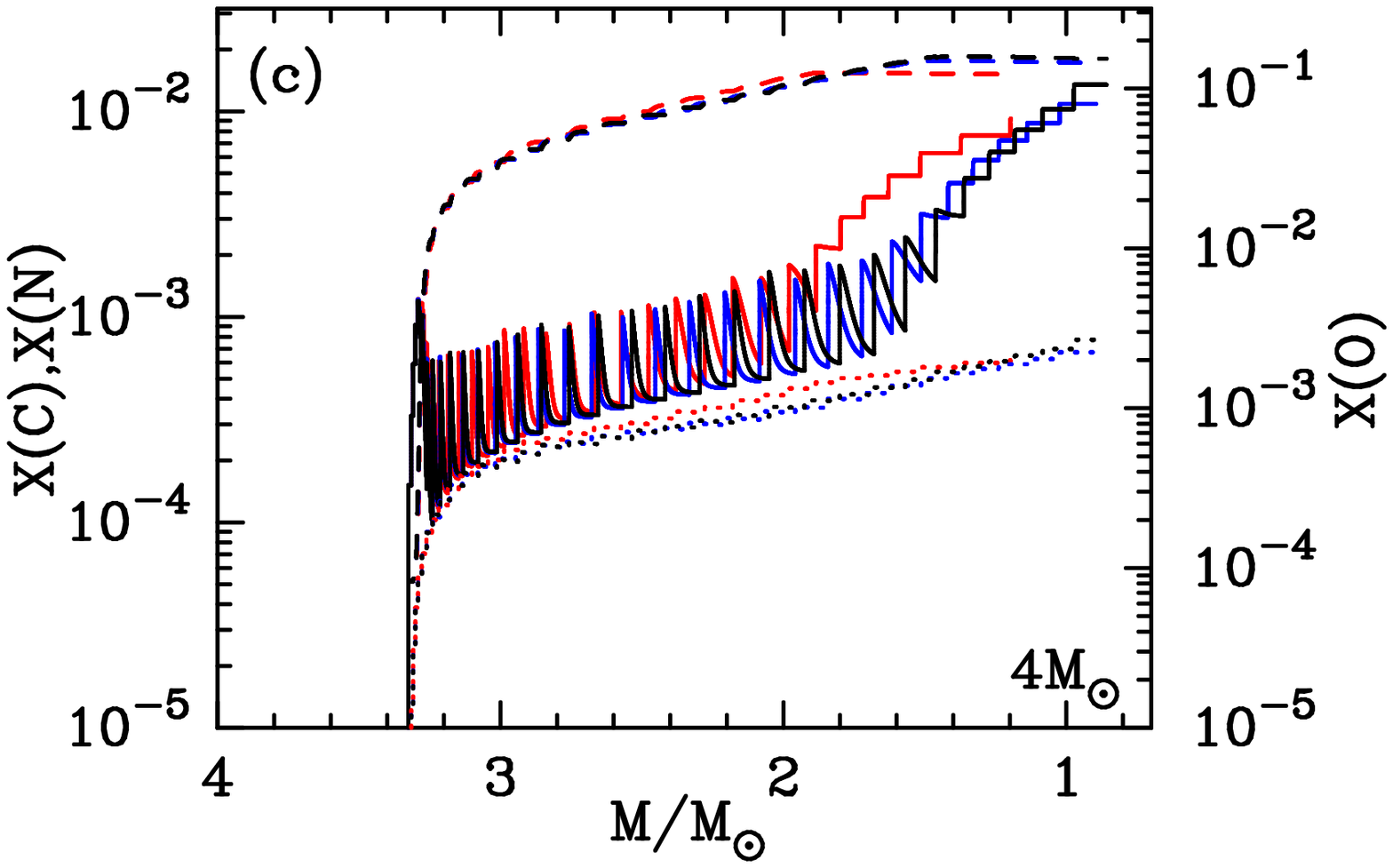}
 \end{minipage}
 \begin{minipage}[b]{0.5\linewidth}
  \centering
  \includegraphics[keepaspectratio, scale=0.5]{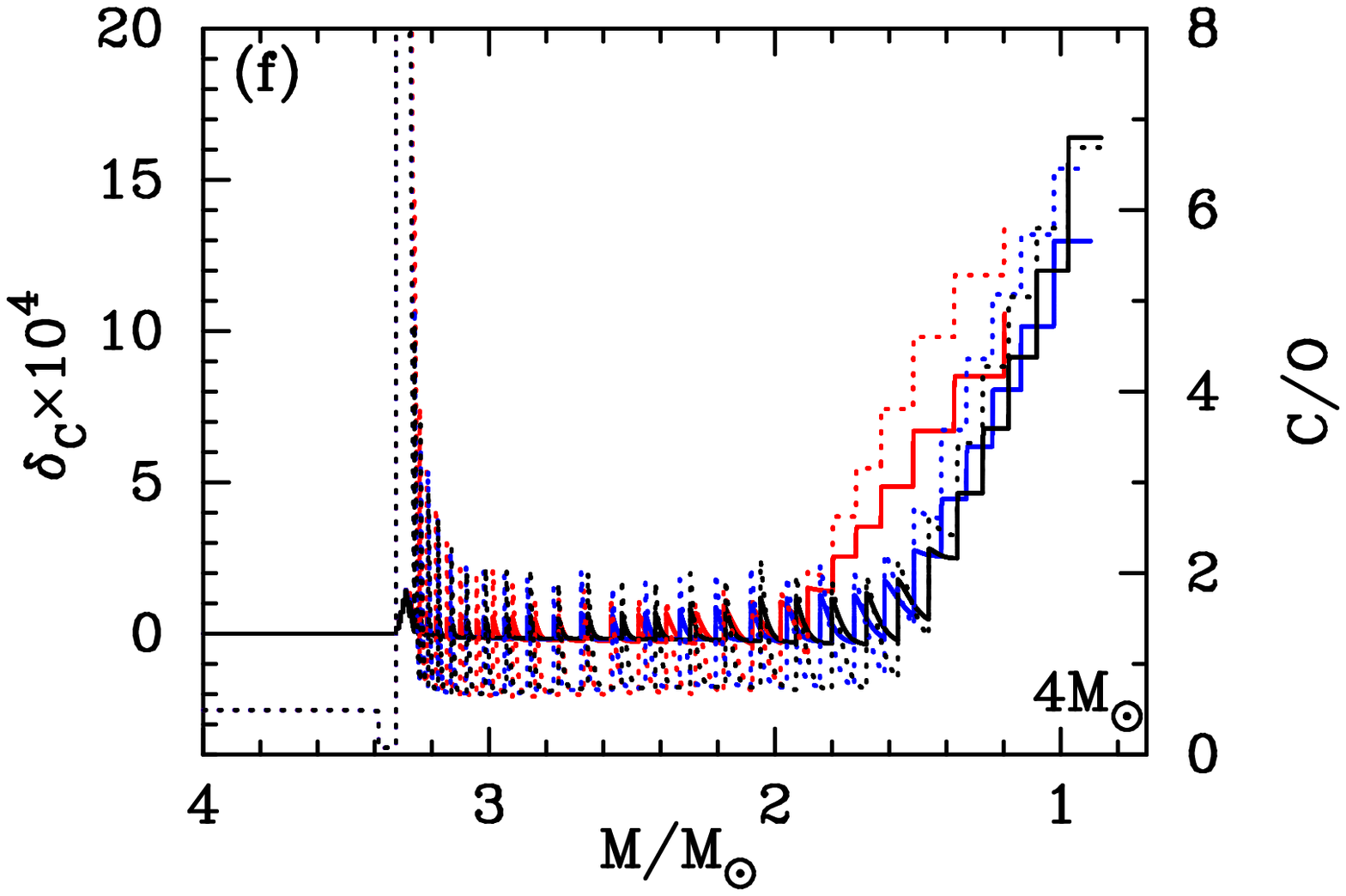}
 \end{minipage}
\end{tabular}
\caption{The effect of low-temperature opacity on the
  temporal evolutions of mass fractions of C, N, and O (left panel),
  and carbon excess $\delta_{\rm C}$ and C/O ratio (right panel) in the
  surface layer for $M_{\rm ini}=$ 2 (top), 3 (middle) and 4  $M_{\sun}$ (bottom) 
       stars with $Z_{\rm ini}=10^{-7}$. The scaled-solar,
      the CO-enhanced, and the CNO-enhanced
      models are colored in black, blue, and red, respectively. The mass
      fraction of C (N, O) is denoted by the solid (dashed, dotted) line
      in the left panel, and $\delta_{\rm C}$ (C/O ratio) by the solid
      (dotted) line in the right panel.}      
\label{fig1}
\end{figure*}

How the difference in the treatment of low-temperature opacity affects the
evolution and structure of low-metallicity stars has been analyzed in the work by 
Constantino et al. (2014), aimed at clarifying whether use of
composition-dependent low-temperature opacities is necessary for modeling 
  the evolution of metal-poor AGB stars.
  Here, we briefly address the effects of 
the treatment of low-temperature opacity on the evolution of AGB stars in
relation to dust formation, primarily referring to the result of
  calculations for $Z_{\rm ini} = 10^{-7}$.  
In the C-rich envelope, the contribution of C$_2$ and CN molecules on the
opacity is enhanced around $T = 3500$ K (see Fig. 13 in Marigo \& Aringer 2009).
Thus, for a given set of gas density and temperature, the value of low-temperature opacity 
increases in the order of the scaled-solar, the CO-enhanced, and the CNO-enhanced 
opacities.
  In what follows, a model star with the scaled-solar
  (the CO-enhanced, the CNO-enhanced) opacity is referred to as the
  scaled-solar (the CO-enhanced, the CNO-enhanced) model.

  \subsection{CNO abundances and carbon excess $\delta_{\rm C}$ in
    surface layer}
 
  Fig. 1 shows the effects of the treatment of
  low-temperature opacity on the evolution of the mass fractions of C, N, and O
  (left panel) and the carbon excess $\delta_{\rm C}$
  and the C/O ratio (right panel) in the surface 
    regions of the stars of $M_{\rm ini} =$ 2 (top), 3 (middle),
    and 4 $M_{\sun}$ (bottom) with $Z_{\rm ini}=10^{-7}$.

The star of $M_{\rm ini} =$ 2 $M_{\sun}$ with $Z_{\rm ini} = 10^{-7}$ 
becomes C-rich after the first TDU and the final carbon excess
$\delta_{\rm C}$ exceeds 0.001, regardless of the treatment 
of low-temperature opacity. This holds for all the 2 $M_{\sun}$
models with $Z_{\rm ini} \leq 10^{-4}$ (see Table 1).
$\delta_{\rm C}$ and the surface mass fractions of C and O increase with time and get larger in decreasing order of the values of low-temperature opacities. 
On the other hand,  without HBB,  the surface mass fraction of N during
the TP-AGB phase is not affected by the treatment of low-temperature opacities. 
 Also, the C/O ratio declines quickly after the
  1st TDU and then converges to a constant, larger than 10,
being almost independent of the type of low-temperature opacity.

The 4 $M_{\sun}$ models with $Z_{\rm ini} = 10^{-7}$, regardless of the treatment of low-temperature opacity,   undergo HBB after several TDU events (see left-bottom panel of Fig.1). 
While $\delta_{\rm C}$ increases with time during the initial AGB phases following the first TDU,  the mass fraction of N (C) in the surface regions increases (decreases) 
quickly after the onset of HBB. Then, the combination of TDU and HBB makes the elemental composition in 
the surface regions C-rich and O-rich
alternatively. However, HBB ceases after the stellar
mass reduces to 1.57, 1.62, and 1.98 $M_{\sun}$ for the scaled-solar,
the CO-enhanced, and the CNO-enhanced models, respectively.  
Consequently, the subsequent TDUs cause the stars to be C-rich again, and
  $\delta_{\rm C}$ as well as the C/O ratio to increase monotonically with time.
The carbon excess $\delta_{\rm C}$ in the surface region is more enhanced in 
the CNO-enhanced model with the largest threshold mass $M_{{\rm C/O}>1}$ than in other opacity models at the same current stellar mass.
Although the details depend on the initial metallicity, this behaviour
holds for all the 4 $M_{\sun}$ models with
$Z_{\rm ini} \leq 10^{-4}$ and the 5 $M_{\sun}$ models with $Z_{\rm ini} \geq 10^{-6}$; TDU is not experienced during the TP-AGB phase
in the models of $M_{\rm ini}=$ 5 $M_{\sun}$ with $Z_{\rm ini}=0$ and $10^{-7}$,
thus both the models are excluded from Table 1 and the following discussions.

The evolution of the 3 $M_{\sun}$ model with $Z_{\rm ini} = 10^{-7}$ is sensitive to the treatment of low-temperature opacity. The CNO-enhanced model is always C-rich in the TP-AGB phase with very weak HBB in
the early phase. On the other hand, the scaled-solar and
the CO-enhanced models experience a stronger HBB until the stellar
masses are reduced to 1.66 and 1.77 $M_{\sun}$, respectively,
and then turn out to be C-rich. These behaviours are common to all the 
$M_{\rm ini} =$ 3 $M_{\sun}$ models with $Z_{\rm ini} \geq 10^{-7}$; the $M_{\rm ini}=$ 3 $M_{\sun}$ models with $Z_{\rm ini}=0$ are always C-rich in 
their TP-AGB phases 
since the scaled-solar, the CO-enhanced, and the CNO-enhanced models
experience weak, very weak, and no HBB, respectively, as shown in Fig. 2.

  We note that, in the case of $M_{\rm ini} =$ 3 $M_{\sun}$,
  the value of $M_{\rm C/O>1}$ is significantly larger in the CNO-enhanced
  model than in the other two models. Thus, it can be expected that the 
  model  
  of $M_{\rm ini} =$ 3 $M_{\sun}$ with the CNO-enhanced opacity  
   could start to form carbon dust at a significantly larger stellar mass. 
  In addition, it should be emphasized that 
    the carbon excess $\delta_{\rm C}$ and the C/O ratio in C-rich AGB stars
    of $M_{\rm ini}=$ 2 and 3 $M_{\sun}$ 
    with $Z_{\rm ini} \leq 10^{-4}$ are 
    much larger than the values considered  
    in the dust-driven wind models for C-rich
    AGB stars with solar and subsolar metallicities
    (e.g., Winters et al. 2000;  Wachter et al.
    2002, 2008; Mattsson, Wahlin \& H\"ofner  2010).

    \begin{figure}
\begin{center}
\includegraphics[width=\linewidth]{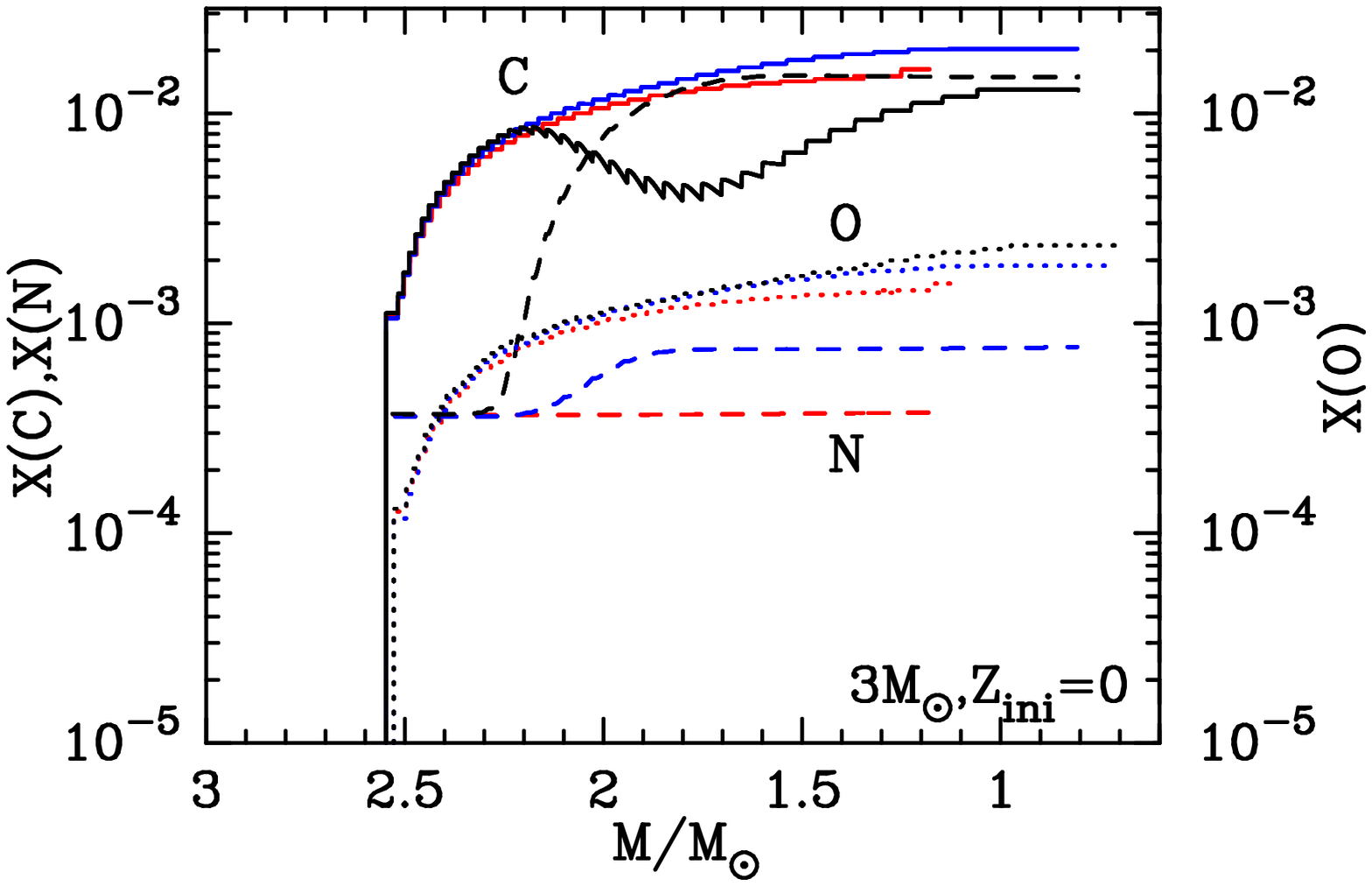}
\caption{Same as the left panel of Fig 1, but for
  $M_{\rm ini}=$ 3 $M_{\sun}$ with $Z_{\rm ini} = 0$.} \label{fig2}
\end{center}
\end{figure}

\begin{figure}
\begin{center}
\includegraphics[width=\linewidth]{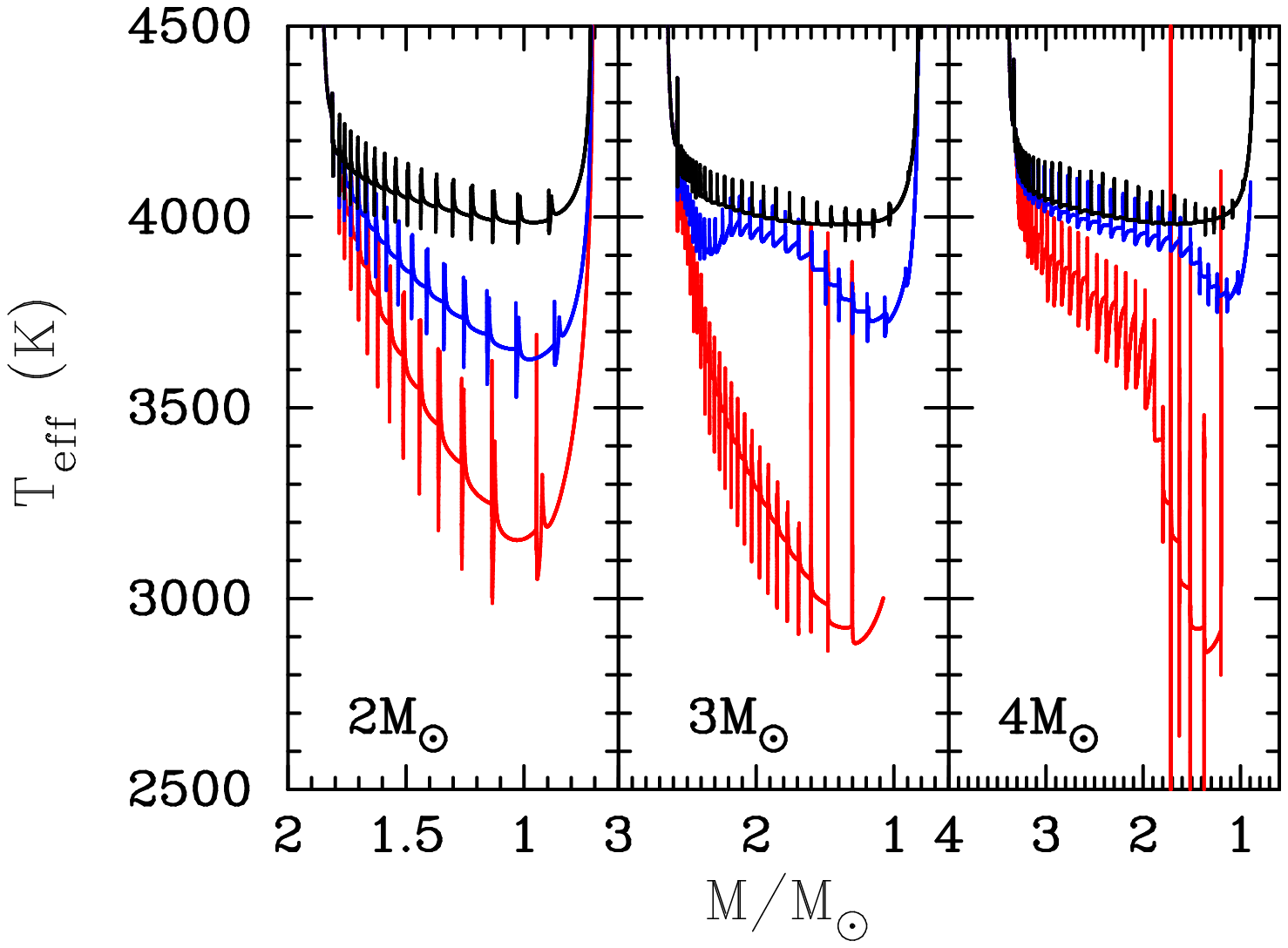}
\caption{The effect of the low-temperature opacity on the time evolution of
  effective
  temperature for the same models with $Z_{\rm ini}=10^{-7}$ in Fig. 1;
  $M_{\rm ini}=2$ (left),
  3 (middle),  and
    4 $M_{\sun}$ (right)  models 
    with the scaled-solar (black line), the CO-enhanced
    (blue line), and the CNO-enhanced
    (red line) opacities.\label{fig3}} 
\end{center}
\end{figure}

\subsection{Effective temperature}

Fig. 3 shows the time evolution of the effective temperature
for the same models presented in Fig. 1; note
that the huge spikes of effective temperature during the TPs in the CNO-enhanced models 
with smaller current stellar mass are artificial, being associated with 
the convergence problems. Owing to the very short duration of TPs, on which the effective temperature changes quickly, we will focus on the behaviour of effective temperature in 
the interpulse phases, in what 
follows.

The effective temperature
does not decrease below 3900 K in the scaled-solar models, regardless of the
initial mass and metallicity (see Table 1); conversely, in the CO-enhanced and
the CNO-enhanced models, the time evolution of the effective
temperature strongly depends on the initial mass, which determines
the change in the elemental compositions in the surface regions during the TP-AGB phase.
The effective temperature
of the $M_{\rm ini}=$ 2 $M_{\sun}$ star decreases with time more
rapidly in the CNO-enhanced model than in the CO-enhanced model; the minimum values reached are $T_{\rm eff,min}=$ 3627 K and 3153 K
in the CO-enhanced and the CNO-enhanced models
with $Z_{\rm ini}=10^{-7}$, respectively. Generally speaking, the minimum effective temperature reached by a model of a given mass is larger the smaller is the metallicity. The only exception to this trend is the $Z_{\rm ini} = 10^{-4}$ CNO-enhanced model. It should be
remarked here that, even if the abundance of N in the surface layer is
not enhanced without HBB, CN molecule dominates the low-temperature opacity
and efficiently decreases the effective temperature since the 1st TDU and
carbon ingestion (see Siess et al. 2002, Lau et al. 2009) increases
the surface abundance of N in AGB stars evolved from extremely metal-poor
stars considered in this paper. 
We note that the higher $T_{\rm eff,min}$ of the
  $Z_{\rm ini}=10^{-4}$ model comes from the fact that the
  model does not experience carbon ingestion.

In the CO-enhanced models with $M_{\rm ini}=$ 3 $M_{\sun}$, we can see
from Fig. 3 that the effective temperature decreases with
time in the initial C-rich phase, but increases after the onset of HBB.
In the O-rich phase when stronger HBB operates together with TDU, the
effective temperature in the CO-enhanced models with $M_{\rm ini}=$ 3 
and 4 $M_{\sun}$ are almost the same as that in
the scaled-solar models at the same current stellar mass. 
After HBB ceases, the effective temperature decreases with
time in the C-rich phase, until the minimum value is reached. On the other hand,
in the CNO-enhanced model, the effective temperature decreases with time
efficiently even if HBB makes the surface layer O-rich, as can be seen from
the time evolution of effective temperature for
$M_{\rm ini}=$ 4 $M_{\sun}$ with the CNO-enhanced opacity. This is because even in O-rich environments the enhancement of N increases the opacity through the CN molecule as demonstrated by 
Lederer \& Aringer (2009), and thus the stellar radius.

Thus, even in the extremely metal poor stars considered in this paper, the
employment of the low-temperature opacity appropriately taking into account
the change of elemental composition, such as the CNO-enhanced opacity, is
inevitable to investigate the evolution of star during TP-AGB phase.
Furthermore, as demonstrated in the next section, the treatment of the
low-temperature opacity definitely influences the formation
of carbon dust and the resulting gas outflow around AGB stars with
$Z_{\rm ini} \leq 10^{-4}$.

\section{Formation of carbon dust and resulting mass loss}
\label{sec:dust production}

As presented in the previous section,
all the models other than $M_{\rm ini}=5$ $M_{\sun}$ with $Z_{\rm ini}=0$ 
and $10^{-7}$ satisfy the minimum requirement for formation of carbon dust  
on the AGB after the stellar mass decreases 
  below $M_{\rm{C/O}>1}$.
However, not only $\delta_{\rm C}$,
  but also the effective temperature during the TP-AGB phase,
  strongly influences the formation of carbon dust and the consequent DDW (see Gail \& Sedlmayr 2013). In addition, Winters et al. (2000), based on hydrodynamical calculations of DDW, showed that C-rich stars with
    stable gas outflows dominated by the effects of the radiation pressure on dust 
    with the time averaged radiative acceleration
     $\langle \alpha \rangle >1$ experience mass-loss rates $\dot{M} \ga 3\times 10^{-7}$ $M_{\sun}$ yr$^{-1}$. 

  The present results, based on  hydrodynamical calculations,
   show that the CO-enhanced and
    the CNO-enhanced models with $T_{\rm eff} \la 4000$ develop the
    DDW with $\dot{M} > 10^{-7}$ $M_{\sun}$ yr$^{-1}$ and
     $\langle \alpha \rangle > 1$ (hereinafter the stable DDW), except for  
    the CO-enhanced models of $M_{\rm ini} =$ 4 $M_{\sun}$ with
    $Z_{\rm ini}=$ 0 and 5 $M_{\sun}$ with $Z_{\rm ini}=10^{-5}$. 
  On the other hand, the mass-loss rate of
  almost all the scaled-solar models with $T_{\rm eff} > 4000$ K,  excluding a few model stars,   
  is limited to less than $10^{-7}$ $M_{\sun}$
  yr$^{-1}$ though $\langle \alpha \rangle >1$. The scaled-solar opacity not
  reflecting the change of elemental
  composition in the surface regions during the TP-AGB phase
  could be inadequate in the low-temperatures regime.
  Thus, in this section,  focusing on 
  the CO-enhanced and the CNO-enhanced models with
  $\dot{M} \geq 10^{-7}$ $M_{\sun}$ yr$^{-1}$ (stable DDW),
  we shall show the dependence of the formation of carbon dust  
  and consequent DDW around AGB stars on the treatment of
     low-temperature opacity as well as on the initial mass and metallicity.
     The input parameters used in the hydrodynamical calculations and the
      derived properties of DDW for the
     CO-enhanced and the
     CNO-enhanced
     models are summarized 
      in Appendix A; Table A.1 for $Z_{\rm ini}=10^{-7}$ and 
       Table A.2 for the other initial metallicities.
  \begin{figure*}
\begin{tabular}{cc}
 \begin{minipage}[b]{0.5\linewidth}
  \centering
  \includegraphics[width=0.9\linewidth]{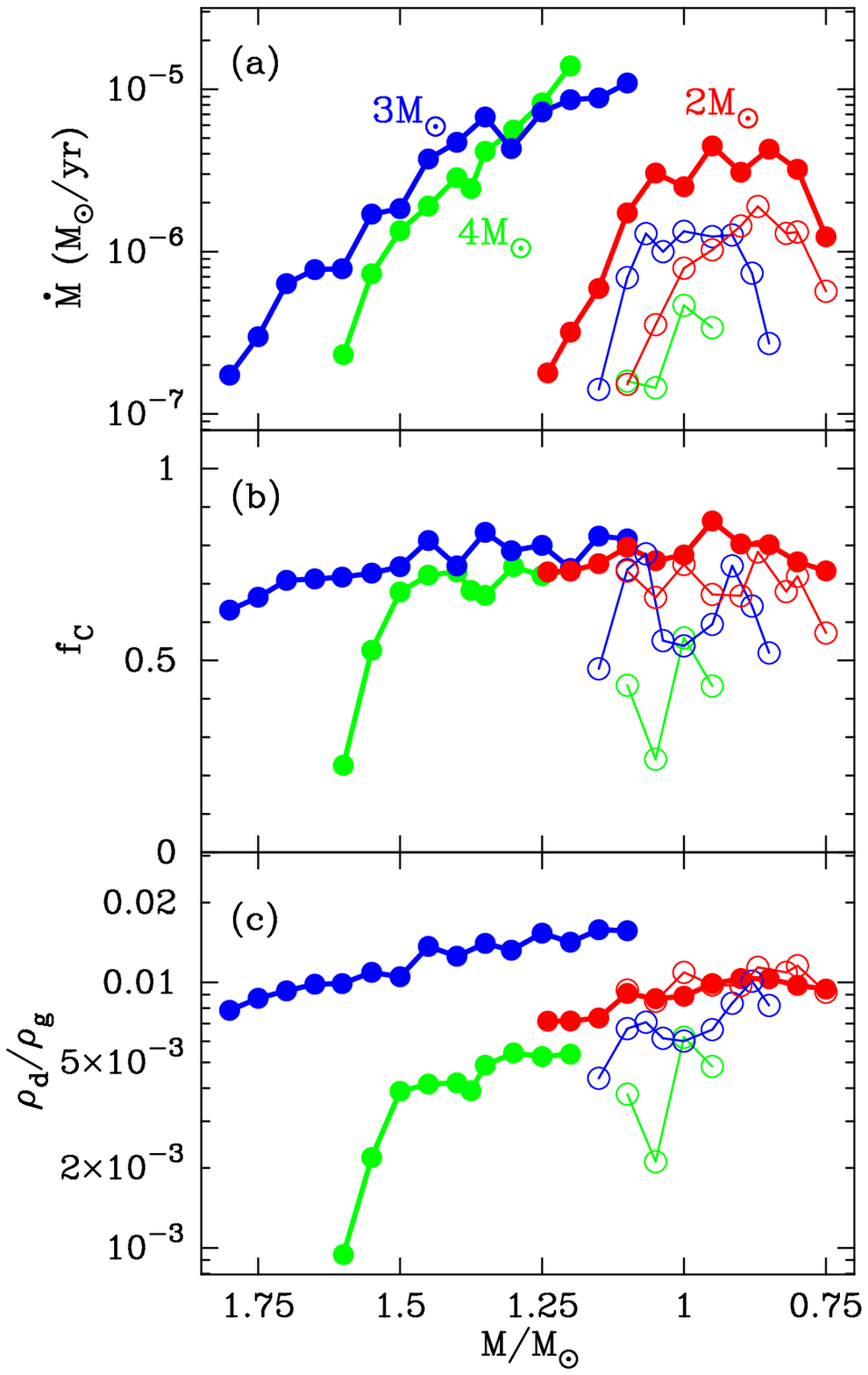}
 \end{minipage}
 \begin{minipage}[b]{0.5\linewidth}
  \centering
  \includegraphics[width=0.9\linewidth]{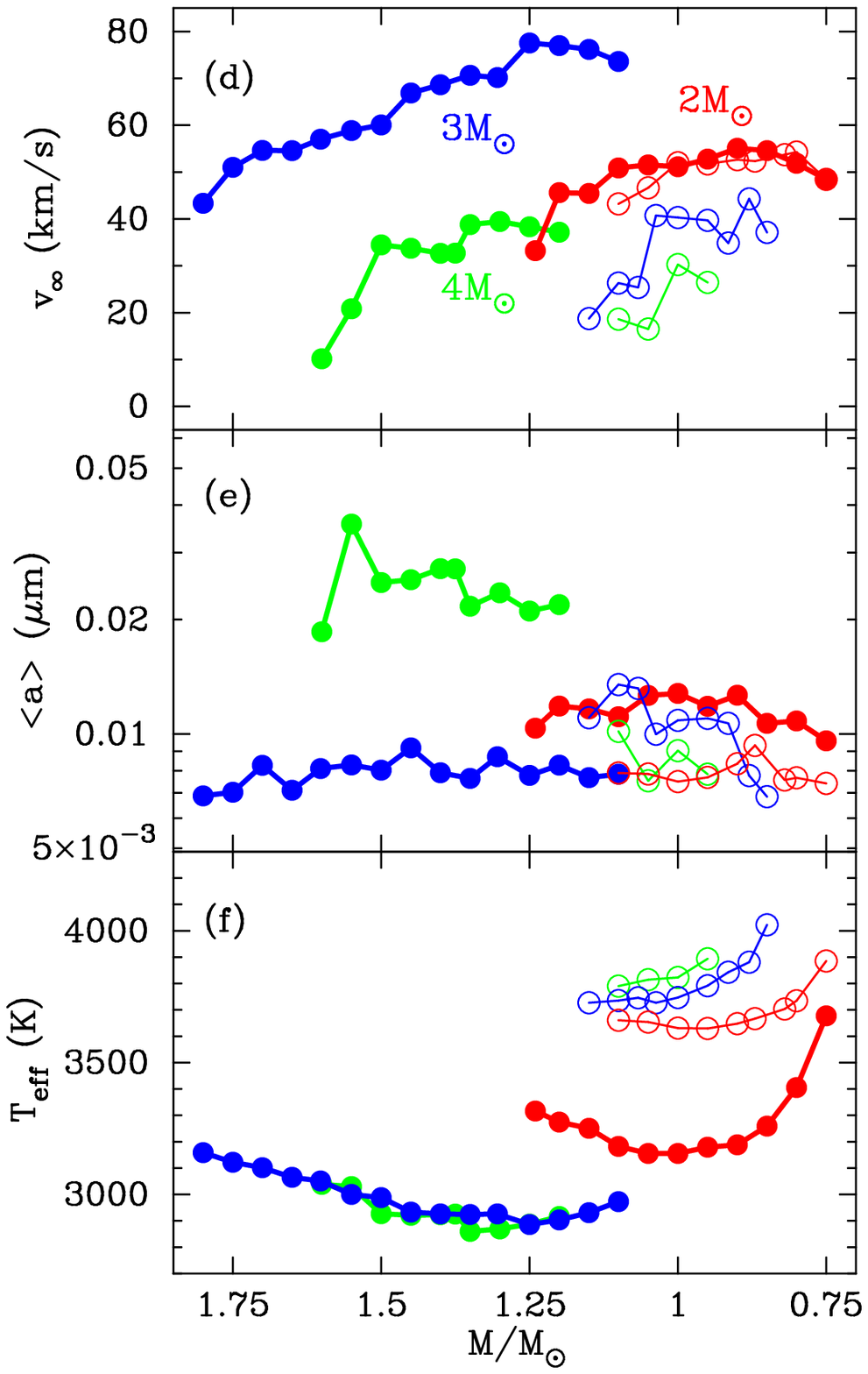}
 \end{minipage}
\end{tabular}
\caption{The effect of low-temperature opacity on dust formation
    and dust-driven wind
      and its dependence on
      the initial mass as a function of current stellar mass;
      from the top to the bottom,   
      (a) mass-loss rate $\dot{M}$,
      (b) condensation efficiency of carbon $f_{\rm c}$, 
    (c) dust-to-gas mass ratio $\rho_{\rm d}/\rho_{\rm g}$,   
      (d) terminal wind velocity $v_\infty$, and 
    (e) mass-averaged radius of dust $\langle a \rangle$
    during C-rich TP-AGB 
    phase calculated by the hydrodynamical model of DDW 
    and (f) effective temperature $T_{\rm eff}$ for the CO-enhanced 
    (open circle - thin solid line) and the CNO-enhanced 
    (filled circle - thick solid line) models of  $M_{\rm ini}=$ 2 (red), 3 (blue),  
    and 4  $M_{\sun}$ (green) with $Z_{\rm ini}=10^{-7}$ \label{fig4}}
\end{figure*}

\subsection{Effect of the low-temperature opacity and its
    dependence on the initial mass}

Fig. 4 displays the time-averaged physical
  quantities characterizing the dust formation and consequent 
  DDW together with the effective temperature as a
  function of the current stellar mass for the CO-enhanced and the
  CNO-enhanced models of $M_{\rm ini}=$ 2,  3, and 4 $M_{\sun}$ with
  $Z_{\rm ini}=10^{-7}$.

First, it should be pointed out that the formation of carbon  dust 
and resulting mass-outflow do not operate in the C-rich phases alternating
with O-rich phases associated with HBB, since the carbon excess
$\delta_{\rm C} <10^{-4}$ is insufficient, and the effective temperature is too
high to form carbon dust in a dense gas region close to the
photosphere; formation of carbon dust occurs in the regions where the temperature is below $\sim 1500$ K (e.g., Yasuda \& Kozasa, 2012). 
Also, in the case of $M \ga$ 2 $M_{\sun}$, the larger gravitational force   
 could prevent 
the star from driving gas-outflow stably through the radiation pressure
force acting on dust. 
  Thus, the effective formation of carbon dust to drive the mass loss is activated only after the stellar mass is significantly reduced from the
  threshold stellar mass $M_{{\rm C/O}>1}$, as described below.

The general trend shown in Fig.4 is that as the mass of the star decreases the mass-loss rate becomes larger, while the effective temperature decreases. This behaviour continues for a while even after the minimum effective temperature is reached. During the very final evolutionary phases, when the effective temperature increases rapidly, owing to the peeling of the external layers, the mass-loss rate and the dust condensation efficiency decline.

  This trend of mass-loss rate holds irrespective of the treatment
  of low-temperature opacity. However, the value of the
  mass-loss rate as well
  as the current stellar mass at which the stable DDW onsets is
  heavily influenced by the treatment of low-temperature opacity,
  depending on the initial mass. 

  In the case of $M_{\rm ini}=$ 3 $M_{\sun}$, the current stellar mass at
  the onset of stable DDW and the
  maximum value of mass-loss rate are remarkably
  different between
  the CNO-enhanced and the CO-enhanced models (see Fig. 4a); 
 as shown
in Section 3.2, the large value of the surface opacity
    causes the photosphere to expand and suppresses the increase of the temperature
    in the innermost layers of the convective envelope.
    Although HBB occurs in both 
    models, the CNO-enhanced model undergoes a much weaker HBB than
    the CO-enhanced model, and HBB ceases at
    $\sim$ 2.0 $M_{\sun}$ in the
    CNO-enhanced and at $\sim$ 1.4 $M_{\sun}$ in the
      CO-enhanced model. Then, the CNO-enhanced model evolving at smaller 
effective temperatures reaching higher values of $\delta_{\rm C}$, activates the formation of 
carbon dust in denser regions to drive the stable gas outflow
    at $M=$ 1.8 $M_{\sun}$; in the CO-enhanced model dust formation begins only after the mass of 
    the stars decreases to $M \sim$ 1.2 $M_{\sun}$.
    The mass-loss rate at the maximum is almost
      one order of magnitude
     smaller in the CO-enhanced model than in
      the CNO-enhanced model.
  On the other hand, the average radius of carbon dust is slightly smaller
  for the CNO-enhanced model since more seed nuclei are produced in the 
  outflowing gas being accelerated efficiently owing to larger values of $\rho_{\rm d}/\rho_{\rm g}$.

  \begin{figure*}
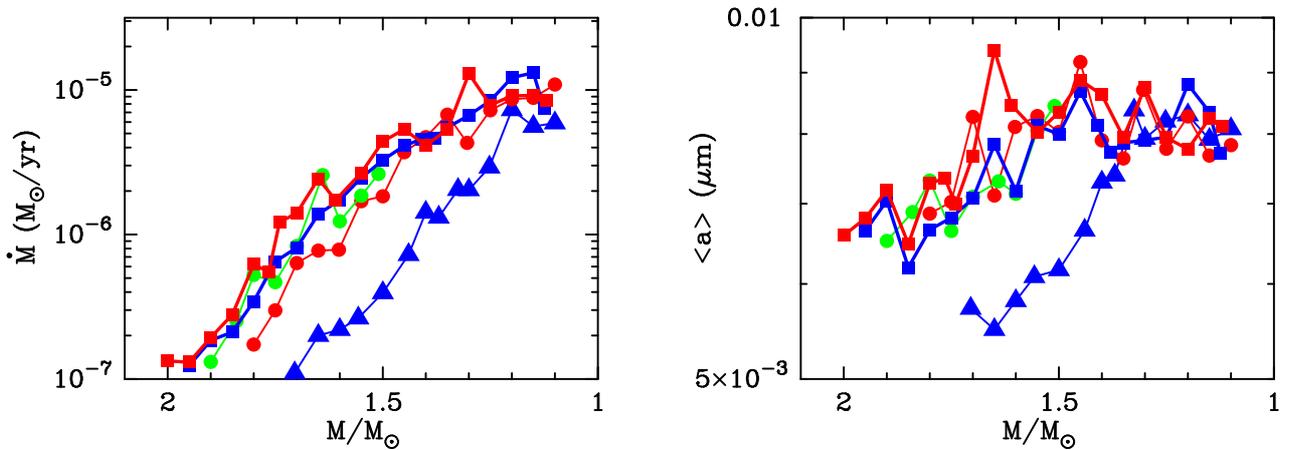

\begin{tabular}{cc}
 \begin{minipage}[b]{0.5\linewidth}
  \centering
  \includegraphics[keepaspectratio, scale=0.5]{fig5-dM.eps}
 \end{minipage}
 \begin{minipage}[b]{0.5\linewidth}
  \centering
  \includegraphics[keepaspectratio, scale=0.5]{fig5-ar.eps}
 \end{minipage}
\end{tabular}
\caption{The dependence of 
  on the mass-loss rate (left panel) and the time-averaged 
  mass-weighted radius
    of carbon dust (right) on the initial metallicity
    for the CNO-enhanced model of $M_{\rm ini}=$3 $M_{\sun}$.
    The symbol and colour denote the metallicity;
    filled red square, blue square, green circle, red circle,
      and blue triangle for $Z_{\rm ini}=$ $10^{-4}$, $10^{-5}$, $10^{-6}$, $10^{-7}$, and 0,
      respectively}
\label{fig5}
\end{figure*}  

The behaviour of the 4 $M_{\sun}$ models is different in comparison with their 
$3 M_{\sun}$ counterparts: both models experience active HBB since
  the core grows massive ($\sim 0.85 M_{\sun}$), and HBB ceases at
  $\sim 1.8$ (1.4)  $M_{\sun}$  
  for the CNO (CO)-enhanced model. In the following phases in the CNO-enhanced
  model, the decrease in the effective temperature leads to the onset of
  stable DDW when the mass is $\sim 1.6$ $M_{\sun}$, and the mass-loss rate
  sharply increases above $10^{-5}$ $M_{\sun}$ yr$^{-1}$. On the other hand, 
  in the CO-enhanced model evolving at higher effective temperatures, the onset of stable DDW is delayed to $M \sim 1.1$ $M_{\sun}$, and the largest mass-loss rate experienced is
  $\sim 5\times 10^{-7}$ $M_{\sun}$ yr$^{-1}$. It should be noted that the terminal gas
  velocity and the dust-to-gas mass ratio in
  the CNO-enhanced model
  are considerably smaller in comparison with the CNO-enhanced model  of 
  $M_{\rm ini}=$3 $M_{\sun}$, despite the fact that the mass-loss rate is
  comparable in both models for
   $M \la 1.5$ $M_{\sun}$. This
  arises from the smaller value of $\delta_{\rm C}$
  due to the delayed onset 
  of effective dredge up of carbon starting at $M \sim 1.8$ $M_{\sun}$
  Thus, the slower gas outflow velocity allows dust grains to
  grow up larger, and results in the mass-weighted average radius 
  more than a factor of 2 larger than that in the other models.

In the $M_{\rm ini}=2$ $M_{\sun}$ models evolving 
    without suffering HBB, the  carbon excess $\delta_{\rm C}$ increases
    monotonically and gets larger in the CO-enhanced model
    than in the CNO-enhanced model as the stellar mass decreases (see Fig.1). On the
    other hand, the effective temperature at the same current stellar mass is
    significantly lower in the CNO-enhanced model, and the difference increases 
    during the evolution.  
    Thus, the gas density in the region of 
    carbon dust formation
    as well as gas acceleration is higher in the CNO-enhanced model than in
    the CO-enhanced model. 
Although the condensation
  efficiency is a little smaller in the CO-enhanced model, the larger value
  of $\delta_{\rm C}$ makes the dust-to-gas mass ratio
  almost comparable in both models, as well as the gas terminal velocity
  being roughly
  proportional to $\rho_{\rm d}/\rho_{\rm g}$. 
  The higher gas density of dust formation region results in the larger
  mass-loss rate in the CNO-enhanced model than in the CO-enhanced model
  (see Fig. 4a). However, being different from the cases of $M_{\rm ini}=$ 3 
  and 4 $M_{\sun}$, the difference of the mass-loss rate between the
  CO-enhanced and the CNO-enhanced models keeps less than
  by a factor of 3 at $M \la 1$ $M_{\sun}$. 

  \subsection{Dependence on the initial metallicity}

  The properties of DDW
   as well as newly-formed carbon dust around AGB stars with
    $Z_{\rm ini} \leq 10^{-4}$ are expected not to depend directly on
    the initial metallicity, since the dredge-up carbon is of 
    secondary origin, thus independent on $Z_{\rm ini}$. 
    Fig.5 displays the dependences of the mass-loss
    rate (left panel) and the mass-weighted average radius of dust
    (right panel) on the initial metallicity for the CNO-enhanced models 
    of $M_{\rm ini}=$ 3 $M_{\sun}$
    with $Z_{\rm ini}=10^{-4}$ (red square), $10^{-5}$
    (blue square), $10^{-6}$ (green circle), $10^{-7}$
    (red circle),  and 0 (blue triangle).

    We can see from Fig.5 that, except for $Z_{\rm ini}=0$,
    the mass-loss rates are almost the same
    at a given current stellar mass and any clear dependence on the
    initial metallicity is not recognized, apart from some fluctuations.
    This is also true for the mass-weighted radius of carbon
    dust; regardless of the initial metallicity,
    the radii, with a few exceptions,
    have almost the same value at a given
    current stellar mass and tend to slightly increase with decreasing the
    current stellar mass. On the other hand, in the $Z_{\rm ini}=0$ case,
    the mass-loss rate at $M \ga 1.2$ $M_{\sun}$ as well as the radius at
    $M \ga 1.4$ $M_{\sun}$ is
    significantly smaller in comparison with the values for $10^{-7} \leq Z_{\rm ini} \leq 10^{-4}$. 
    This difference reflects the fact that the higher effective temperature of $Z_{\rm ini}=0$ star without
    enrichment of N due to HBB (see tables A.1, A,2, Section 3.2.2 and
    Fig. 2) during the AGB phase prevents carbon dust from forming in a
    dense region close to the photosphere. Thus, although the initial
    metallicity may subtly influence the properties of DDW and newly-formed
    carbon dust through its effects on the stellar evolution,  
    the present results demonstrate that cabon dust formation and
    the DDW do not show any significant dependence on the initial
    metallicity as long as $10^{-7} \leq Z_{\rm} \leq10^{-4}$.

    In summary, the treatment of
    low-temperature opacity  
  strongly affects dust formation and consequent DDW 
  on TP-AGB through its effect on the surface elemental
  composition and the effective temperature, 
  depending on the initial stellar mass.
  The current stellar mass at the onset of stable DDW is considerably smaller
  ($\sim 1$ $M_{\sun}$) in the CO-enhanced model in comparison with
  that in the CNO-enhanced model. The largest mass-loss rate
  in the CO-enhanced model is at least one-order magnitude smaller 
than in the CNO-enhanced model, except for the case of
  $M_{\rm ini}=2$ $M_{\sun}$, which does not experience HBB. 
  Thus, the adoption of low-temperature opacity 
  varying with the change of elemental composition in the surface during TP-AGB phase is 
  inevitable to investigate the dust formation and the mass loss around AGB stars with 
  extremely low initial metallicity, considered in this paper.
  
 The mass-weighted radius of carbon
  dust formed in the outflowing gas is of the order of 0.01 $\micron$,   
   regardless of the treatment of low-temperature 
  opacity as well as the initial mass and metallicity, except for  the models  of $M_{\rm ini}=$ 4
   and 5 $M_{\sun}$ with 
  the CNO-enhanced opacity developing 
  the slow and denser winds ($\langle a \rangle \sim 0.03 \micron$). 
  The derived radius of carbon dust is significantly smaller than the typical radius of 
  carbon dust necessary for reproducing the colours of obscured C-rich AGB stars observed in the Magellanic Clouds;  
   based on the stellar evolution calculations and the dust formation calculations employing the scheme developed by 
  Ferrarotti \& Gail (2006), 
  the typical radii of carbon dust grains are 0.06-0.2 $\micron$ 
  in the Magellanic Clouds, by 
  assuming the number ratio of seed particles to hydrogen nuclei $n_{\rm s}/n_{\rm H}=10^{-13}$ 
  (Dell'Agli et al. 2015a, 2015b), 
  and are 0.035-0.06 $\micron$ in the Small Magellanic Cloud (SMC), 
  by varying $n_{\rm s}/n_{\rm H}$ up to $10^{-11}$ 
  (Nanni et al. 2016).  The assumed/considered values of $n_{\rm s}/n_{\rm H}$ in their models 
  are considerably smaller than the values calculated in the present DDW models 
  ($7\times 10^{-12} \le n_{\rm s}/n_{\rm H} \le 10^{-8}$ depending on the initial mass as well 
  as on the input stellar parameters). Accordingly, the derived size of carbon 
  dust is smaller,  being roughly proportional to $(n_{\rm H}/n_{\rm s})^{1/3}$.   
  Since the aim of this paper is not to construct a self-consistent model with stellar evolution,  
  the comparison with the observations is beyond the scope of this paper.
   
\section{Discussion}

The hydrodynamical calculations of DDW in the previous
  section clearly demonstrate that the treatment of low-temperature 
  opacity strongly affects dust formation and the resulting mass loss.   
  Although the CNO-enhanced opacity is the most appropriate
  one among the three types of opacities considered,  
    it should be remarked
  that the hydrodynamical model of DDW employed in this paper derives the
  properties of DDW once a set of stellar parameters is
  given,  as mentioned in
  Section 2.2 and presented in Section 3. Thus, irrespective of the mass-loss
  rate and the low-temperature opacity assumed in the stellar evolution
  calculations, the results of hydrodynamical calculation of DDW along the
  evolutionary track on the AGB enable us to investigate the dependence of the
  properties of DDW on the input stellar parameters; 
   the ranges covered by the
  CNO-enhanced and the CO-enhanced models with  
  $\dot{M} \ge 10^{-7}$ $M_{\sun}$ yr$^{-1}$ are 
   2693 $ < T_{\rm eff}/{\rm K} < $ 4037,
      $1.23 < L/10^4L_{\sun} < 3.23$,  
      $0.7 \leq M/M_{\sun} \leq 2.04$, 
      $3.28 < \delta_{\rm C}\times 10^4 < 27.0$, and   
      $1.50 < \kappa_{\rm R}/{\rm cm}^2 {\rm g}^{-1} \times 10^4 < 90.0$. 
  
  In this section,  based on the results of DDW calculation
  presented in the previous section, we shall derive and discuss a 
  condition necessary for the efficient DDW with
  $\dot{M} \geq 10^{-6} M_{\sun}$ yr$^{-1}$ and
   the analytic formulae of
  gas and dust mass-loss rates in
    terms of the input stellar
    parameters. Furthermore, the implication on the
evolution of intermediate stars with $Z_{\rm ini} \leq 10^{-4}$ in the
early Universe is discussed in connection
with the dust formation and the mass loss.

\subsection{A necessary condition for efficient dust-driven wind} \label{subsec:necess.con}

  The thresholds of the stellar parameters for the stable DDW
  around C-rich AGB stars with solar metallicity  
  have been investigated by means of hydrodynamical calculations; 
  based on 
  the range of stellar parameters inferred from the observations of
  galactic carbon stars, Winters et al. (2000) found critical 
  values of the various parameters for producing the stable DDW, 
  depending on a combination of all the other parameters used in the
  hydrodynamical model. However,  no attempt has been done to
    express the dependence explicitly in terms of the other parameters.
    Also, the hydrodynamical calculations by Winter et al. (2000) are confined in a narrower range of
  stellar parameters,
  especially for $T_{\rm eff}$ and $\delta_{\rm C}$, compared with the
   range covered by the present calculations. Thus,
  it is instructive to attempt constraining the conditions   
  necessary for driving DDW as a combination of stellar 
  parameters, based on the present results.  
  
  First, it should be noted that the assumption of the position coupling
  (drift velocity of dust is set to be 0) and 
  the setting of the velocity amplitude $\Delta u_{\rm P}= 2$ km s$^{-1}$
  in the hydrodynamical model may influence the calculated mass-loss
  rates. Although Winters et al. (2000) have shown that the
  dependence of $\Delta u_{\rm P}$ on the mass-loss rate is weak as long as
  $\dot{M} \ga 3 \times 10^{-6}$ $M_{\sun}$ yr$^{-1}$, 
    at present time, little is known about the value of $\Delta u_{\rm P}$ 
    allowed for C-rich AGB stars (Gail \& Sedlmayr 2013).   
     As for the assumption of position coupling, the
      recent two-fluid hydrodynamic 
  model of DDW considering the dust formation as well as 
  the interaction between gas and dust has demonstrated 
  that the properties of DDW are well reproduced
    by assuming the position coupling in the case that
    $\dot{M} \ga 10^{-6}$ $M_{\sun}$ yr$^{-1}$ (Yasuda et al. 2016, in
    preparation).  Thus, the calculated
    properties of DDW with $\dot{M} \ga 10^{-6}$ $M_{\sun}$ yr$^{-1}$ would not
    seriously suffer from the uncertainties arising from the assumption of
    position coupling 
    underlying in the hydrodynamical model used in the present paper.
    Here, referring to the DDW with 
    $\dot{M} \geq 10^{-6}$ $M_{\sun}$ yr$^{-1}$ as the efficient DDW in the
    following, 
    we shall constrain the condition for producing the efficient DDW.
 
 Among the  stellar parameters used in the hydrodynamical model, 
      the effective temperature
      is the most relevant parameter for DDW, as discussed in
      previous studies (e.g., Winters et al. 2000, Wachter et al. 2002). 
      The efficiency of gas acceleration due to radiation 
pressure on dust grains is roughly proportional to
      $\delta_{\rm C} L/M$.  Also, the Rosseland mean opacity
      $\kappa_{\rm R}$ at the photosphere, that controls the density
      structure of the surface regions, is
      considered to have a significant effect on DDW in
      connection  
      with the density of gas levitated by the pulsation shock.
      
 Fig. 6 shows $\Lambda=\delta_{\rm C} L/\kappa_{\rm R} M$ versus
  $T_{\rm eff}$  for the CNO-enhanced and
  CO-enhanced models tabulated in Appendix A;
  The dotted lines indicate the boundaries on the 
  $\log T_{\rm eff}-\log \Lambda$  plane for
  the possible
  formation of efficient DDW.
  From this plot, we can see that efficient DDW is
  possible only if  $T_{\rm eff} \la 3850$ K and 
  $ \log \Lambda \ga 10.34\log T_{\rm eff}-32.33$,
 though the boundary seems to somewhat reflect the TP-AGB  tracks of the models.
   Although the derived constraint
    condition is only a necessary condition,
    the condition could be useful for 
    judging when the efficient DDW resulting from the 
    formation of carbon dust onsets  
      in the course of evolution of C-rich AGB stars, by
  referring the stellar parameters along the
  evolutionary track derived by stellar evolution calculation.

\begin{figure}
\begin{center}
\includegraphics[width=\linewidth]{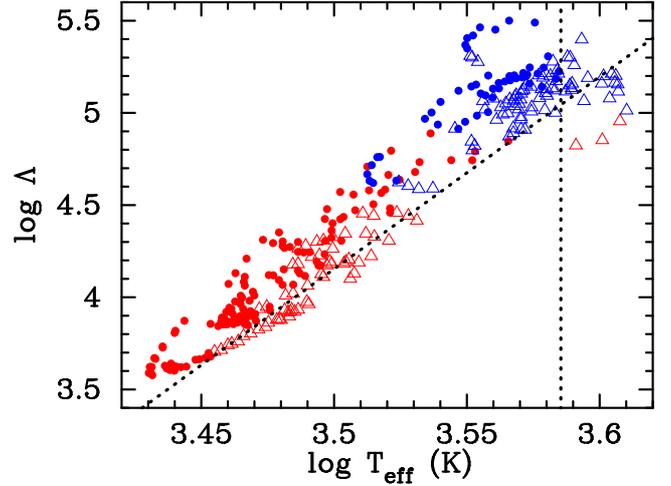}
\caption{The plot of
  $\Lambda=\delta_{\rm C}L/\kappa_{\rm R}M$
    versus $T_{\rm eff}$ with $M$ and $L$ in solar units, and $\kappa_{\rm R}$ in units of cm$^2$ g$^{-1}$;
    the red (blue) filled
    circle for the CNO (CO)-enhanced model with $\dot{M} \ge
    10^{-6}$ $M_{\sun}$ yr$^{-1}$ and red (blue) open triangle for the
    CNO (CO)-enhanced model with
    $\dot{M} < 10^{-6}$ $M_{\sun}$ yr$^{-1}$. The
    dotted lines represent the boundaries for the possible formation of DDW with
    $\dot{M} \ge 10^{-6}$ $M_{\sun}$ yr$^{-1}$.}\label{fig6}
\end{center}
\end{figure}

\begin{table*}
\begin{center}
\caption{The coefficients of the mass-loss formulae (Eq. 3) of gas ($\dot{M}^{\rm g}_{\rm fit}$) and dust ($\dot{M}^{\rm d}_{\rm fit}$) for the cases with and without $Z_{\rm ini}$ and their correlation coefficient $R$ and the maximum deviation $D$ in \% from the values calculated by the hydrodynamical model.}
\label{tab3}
\begin{tabular}{lcccccccccc} \hline \hline
 & $a$ & $b$ & $c$ & $d$ & $e$ & $f$ & $g$ & $h$ & $R$ & $D$ (\%)\\\hline 
$\dot{M}^{\rm g}_{\rm fit}$ without $Z_{\rm ini}$ & -5.733 & -19.13 & 3.164 & -5.254 & 0.7768 & -0.7089 & -0.8955 & 0 & 0.88 & 39\\
$\dot{M}^{\rm g}_{\rm fit}$ with $Z_{\rm ini}$ & -5.590 & -20.07 & 3.221 & -5.182 & 0.8836 & -0.8476 & -0.8138 & 0.02220 & 0.90 & 34\\ \hline
$\dot{M}^{\rm d}_{\rm fit}$ without $Z_{\rm ini}$ & -8.991 & -19.21 & 2.874 & -5.361 & 1.843 & -0.6417 & -0.8834 & 0 & 0.92 & 51\\
$\dot{M}^{\rm d}_{\rm fit}$ with $Z_{\rm ini}$ & -8.822 & -19.41 & 2.696 & -5.075 & 1.977 & -0.8163 & -0.5075 & 0.02774 & 0.94 & 46\\ \hline
\end{tabular}
\end{center}
\end{table*}

\subsection{Analytic formulae of gas and dust mass-loss rates}
\label{subsec:formulae}

The amount of gas and dust that C-rich AGB stars supply to the interstellar space is
  crucial not only to reveal the origin of dust but also to  
  investigate the formation and evolution of stars in  
  galaxies through
  chemical evolution model in the
  early Universe (e.g., Grieco et al. 2014).
  The formulae of mass-loss rate for C-rich AGB stars with 
  solar and subsolar metallicities has been proposed, based on
  the hydrodynamical calculations 
  of DDW (Arndt, Fleischer \& Sedlmayr 1997;
  Wachter et al. 2002, 2008).  Although
 Weiss \& Ferguson (2009) applied the formula by
  Wachter et al. (2002) to investigate the evolution of stars with
  $Z_{\rm ini}=5\times 10^{-4}-0.04$, it is questionable 
  whether the same formula can be applied to C-rich AGB stars of metallicity $Z_{\rm ini} \leq 10^{-4}$. 
  Here, we shall derive the analytic formulae for gas and dust mass-loss rates
  in terms of the input stellar parameters employed in the hydrodynamical
  calculations for the CNO-enhanced and CO-enhanced models with the efficient
  DDW. 

  For simplicity, we derive the formulae under the assumption that
  the mass-loss rate is simply approximated by a 
  linear function of the logarithms of the
  input parameters
  ($M$, $L$, $T_{\rm eff}$, $\kappa_{\rm R}$, $\delta_{\rm C}$, and
  $P$).  Also we shall consider the initial metallicity as a parameter, since 
  the mass-loss rates of the $Z_{\rm ini}=0$ models deviate from 
the others (see the left panel of
  Fig. 5),  though the mass-loss rates of stars with
  $10^{-7} \leq Z_{\rm ini} \leq 10^{-4}$ do not show a clear dependence on the
  initial metallicity. 
  Applying the least-square method to the mass-loss rates and the
  dust-to-gas mass ratios tabulated in Appendix A, the fitting formula
  is expressed as 
  \begin{eqnarray}
  & &    \log\dot{M}_{\rm fit} = a +
      b \log\left(\frac{T_{\rm eff}}{3000 {\rm K}}\right)
      + c \log \left(\frac{L}{10^4L_{\sun}}\right) \nonumber \\
  & &    +  d \log\left(\frac{M}{M_{\sun}}\right) 
     +  
   e \log\left(\frac{\delta_{\rm C}}{10^{-4}}\right) 
  + f \log \left(\frac{\kappa_{\rm R}}{10^{-4} {\rm cm}^2 {\rm g}^{-1}}
   \right) \nonumber \\
 & &  + g \log \left(\frac{P}{650 {\rm d}}\right)
   + h \log Z_{\rm ini}. 
  \end{eqnarray}
Note that we adopt $Z_{\rm ini}=10^{-12}$ as a representative of $Z_{\rm ini}=0$
          when fitting.
  The numerical coefficients from $a$ to $h$ for the formula of
  gas (dust) mass-loss rate $\dot{M}^{\rm g}_{\rm fit}$
   ($\dot{M}^{\rm d}_{\rm fit}$) with and
      without including the initial metallicity are provided in Table 2, 
      with the correlation coefficient $R$ and the maximum deviation $D$ 
      from the calculated values. The formula of gas (dust)
      mass-loss rate without including metallicity fits the values calculated
      by the DDW model with the correlation coefficient
      0.87 (0.92) 
      and the maximum deviation 38 (50) \%. 
      The fittings are only
      slightly improved by including the metallicity, reflecting the fact that
      the mass-loss rate does not sensitively depend on the initial metallicity,
      with the coefficient $h \sim 0.02-0.03$.
      
      The power of the effective temperature in the gas mass-loss formula is huge 
      (e.g.,  $b=$-19.13 for the case without $Z_{\rm ini}$) in comparison with that in the formulae  by Wachter et al. (2008, 
      hereafter W08). The huge power arises from the inclusion of $\delta_{\rm C}$ and $\kappa_{\rm R}$ in the fitting formula. 
      In fact, the gas mass-loss rate being fitted by using $M$, $L$, $T_{\rm eff}$ in the same manner as W08, 
      the formula is given by  $\dot{M} ^{\rm g}_{\rm fit}=7.62\times10^{-6} (M/M_{\sun})^{-4.28} 
      (T_{\rm eff}/2600 {\rm K})^{-7.64}(L/10^4L_{\sun})^{1.66}$ with the correlation coefficient 0.80 and 
      the maximum deviation 42 \%. 
      Thus, the power of the effective temperature is reduced to -7.64 and is comparable to the value in 
      the formula by W08.
      Here it should be noted that the inclusion of $\delta_{\rm C}$ and $\kappa_{\rm R}$ is inevitable in the fitting formula since 
      $\delta_{\rm C}$ and $\kappa_{\rm R}$ control the amount of carbon available for dust formation and the density of gas 
      levitated by the pulsation shock, respectively.

  \begin{figure*}
\begin{tabular}{cc}
 \begin{minipage}[b]{0.5\linewidth}
  \centering
  \includegraphics[keepaspectratio, scale=0.5]{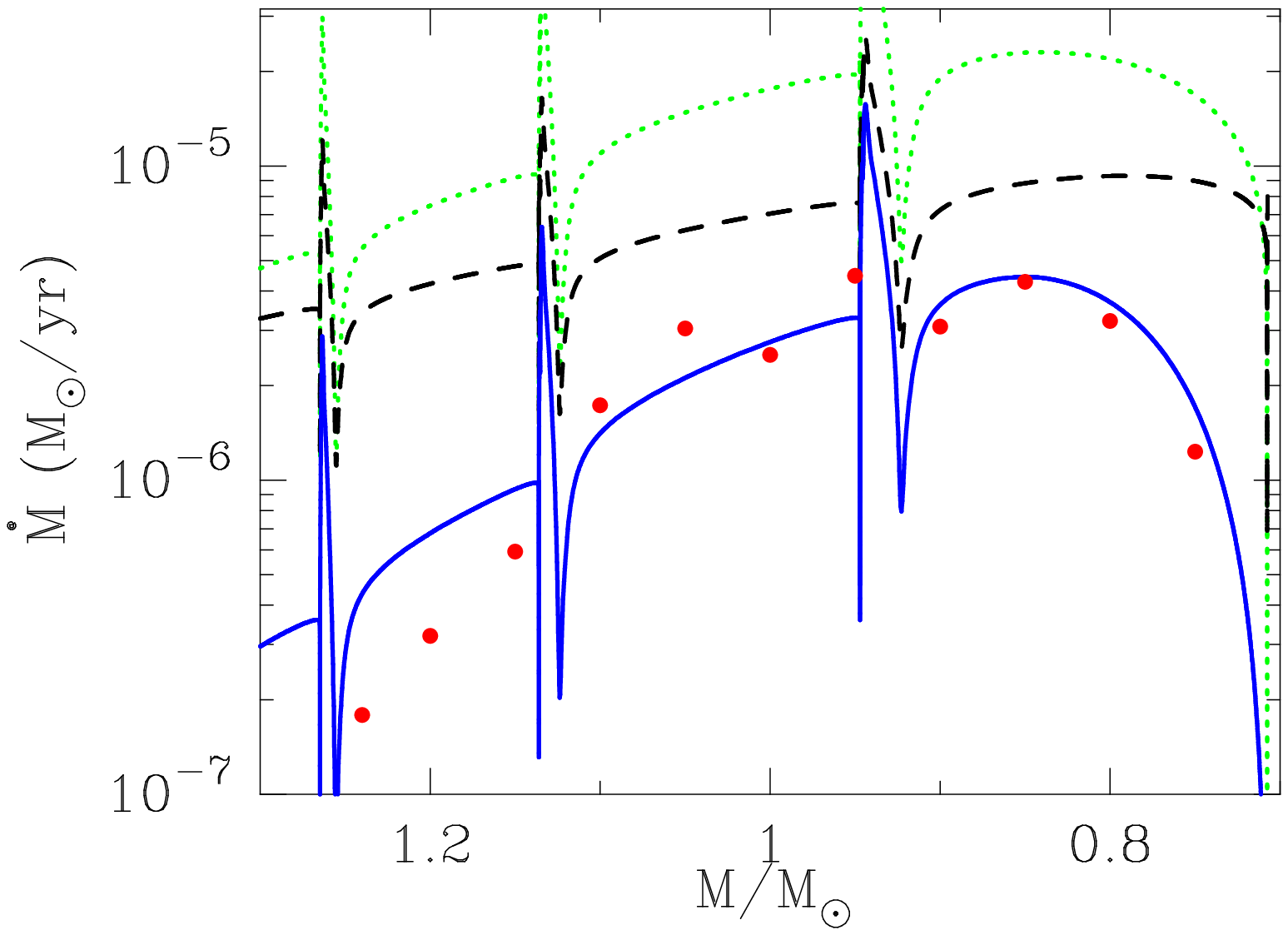}
 \end{minipage}
 \begin{minipage}[b]{0.5\linewidth}
  \centering
  \includegraphics[keepaspectratio, scale=0.5]{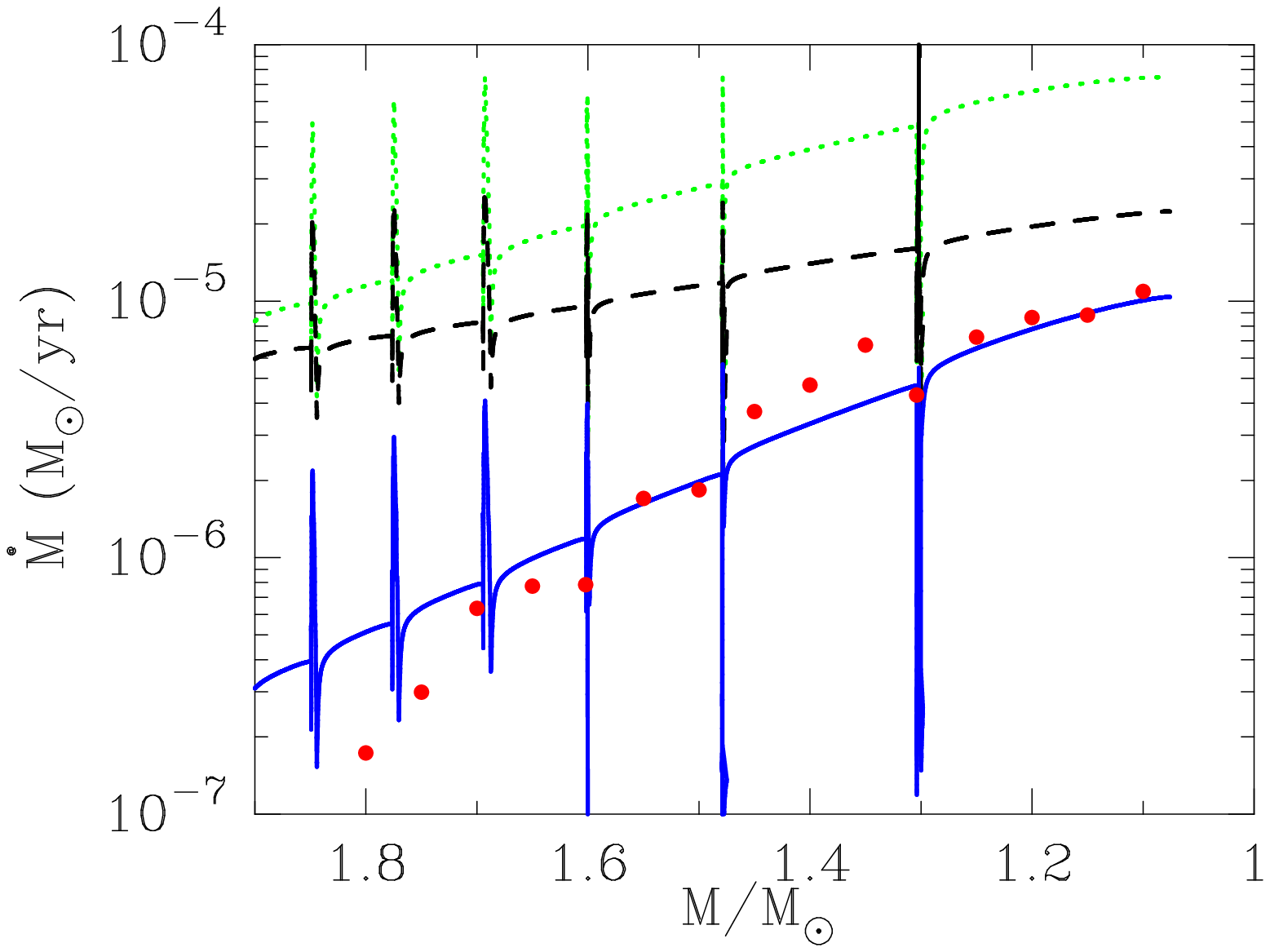}
 \end{minipage}
\end{tabular}
\caption{Time evolution of mass-loss rates for the CNO-enhanced models of
  $M_{\rm ini}=$ 2 (left panel) and 3 (right panel) $M_{\sun}$ with 
  $Z_{\rm ini}=10^{-7}$. The mass-loss rate calculated by
  the DDW model and the fitting
  formula (Eq. 3) are denoted by red circle and solid line, respectively.
  The dotted (dashed) line shows the mass-loss rate calculated by
  the formula for the SMC models of W08 (SC05 assumed in the stellar
  evolution calculation) for a reference.}
\end{figure*}

  Fig. 7 shows the evolution of the gas 
  mass-loss rate of the CNO-enhanced models with $Z_{\rm ini}=10^{-7}$;  
  $M_{\rm ini}$= 2 $M_{\sun}$ (left) and 3 $M_{\sun}$ (right).  
  The mass-loss rates
  calculated by the DDW model and the fitting formula
  without
  including metallicity are denoted by the filled circles and the solid line, 
respectively, with the dotted and dashed lines indicating the 
results obtained by assuming, respectively, the mass-loss formulae by W08 for the SMC models 
and SC05. 
We can see that the fitting formula Equation (3)  reasonably
reproduces the mass-loss rates derived from the DDW model,
as long as $\dot{M} \ga 10^{-6}$ $M_{\sun}$ yr$^{-1}$ 
in both models.  

  As shown in Fig. 7,  the mass-loss rate 
  assuming the formula for the SMC models by W08  
  is more than one order of magnitude larger than 
  the rate calculated for C-rich AGB stars with $Z_{\rm ini} = 10^{-7}$.  
  Although the difference in the stellar parameters used in 
  the calculations makes it difficult to compare the results directly, the gap in the calculated mass-loss rate is 
  caused by the differences in the Rosseland mean opacity $\kappa_{\rm R}$ at the surface and the amplitude of
   pulsation $\Delta u_{\rm P}$;  $\kappa_{\rm R}= 5\times 10^{-5}$ cm$^2$g$^{-1}$ and 
   $\Delta u_{\rm P}$=5 km s$^{-1}$ 
    for the SMC models in W08; $\kappa_{\rm R}=\sim 10^{-3}$ cm$^2$g$^{-1}$ 
    as a  typical 
    value (see Table A1) and $\Delta u_{\rm P}=2$ km s$^{-1}$ in the CNO-enhanced 
    models with $Z_{\rm ini}=10^{-7}$.  
   The difference in the value of $\kappa_{\rm R}$  implies that the gas 
   density in the surface region, being roughly proportional to 
   $\kappa_{\rm R}$,  is 
   a factor of 20 larger in the SMC models  than in the  CNO-enhanced models.   
   Also the carbon excess $\delta_{\rm C}=8.57 \times 10^{-5}$  in the SMC models assuming  C/O=1.8 and taking the oxygen abundance from Russell \& Dopita (1992), 
   while $\delta_{\rm C} \sim 10^{-3}$ as a typical value in the 
   CNO-enhanced models investigated here.  Although the amount of carbon in 
   the surface layer is comparable, the higher  gas density in 
   the surface region as well as the enhanced  density of gas levitated by the pulsation 
   shock with larger $\Delta u_{\rm P}$ leads 
   to larger mass-loss rate in the SMC models. Here it should be addressed that 
   the values of $\kappa_{\rm R}$ and $\delta_{\rm C}$ used in W08 seem to be unrealistic in comparison with 
   the values derived from stellar evolution calculations, and the applicability of their formula 
    for C-rich AGB stars with $Z_{\rm ini} \le 10^{-4}$ should be checked by 
   using the appropriate values.  Also, 
    the dependence of the mass-loss rate on the velocity amplitude of pulsation should be investigated since the 
    dependence is considered to be more sensitive for the stars with larger $\delta_{\rm C}$ and $\kappa_{\rm R}$.
    
The mass-loss rate derived from the DDW model  
  is significantly smaller than the mass-loss rate assumed in the stellar evolution
calculations. If we use the derived mass-loss formula after  the necessary condition for the efficient DDW 
is satisfied on the TP-AGB, the dredged-up carbon accumulates 
in the surface regions, and accordingly the effective temperature decreases and the mass-loss rate could increase. Although in the present calculations the derived mass-mass loss rate is inconsistent with the assumed mass-loss rate, 
it should be reminded here again that the hydrodynamical model
can derive the properties of DDW by specifying a set of 
the input parameters, being
independent of stellar evolution model. Thus, the derived formulae,  
together with
a necessary condition for the efficient DDW presented in Section 5.1,
could enable us to evaluate the mass-loss rate and the dust yield during C-rich
AGB phase of stars with $Z_{\rm ini} \leq 10^{-4}$ in a manner consistent
with the stellar evolution, including whether the efficient DDW can operate
on C-rich AGB. 

\subsection{Implication for evolution of C-rich AGB stars and
  the dust-driven wind 
  in the early Universe}

The investigations on the formation of stars in low-metallicity environments
have revealed that the critical metallicity $Z_{\rm cri}$ for
the transition from Population III  to Population II stars is
as low as $\sim 10^{-9}-10^{-7.5}$, depending on the depletion factor of
metal into dust (e.g., Omukai et al. 2005, Schneider et al. 2006,
Chiaki et al. 2015).  Although there is no information available for the
initial mass function at present time, the intermediate-mass AGB stars
with $Z_{\rm ini} \leq 10^{-4}$ can contribute 
 to the enrichment of dust in the early Universe if the condition for the efficient DDW 
 derived in the previous subsection is satisfied during the TP-AGB phase.  
 However, the
possibility of developing dust formation and resulting DDW on the AGB is 
strongly influenced by the mass-loss history during
the evolution, on which the time evolution of effective
temperature as well as the elemental composition and opacity in the
surface layer strongly depend through the number of TDU episodes and/or 
the occurrence of HBB.  At present, we have no knowledge
on the mass-loss mechanism and the mass-loss rate before the onset of DDW
for extremely low-metallicity stars considered in this paper.

In recent investigations focused on low-metallicity ($-1.6 <$ [Fe/H] $< -0.5$) AGB stars 
observed in distant galaxies, Rosenfield et al. (2014, 2016) showed that a mass-loss rate higher than SC05 is required 
on the AGB phases previous to the onset of DDW, to reproduce the observed TP-AGB luminosity function as well as the number ratio of TP-AGB to red giant stars. If this is true for the
AGB phase of stars with $Z_{\rm ini} \leq 10^{-4}$, the so-called pulsation
enhanced DDW would not operate as the mass-loss mechanism; in stars losing
the mass efficiently on the AGB, the number of TDU is reduced, and the inefficient
decrease of effective temperature as well as the insufficient carbon 
excess ($\delta_{\rm C}$) prohibits the onset of dust formation
and the efficient DDW. On the other hand, the mass-loss rate  in
the pre-dust phase on the AGB being depressed,
it is possible for the DDW to dominate the mass loss after the stellar
mass is substantially reduced below $M_{\rm C/O>1}$,  as 
  demonstrated in
Section 4; in cases such as $M_{\rm ini} =$ 2 $M_{\sun}$ not experiencing any HBB,  
the increase in 
the carbon excess and the decrease in effective temperature could make the DDW
more efficient. For stars of mass $M_{\rm ini} \ge$ 3 $M_{\sun}$ which experiences HBB,  a
smaller mass-loss rate on the AGB results in more active HBB that   
seems to decrease the threshold mass $M_{\rm C/O>1}$ and the carbon 
excess in the C-rich phase. However, the enrichment of N in the
surface layer associated with HBB can counteract HBB itself, 
 since the enhanced surface opacity depresses 
the increase of gas temperature in the innermost layer of convective
envelope and  makes HBB weaker. Thus, it can be expected
that, even in massive stars experiencing HBB, the mass loss by DDW
could operate, although the details depend on the initial mass and the
mass-loss rate in the pre-dust phase. Anyway, the present results of
DDW calculations demonstrate that  the formation  of carbon dust and 
resulting DDW is possible even in
the low-metallicity environments with $Z_{\rm ini} \leq 10^{-4}$ as
long as the mass-loss rate in the pre-dust phase on the AGB is reduced 
to some extent from the rate given by SC05.

Finally, it is useful to note the followings in connection with the
uncertainties inherent in the present DDW model; In section 4, the DDW 
with $\dot{M} \geq 10^{-7}$ $M_{\sun}$ yr$^{-1}$ is referred to
as the stable DDW according to Winters et al. (2000), since 
the time averaged value $ \langle \alpha \rangle \ga 1$.
  However, in the 
present calculations,  contrary to the results of Winters et al. (2000), 
we have not found any sustainable wind 
with $\langle \alpha \rangle < 1$ ,  since the carbon excess 
$\delta_{\rm C}$ of C-rich AGB stars with $Z_{\rm ini} \leq 10^{-4}$ is
significantly larger than the value inferred from the observations of
galactic C-rich AGB stars; $\delta_{\rm C} \sim 6.76 \times 10^{-4}$ 
corresponding to C/O=2.0 for solar metallicity.  
Also the recent investigation by the two-fluid hydrodynamic calculation for
DDW has shown that the assumption of position
coupling will break down
around $\dot{M} \sim \textrm{several} \times  10^{-7}$ $M_{\sun}$ yr$^{-1}$
(Yasuda et al. in preparation).
Thus, the application of
two-fluid hydrodynamic model is inevitable to explore the constraint
conditions for the onset of stable DDW. In addition, although
the velocity amplitude of pulsation $\Delta u_{\rm p}$ is set to be
2 km s$^{-1}$,  the large $\delta_{\rm C}$ and the $\kappa_{\rm R}$ in the 
surface region may make the dependence of
the value of $\Delta u_{p}$ on the formation of carbon dust and resulting
DDW more sensitive than the case for the galactic C-rich stars; the increase
of $\Delta u_{\rm p}$ up to 8 km s$^{-1}$ (Winters et al. 2000) may substantially enhance
the mass-loss rate by DDW. These aspects should be investigated 
systematically in the future works to explore the properties of DDW and
the nature of carbon dust formed around AGB stars in the early Universe,
being consistent with the stellar evolution calculations.

\section{Summary}

In order to explore dust formation and  resulting
mass loss around intermediate-mass AGB stars with initial metallicity
$Z_{\rm ini} \leq 10^{-4}$ in the early Universe,
the hydrodynamical calculations of dust-driven wind (DDW)  are carried out 
for stars with initial mass in the range of  $2 \leq M_{\rm ini}/M_{\sun} \leq 5$.  
The input stellar parameters necessary for the hydrodynamical calculation 
are  calculated by MESA code, assuming the mass-loss rate 
given by Schr\"oder  \& Cuntz (2005) in the post-main sequence phase,  
as a first  step for this study.  
In addition, three types of
low-temperature opacities (scaled-solar, CO-enhanced, and CNO-enhanced 
opacities)  are considered to elucidate the effect of the
treatment of low-temperature opacity on the time evolution of stellar
parameters related to the dust formation and consequent DDW.

We confirm that all model stars except for $M_{\rm ini}=5$ $M_{\sun}$ with
$Z_{\rm ini}=0$ and $10^{-7}$ finally turn to
be C-rich  and  satisfy 
the minimum condition for the formation of carbon dust, regardless of the
treatment of low-temperature opacity. However, the effective 
temperature, the quantity most sensitive to the dust formation 
process,  is strongly 
affected by the treatment of low-temperature opacity; the minimum effective
temperature $T_{\rm eff, min}$ in the interpulse phases does not 
decrease below 3900 K for the stars with the scaled-solar opacity, while
$T_{\rm eff, min}$ decreases below 3100 K for the stars of
$M_{\rm ini} \geq 3M_{\sun}$ with the CNO-enhanced opacity. 

The hydrodynamical calculations of DDW  along the  
evolutionary track of C-rich AGB  stars simulated with the CO-enhanced  
and the CNO-enhanced
opacities show the followings; 
The stellar mass at which the stable DDW with
$\dot{M} \geq 10^{-7}$ $M_{\sun}$ yr$^{-1}$  onsets is significantly smaller in
the CO-enhanced model than in the CNO-enhanced models, and the maximum
mass-loss rate on C-rich AGB is more than one order of magnitude
smaller in the CO-enhanced models than in the CNO-enhanced models for
$M_{\rm ini} \geq$ 3 $M_{\sun}$. Thus, the employment of composition-dependent
low-temperature opacity,  such as the CNO-enhanced opacity,  is inevitable to
investigate the formation of dust and resulting mass loss in low-metallicity
AGB stars. Also, we find that, given the initial mass,  the time evolution of mass-loss rate 
as well as the time-averaged mass-weighted
radius of carbon dust is almost independent of the initial metallicity, 
as long as $10^{-7} \leq Z_{\rm ini} \leq 10^{-4}$.

The results of DDW calculation covering a wide range of stellar parameters, 
regardless of the treatment of low-temperature opacity and the mass-loss
rate assumed in the stellar evolution calculations, 
enable us to derive a necessary condition for driving 
the efficient DDW with $\dot{M} \geq 10^{-6}$ yr$^{-1}$ as a combination of
stellar parameters, and the
fitting formulae of gas and dust mass-loss rates in terms of input stellar
parameters; the fitting formula of gas mass-loss rate reasonably 
reproduces the mass-loss rate calculated by DDW model. 
The derived necessary condition and the fitting formulae 
would enable us to evaluate when the efficient DDW onsets
and how much amount of dust is produced in the intermediate AGB stars with
$Z_{\rm ini} \leq 10^{-4}$, being coupled with the stellar evolution
calculations.

The present results of calculations employing the mass-loss
rate by SC05 in the post main-sequence phase suggest that the
efficient DDW being consistent with the stellar evolution could be
possible if the mass-loss rate during evolution of star 
is somewhat enhanced before entering into the AGB and  
depressed on the AGB before the onset of DDW from the rate given by SC05.
Also, it should be emphasized here that the assumption of 
position coupling  is not valid for the case of  low mass-loss rate 
such as $\dot{M} \sim \textrm{several} \times 10^{-7}$ $M_{\sun}$ yr$^{-1}$;
the assumption of position coupling results in overestimating the mass-loss
rate of C-rich AGB stars with larger $\delta_{\rm C}$  considered in this paper 
since $\dot{M} \propto \delta_{\rm C}$.
Thus,  a two-fluid hydrodynamical model 
calculation of DDW is necessary  for  clarifying 
when and in what conditions the DDW really onsets during  
the course of evolution of AGB stars.   
Also the large values of $\delta_{\rm C}$ and $\kappa_{\rm R}$ may results in a sensitive dependence of
mass-loss rate on the velocity amplitude of pulsation.
These subjects  are left for the
future investigations.

\section*{Acknowledgements}
We thank the anonymous referee for detailed and thorough comments which 
are very useful and helpful in improving  the manuscript. 
This research has been partly supported by the Grant-in Aid  for Scientific Research of the Japan 
Society for the Promotion of Science (23224004, 16H0218). 








\appendix

\section{Input stellar parameters for hydrodynamical calculations and the derived properties of dust-driven winds}

The input stellar parameters of hydrodynamical calculations and the derived properties of  dust-driven winds 
are tabulated for the CO-enhanced and the 
CNO-enhanced models with the mass-loss rate $\dot{M} \ge 10^{-7}$ $M_{\sun}$ yr$^{-1}$; Table A1 for $Z_{\rm ini}=10^{-7}$ and 
Table A2 for $Z_{\rm ini}=$ 10$^{-4}$, 10$^{-5}$, 10$^{-6}$, and 0.
 
\begin{landscape}
  \begin{table}
    \caption{Input parameters  and  the derived properties of dust-driven wind for $M_{\rm ini} =$ 2, 3, and 4 $M_{\sun}$ with $Z_{\rm ini} = 10^{-7}$;
      the current stellar mass $M$ ($M_{\sun}$),
      the effective temperature $T_{\rm eff}$ (K),
      the luminosity $L$ ($L_{\sun}$),
      the pulsation period $P$ (days),
      the opacity (cm$^2$ g$^{-1}$),
      the number ratio at the surface of He, C, O, and N to H,
      the current stellar mass $M$ ($M_{\sun}$),
      the time averaged acceleration ratio $\langle \alpha \rangle$,
      the mass-loss rate $\dot{M}$ ($M_{\sun}$ yr$^{-1}$),
      the terminal velocity $v_\infty$ (km s$^{-1}$),
      the dust-to-gas mass ratio $\rho_{\rm d}/\rho_{\rm g}$,
      the condensation efficiency $f_{\rm C}$,
      the mass-weighted average radius of dust $\langle a \rangle$ (${\rm \mu}$m).}
    \label{tabapp1}
    \centering
    \begin{tabular}{cccccccccccccccc}\hline \hline
$M$ & $T_{\rm eff}$ & $L$ & $P$ & $\kappa$ & He/H & C/H & O/H & N/H & $\delta_{\rm C}$&$\alpha$&$\dot{M}$&$v_{\rm \infty}$&$\rho_{\rm d}/\rho_{\rm g}$&$f_{\rm C}$&$\langle a \rangle$\\ \hline
      \multicolumn{3}{l}{$M_{\rm ini} = 2 M_{\rm \odot}$, $Z_{\rm ini} = 10^{-7}$}&&$\times10^4$&&$\times10^3$&$\times10^5$&$\times10^5$&$\times10^3$&&$\times10^7$&&$\times10^3$&&$\times10^3$ \\ \hline
      \multicolumn{2}{l}{CNO-enhanced}&&&&&&&&&&&&&&\\
      \hline
1.240&3316.05&12828.2&535.866&5.80808&0.107852&1.23789&9.74055&1.93039&1.14048&11.7731&1.78877&33.2258&7.14138&0.730532&10.353\\
1.200&3274.39&13458.4&602.060&6.00227&0.107852&1.23789&9.74055&1.93039&1.14048&16.0197&3.19385&45.6061&7.16348&0.732794&11.839\\
1.150&3250.59&13613.8&643.431&6.08223&0.107852&1.23789&9.74055&1.93039&1.14048&17.6043&5.92034&45.4434&7.35327&0.752208&11.637\\
1.100&3182.16&13187.4&700.316&7.89849&0.108914&1.44937&11.4939&1.97024&1.33443&20.9863&17.3274&50.8977&9.09215&0.794908&11.096\\
1.050&3155.38&13689.5&773.749&8.24118&0.108914&1.44937&11.4939&1.97024&1.33443&24.5245&30.4119&51.5424&8.69243&0.759962&12.622\\
1.000&3155.38&13832.9&807.880&8.01272&0.108914&1.44937&11.4939&1.97024&1.33443&26.8692&25.0564&51.1623&8.86184&0.774773&12.781\\
0.950&3179.18&13817.0&815.120&7.31592&0.108914&1.44937&11.4939&1.97024&1.33443&28.6445&44.7127&52.7775&9.87456&0.863313&11.826\\
0.900&3188.11&12988.2&792.366&7.94418&0.109937&1.63186&12.9682&2.04994&1.50218&32.6010&30.8123&55.1010&10.3428&0.803272&12.642\\
0.850&3259.52&13717.4&800.640&6.37923&0.109937&1.63186&12.9682&2.04994&1.50218&38.0782&42.7627&54.5666&10.3159&0.801186&10.661\\
0.800&3405.31&13932.6&722.035&4.83712&0.109937&1.63186&12.9682&2.04994&1.50218&40.6052&32.1355&51.8507&9.74715&0.757013&10.821\\
0.750&3677.66&14040.2&572.066&3.99576&0.109937&1.63186&12.9682&2.04994&1.50218&37.5421&12.3182&48.4256&9.44232&0.733338&9.5880\\
 \hline
\multicolumn{2}{l}{CO-enhanced}&&&&&&&&&&&&&&\\
\hline
1.100&3660.61&14067.0&440.693&1.94088&0.109832&1.61924&12.8129&2.02641&1.49111&19.5898&1.52120&43.2057&9.36897&0.733041&7.88827\\
1.050&3653.94&14255.7&464.594&1.88401&0.109832&1.61924&12.8129&2.02641&1.49111&23.1582&3.55201&46.6300&8.48705&0.664039&7.84537\\
1.000&3630.72&13801.7&478.038&1.95585&0.111082&1.84235&14.4681&2.08309&1.69767&28.4534&7.89617&52.0096&10.9169&0.750228&7.49216\\
0.950&3629.56&14320.5&514.884&1.89000&0.111082&1.84235&14.4681&2.08309&1.69767&32.3826&10.2463&51.7381&9.77730&0.671912&7.67924\\
0.900&3647.55&14438.4&530.320&1.87504&0.111082&1.84235&14.4681&2.08309&1.69767&36.3046&14.3795&52.6394&9.72441&0.668278&8.34448\\
0.870&3666.71&14364.7&528.826&1.89299&0.111082&1.84235&14.4681&2.08309&1.69767&38.0084&18.9908&52.3601&11.3945&0.783048&9.32786\\
0.820&3703.06&13872.5&517.824&2.05761&0.112242&2.03479&15.9193&2.21914&1.87560&42.4604&12.9375&53.7324&10.9197&0.679230&7.56001\\
0.800&3734.30&14190.8&520.362&2.11148&0.112242&2.03479&15.9193&2.21914&1.87560&43.4227&13.1908&54.2522&11.5639&0.719300&7.66789\\
0.750&3885.49&14473.8&479.532&2.64889&0.112242&2.03479&15.9193&2.21914&1.87560&43.9293&5.68844&48.4348&9.18683&0.571443&7.40766\\
\hline
      \multicolumn{3}{l}{$M_{\rm ini} = 3 M_{\rm \odot}$, $Z_{\rm ini} = 10^{-7}$}&&$\times10^4$&&$\times10^3$&$\times10^4$&$\times10^5$&$\times10^3$&&$\times10^7$&&$\times10^3$&&$\times10^3$ \\ \hline
      \multicolumn{2}{l}{CNO-enhanced}&&&&&&&&&&&&&&\\
      \hline
1.800&3158.40&22209.4&816.605&9.75781&0.120313&1.56544&1.14366&3.71391&1.45107&15.0468&1.73344&43.3230&7.85202&0.631304&6.8665\\
1.750&3122.27&21984.7&860.115&11.3645&0.120741&1.64902&1.19679&3.72209&1.52934&17.9789&2.98782&51.0273&8.71237&0.664629&7.0237\\
1.700&3101.29&22114.4&903.625&11.9957&0.120741&1.64902&1.19679&3.72209&1.52934&20.8686&6.34790&54.5990&9.29032&0.708717&8.2676\\
1.650&3065.17&22010.6&963.452&14.1761&0.121186&1.73639&1.25258&3.73027&1.61113&22.0531&7.74523&54.5884&9.84134&0.712640&7.1101\\
1.602&3051.19&21993.4&998.803&14.6926&0.121186&1.73639&1.25258&3.73027&1.61113&24.7114&7.84603&57.0181&9.90075&0.716942&8.1086\\
1.550&2999.91&22079.8&1093.98&18.9387&0.121971&1.88454&1.35087&3.76572&1.74945&27.8718&17.0124&58.8561&10.9117&0.727676&8.2811\\
1.500&2988.26&22088.4&1137.49&19.5126&0.121971&1.88454&1.35087&3.76572&1.74945&29.7228&18.3791&60.0656&10.4988&0.744090&8.0306\\
1.450&2933.49&22088.4&1251.71&26.5130&0.123218&2.11247&1.49964&3.88027&1.96251&37.5055&37.1253&66.8836&13.6682&0.812546&9.1855\\
1.400&2926.50&22166.2&1297.94&26.8573&0.123218&2.11247&1.49964&3.88027&1.96251&41.2179&47.1284&68.6298&12.5563&0.746443&7.9037\\
1.350&2924.16&22174.9&1336.01&26.5704&0.123218&2.11247&1.49964&3.88027&1.96251&44.2572&67.4315&70.6540&14.0318&0.834161&7.6342\\
1.304&2926.50&22131.6&1363.20&25.8818&0.123218&2.11247&1.49964&3.88027&1.96251&47.1193&43.0863&70.1846&13.2164&0.785688&8.7076\\
1.250&2885.71&22140.3&1482.85&33.4838&0.124894&2.40498&1.70153&3.89663&2.23483&59.8928&72.3613&77.5628&15.3208&0.799807&7.7771\\
1.200&2903.19&22235.4&1501.89&30.6386&0.124894&2.40498&1.70153&3.89663&2.23483&62.1005&86.2669&76.9654&14.1695&0.739701&8.2756\\
1.150&2931.16&22278.6&1496.45&26.7096&0.124894&2.40498&1.70153&3.89663&2.23483&66.1666&88.3011&76.1598&15.7880&0.824197&7.6751\\
1.100&2973.11&22295.8&1466.54&22.1032&0.124894&2.40498&1.70153&3.89663&2.23483&70.5593&109.164&73.6213&15.6457&0.816768&7.8305\\
 \hline
    \end{tabular}
    \end{table}
\end{landscape}

\begin{landscape}
  \begin{table}
\begin{flushleft}\textbf{Table A1 continued.}\end{flushleft}
    \centering
    \begin{tabular}{cccccccccccccccc}\hline \hline
$M$ & $T_{\rm eff}$ & $L$ & $P$ & $\kappa$ & He/H & C/H & O/H & N/H & $\delta_{\rm C}$&$\alpha$&$\dot{M}$&$v_{\rm \infty}$&$\rho_{\rm d}/\rho_{\rm g}$&$f_{\rm C}$&$\langle a \rangle$\\ \hline
      \multicolumn{3}{l}{$M_{\rm ini} = 3 M_{\rm \odot}$, $Z_{\rm ini} = 10^{-7}$}&&$\times10^4$&&$\times10^3$&$\times10^4$&$\times10^5$&$\times10^3$&&$\times10^7$&&$\times10^3$&&$\times10^3$ \\ \hline
\multicolumn{2}{l}{CO-enhanced}&&&&&&&&&&&&&&\\
\hline
1.150&3727.40&22182.3&609.677&1.53031&0.133332&1.31606&2.54775&206.946&1.06129&11.4187&1.41819&18.7337&4.35226&0.478442&11.0352\\
1.100&3735.40&22297.8&627.496&1.52378&0.133332&1.31606&2.54775&206.946&1.06129&17.1083&6.91223&26.3351&6.70094&0.736632&13.4831\\
1.067&3745.80&22235.9&633.235&1.53684&0.133332&1.31606&2.54775&206.946&1.06129&18.6653&12.9067&25.3804&7.07728&0.778002&13.1789\\
1.037&3727.00&21798.5&646.826&1.61520&0.134749&1.57958&2.75975&206.946&1.30361&28.0799&9.97957&40.7216&6.16531&0.551767&9.97913\\
1.000&3747.00&22198.8&662.531&1.64132&0.134749&1.57958&2.75975&206.946&1.30361&30.8209&13.3295&40.3212&6.00837&0.537722&10.8613\\
0.950&3792.19&22322.6&661.021&1.73928&0.134749&1.57958&2.75975&206.946&1.30361&32.6872&12.3773&39.6887&6.63601&0.593893&10.9893\\
0.915&3842.99&22289.5&645.315&1.89600&0.134749&1.57958&2.75975&206.946&1.30361&32.2771&12.6635&34.8753&8.34001&0.746393&10.6638\\
0.880&3880.59&21914.0&630.516&2.29400&0.137819&2.16406&3.24158&206.946&1.83990&50.5416&7.35495&44.2847&10.1169&0.641506&7.76959\\
0.850&4022.63&22343.2&576.764&2.96042&0.137819&2.16406&3.24158&206.946&1.83990&42.5482&2.71172&37.1135&8.19227&0.519465&6.83243\\
\hline
      \multicolumn{3}{l}{$M_{\rm ini} = 4 M_{\rm \odot}$, $Z_{\rm ini} = 10^{-7}$}&&$\times10^4$&&$\times10^4$&$\times10^4$&$\times10^3$&$\times10^4$&&$\times10^7$&&$\times10^4$&&$\times10^3$ \\ \hline
      \multicolumn{2}{l}{CNO-enhanced}&&&&&&&&&&&&&&\\
      \hline
1.600&3038.62&25018.6&1146.48&9.15177&0.150544&6.58257&1.72841&1.79393&4.85416&3.32040&2.31899&10.1586&9.43326&0.226722&18.558\\
1.550&3029.66&25060.1&1188.53&9.21287&0.150544&6.58257&1.72841&1.79393&4.85416&6.92343&7.29344&20.8258&21.9100&0.526592&35.675\\
1.500&2927.01&24718.9&1363.40&15.4451&0.151631&8.54022&1.85720&1.79393&6.68302&11.4354&13.4321&34.4586&38.8665&0.678500&25.011\\
1.450&2921.31&24986.4&1423.16&15.4145&0.151631&8.54022&1.85720&1.79393&6.68302&11.9103&19.1026&33.7557&41.3796&0.722371&25.439\\
1.400&2922.94&25023.2&1458.57&15.1090&0.151631&8.54022&1.85720&1.79393&6.68302&12.2713&28.5085&32.6575&41.8659&0.730861&27.199\\
1.375&2925.38&24981.7&1471.85&14.8909&0.151631&8.54022&1.85720&1.79393&6.68302&12.6620&24.4443&32.7098&39.1021&0.682613&27.188\\
1.350&2859.69&24669.8&1603.88&22.2599&0.152694&10.4881&1.98536&1.79393&8.50274&17.3814&41.4701&38.8087&48.7693&0.669167&21.668\\
1.300&2869.14&24993.6&1648.81&21.0528&0.152694&10.4881&1.98536&1.79393&8.50274&18.4077&55.9711&39.4340&54.2351&0.744163&23.506\\
1.250&2887.71&25061.9&1661.21&19.4036&0.152694&10.4881&1.98536&1.79393&8.50274&21.4882&82.2577&38.3222&52.5451&0.720974&21.046\\
1.200&2916.39&25022.3&1646.75&17.3974&0.152694&10.4881&1.98536&1.79393&8.50274&21.2623&139.187&37.1721&53.6487&0.736117&21.873\\
 \hline
\multicolumn{2}{l}{CO-enhanced}&&&&&&&&&&&&&&\\
\hline
1.100&3790.17&25251.9&667.445&1.50060&0.159737&12.3140&2.14489&2.09668&10.1691&14.8414&1.59205&18.6041&37.9496&0.435382&10.1558\\
1.050&3814.74&25398.8&677.893&1.54631&0.159737&12.3140&2.14489&2.09668&10.1691&12.9893&1.45142&16.5114&21.1322&0.242442&7.52456\\
1.000&3823.19&24969.1&685.439&1.70630&0.161315&15.3614&2.37780&2.09695&12.9836&27.5113&4.66115&30.2699&62.1904&0.558824&9.03862\\
0.950&3894.23&25450.2&676.442&1.91200&0.161315&15.3614&2.37780&2.09695&12.9836&25.4080&3.40131&26.4524&48.2340&0.433416&7.83191\\
\hline
    \end{tabular}
    \end{table}
\end{landscape}

\begin{landscape}
  \begin{table}
    \caption{Same as Table A1 but for all models except $M_{\rm ini} =$ 2, 3 and 4 $M_{\rm \odot}$ with $Z_{\rm ini} = 10^{-7}$ model.}
    \label{tabapp2}
    \centering
    \begin{tabular}{cccccccccccccccc}\hline \hline
$M$ & $T_{\rm eff}$ & $L$ & $P$ & $\kappa$ & He/H & C/H & O/H & N/H & $\delta_{\rm C}$&$\alpha$&$\dot{M}$&$v_{\rm \infty}$&$\rho_{\rm d}/\rho_{\rm g}$&$f_{\rm C}$&$\langle a \rangle$\\ \hline
      \multicolumn{3}{l}{$M_{\rm ini} = 2 M_{\rm \odot}$, $Z_{\rm ini} = 10^{-4}$}&&$\times10^4$&$\times10^2$&$\times10^3$&$\times10^4$&$\times10^6$&$\times10^3$&&$\times10^7$&&$\times10^3$&&$\times10^3$ \\ \hline
      \multicolumn{2}{l}{CNO-enhanced}&&&&&&&&&&&&&&\\
      \hline
1.250&3338.94&12916.9&522.374&5.13550&9.39632&1.56638&1.37701&3.28089&1.42868&18.1663&3.55304&49.0221&9.26062&0.756227&9.2684\\
1.190&3272.60&12602.6&570.602&6.21670&9.51406&1.78034&1.53969&3.50404&1.62637&22.1714&4.78770&55.8699&10.8788&0.780381&8.4654\\
1.150&3242.08&12965.7&621.543&6.42589&9.51406&1.78034&1.53969&3.50404&1.62637&25.2123&8.59955&57.1860&10.9897&0.788343&8.9031\\
1.100&3220.76&13112.8&665.690&6.53481&9.51406&1.78034&1.53969&3.50404&1.62637&27.5729&16.0594&59.0110&11.5646&0.829580&9.6414\\
1.070&3215.90&13091.2&681.205&6.53279&9.69428&2.10699&1.80110&4.14161&1.92688&35.0746&39.8511&64.4142&15.2790&0.925098&8.6247\\
1.000&3136.99&12925.0&775.883&9.36333&9.69428&2.10699&1.80110&4.14161&1.92688&39.4253&39.2466&64.3599&13.5719&0.821735&9.7659\\
0.950&3143.10&13158.5&813.308&8.86638&9.69428&2.10699&1.80110&4.14161&1.92688&44.2370&78.8438&65.3057&13.9913&0.847132&10.104\\
0.900&3179.10&13264.8&817.389&7.62180&9.69428&2.10699&1.80110&4.14161&1.92688&48.3443&51.9786&64.3025&14.0708&0.851943&8.9385\\
0.850&3254.64&13325.5&783.725&5.91049&9.69428&2.10699&1.80110&4.14161&1.92688&51.8245&63.4992&61.8357&14.0329&0.849648&8.7691\\
0.820&3322.48&13310.3&743.431&5.01595&9.69428&2.10699&1.80110&4.14161&1.92688&54.1433&49.3138&61.7336&14.1852&0.858873&12.918\\
0.750&3438.10&12642.1&668.964&5.56045&10.0821&2.79086&2.37194&5.75815&2.55367&73.7649&49.2142&67.6928&18.6802&0.853423&6.8885\\
 \hline
\multicolumn{2}{l}{CO-enhanced}&&&&&&&&&&&&&&\\
\hline
1.230&3403.20&12514.1&478.131&4.25313&9.50928&1.76904&1.53220&3.41822&1.61582&18.8180&1.75194&43.8845&10.0931&0.728753&7.57766\\
1.200&3372.04&12936.6&519.498&4.33107&9.50928&1.76904&1.53220&3.41822&1.61582&22.1539&5.21534&52.8957&10.0044&0.722342&8.70826\\
1.150&3344.92&13149.8&560.765&4.39201&9.50928&1.76904&1.53220&3.41822&1.61582&23.9172&5.37421&56.3700&10.0490&0.725567&8.37805\\
1.120&3337.66&13124.8&574.796&4.40725&9.50928&1.76904&1.53220&3.41822&1.61582&24.9763&10.2121&56.8877&12.0135&0.867411&9.77634\\
1.070&3270.97&12904.9&630.094&5.42814&9.66725&2.05172&1.75441&4.05309&1.87628&33.0183&15.3307&62.2296&12.4833&0.776208&8.22694\\
1.050&3261.07&13081.8&655.680&5.45861&9.66725&2.05172&1.75441&4.05309&1.87628&34.9095&28.4310&61.9283&13.6059&0.846014&9.37263\\
1.000&3254.46&13275.1&693.645&5.35957&9.66725&2.05172&1.75441&4.05309&1.87628&38.7163&24.3635&62.7492&12.3544&0.768195&9.08380\\
0.950&3267.01&13348.5&713.416&5.05744&9.66725&2.05172&1.75441&4.05309&1.87628&42.6281&30.3574&61.5102&12.9003&0.802139&8.92826\\
0.915&3290.12&13286.0&709.327&4.71961&9.66725&2.05172&1.75441&4.05309&1.87628&44.3248&31.7759&61.5741&13.1414&0.817128&8.98153\\
0.870&3282.86&12408.8&697.772&5.45099&9.86517&2.40456&2.04328&5.00541&2.20023&52.6909&38.6171&66.1579&16.0982&0.853604&7.88866\\
0.850&3303.32&12857.3&716.755&5.03197&9.86517&2.40456&2.04328&5.00541&2.20023&56.6553&59.9712&65.8274&15.6191&0.828196&7.83694\\
0.800&3421.51&13249.5&676.313&3.91966&9.86517&2.40456&2.04328&5.00541&2.20023&62.5362&45.6232&63.6694&15.9021&0.843203&8.09609\\
0.750&3628.09&13416.6&577.272&3.25685&9.86517&2.40456&2.04328&5.00541&2.20023&62.3623&28.5157&62.2468&14.4193&0.764581&7.49792\\
\hline
      \multicolumn{3}{l}{$M_{\rm ini} = 2 M_{\rm \odot}$, $Z_{\rm ini} = 10^{-5}$}&&$\times10^4$&$\times10^2$&$\times10^3$&$\times10^5$&$\times10^5$&$\times10^3$&&$\times10^7$&&$\times10^3$&&$\times10^3$ \\ \hline
      \multicolumn{2}{l}{CNO-enhanced}&&&&&&&&&&&&&&\\
      \hline
1.300&3265.46&12576.9&536.763&7.06323&9.56070&1.32209&9.95511&2.84940&1.22254&13.0549&2.71896&43.6010&8.39711&0.801335&10.676\\
1.240&3194.93&12274.7&589.369&8.80511&9.65633&1.50140&11.4535&2.88266&1.38687&17.3460&4.55029&50.9599&9.28072&0.780718&10.938\\
1.200&3155.75&12681.9&653.168&9.64457&9.65633&1.50140&11.4535&2.88266&1.38687&19.4534&5.88823&54.1910&9.50918&0.799937&10.001\\
1.150&3125.97&12833.0&705.774&10.40010&9.65633&1.50140&11.4535&2.88266&1.38687&21.0090&13.5384&51.3133&10.2159&0.859391&10.213\\
1.130&3119.70&12812.0&718.086&10.56800&9.65633&1.50140&11.4535&2.88266&1.38687&21.4855&12.6548&53.0611&9.64660&0.811497&10.422\\
1.090&3052.30&12312.5&770.083&15.69420&9.79182&1.76374&13.6145&2.92969&1.62760&25.5039&17.3597&57.7088&10.5748&0.758006&9.8957\\
1.050&3024.09&12770.0&849.425&16.94940&9.79182&1.76374&13.6145&2.92969&1.62760&29.7327&24.9733&60.8263&11.6148&0.832556&10.287\\
1.000&3014.68&12958.9&901.493&16.99590&9.79182&1.76374&13.6145&2.92969&1.62760&33.3294&35.2992&60.9806&11.9640&0.857583&10.642\\
0.950&3027.22&13047.0&926.287&15.78720&9.79182&1.76374&13.6145&2.92969&1.62760&36.3065&47.1479&60.9727&12.2716&0.879634&10.943\\
0.900&3064.84&13042.9&921.328&13.36980&9.79182&1.76374&13.6145&2.92969&1.62760&39.2270&64.1142&60.7041&10.9092&0.781979&11.003\\
0.850&2947.29&12333.5&1052.74&29.79860&10.12650&2.38802&18.8011&3.03918&2.20001&54.5560&64.8613&70.8512&16.6264&0.881700&8.2605\\
0.800&3064.84&12975.7&995.711&17.87920&10.12650&2.38802&18.8011&3.03918&2.20001&62.7349&64.2945&68.4908&16.7248&0.886920&8.9058\\
0.750&3314.34&13198.9&787.439&8.39542&10.12650&2.38802&18.8011&3.03918&2.20001&66.1491&62.5999&64.4525&16.4666&0.873227&7.6490\\
0.700&3990.15&13338.4&422.963&5.88501&10.12650&2.38802&18.8011&3.03918&2.20001&44.4785&1.27711&44.1531&12.8502&0.681447&5.2637\\
 \hline
    \end{tabular}
    \end{table}
\end{landscape}

\begin{landscape}
  \begin{table}
  \begin{flushleft}\textbf{    Table A2 continued.}\end{flushleft}
    \centering
    \begin{tabular}{cccccccccccccccc}\hline \hline
$M$ & $T_{\rm eff}$ & $L$ & $P$ & $\kappa$ & He/H & C/H & O/H & N/H & $\delta_{\rm C}$&$\alpha$&$\dot{M}$&$v_{\rm \infty}$&$\rho_{\rm d}/\rho_{\rm g}$&$f_{\rm C}$&$\langle a \rangle$\\ \hline
      \multicolumn{3}{l}{$M_{\rm ini} = 2 M_{\rm \odot}$, $Z_{\rm ini} = 10^{-5}$}&&$\times10^4$&$\times10^2$&$\times10^3$&$\times10^5$&$\times10^5$&$\times10^3$&&$\times10^7$&&$\times10^3$&&$\times10^3$ \\ \hline
\multicolumn{2}{l}{CO-enhanced}&&&&&&&&&&&&&&\\
\hline
1.100&3511.80&13261.9&486.157&2.61727&9.89846&1.93278&14.7245&2.97016&1.78554&27.3280&4.52347&55.9459&10.4870&0.685219&6.99248\\
1.050&3499.03&13420.6&516.886&2.57102&9.89846&1.93278&14.7245&2.97016&1.78554&30.2535&9.34291&55.2315&11.1952&0.731492&7.86904\\
1.000&3459.92&12491.0&521.076&2.97807&10.0670&2.22990&17.0100&3.03546&2.05980&36.2482&14.2681&61.6155&13.6363&0.772358&7.47469\\
0.950&3441.16&13363.9&588.818&2.88555&10.0670&2.22990&17.0100&3.03546&2.05980&43.6019&31.0592&60.4876&14.8818&0.842899&9.24524\\
0.900&3467.90&13545.3&602.785&2.70053&10.0670&2.22990&17.0100&3.03546&2.05980&48.4347&32.7648&61.7021&14.7553&0.835739&9.06443\\
0.850&3522.57&13619.0&595.801&2.50163&10.0670&2.22990&17.0100&3.03546&2.05980&53.3740&37.1340&59.9614&14.0226&0.794238&8.58517\\
0.825&3562.26&13579.4&582.532&2.43688&10.0670&2.22990&17.0100&3.03546&2.05980&53.3422&26.8432&60.3475&14.1118&0.799292&7.86681\\
0.750&3722.03&13256.2&518.282&2.73291&10.2899&2.60988&20.0139&3.23136&2.40974&64.8585&25.6274&62.8058&16.1179&0.780342&6.41660\\
\hline
      \multicolumn{3}{l}{$M_{\rm ini} = 2 M_{\rm \odot}$, $Z_{\rm ini} = 10^{-6}$}&&$\times10^4$&&$\times10^3$&$\times10^5$&$\times10^5$&$\times10^3$&&$\times10^7$&&$\times10^3$&&$\times10^3$ \\ \hline
      \multicolumn{2}{l}{CNO-enhanced}&&&&&&&&&&&&&&\\
      \hline
1.250&3230.78&12885.9&589.111&7.24435&0.102158&1.17419&8.63692&3.23885&1.08782&13.3615&2.88796&40.9209&6.99015&0.749680&12.267\\
1.200&3203.10&13099.1&636.355&7.39577&0.102158&1.17419&8.63692&3.23885&1.08782&15.5386&5.52113&44.5565&7.27887&0.780645&14.110\\
1.150&3126.97&12538.8&689.050&10.4545&0.103290&1.38658&10.4221&3.28136&1.28236&18.9502&9.19304&52.8373&9.20854&0.837776&10.946\\
1.100&3083.72&13158.6&785.354&11.7567&0.103290&1.38658&10.4221&3.28136&1.28236&20.5854&15.0710&50.2585&9.06116&0.824368&11.785\\
1.050&3069.43&13317.3&834.414&12.0596&0.103290&1.38658&10.4221&3.28136&1.28236&23.1650&29.8243&50.8974&9.08085&0.826159&11.675\\
1.000&3072.11&13337.1&863.487&11.6659&0.103290&1.38658&10.4221&3.28136&1.28236&24.9805&33.9343&53.3817&9.76594&0.888488&12.494\\
0.950&3064.75&12345.4&841.683&13.5435&0.104249&1.57282&11.9523&3.36637&1.45330&28.3503&37.9218&57.0333&10.7191&0.860501&11.828\\
0.900&3080.26&13168.5&912.548&12.1504&0.104249&1.57282&11.9523&3.36637&1.45330&33.4796&38.4812&57.1488&10.6013&0.851043&11.276\\
0.850&3158.11&13396.6&879.841&9.09170&0.104249&1.57282&11.9523&3.36637&1.45330&37.3799&57.2383&54.6450&10.1902&0.818037&11.775\\
0.800&3305.18&13510.7&787.171&6.39639&0.104249&1.57282&11.9523&3.36637&1.45330&39.4929&46.2347&53.9596&10.4944&0.842462&11.018\\
0.750&3564.70&13555.3&619.592&4.74218&0.104249&1.57282&11.9523&3.36637&1.45330&37.2452&17.1357&48.8594&8.93152&0.716997&10.179\\
 \hline
\multicolumn{2}{l}{CO-enhanced}&&&&&&&&&&&&&&\\
\hline
1.100&3640.21&13782.6&441.560&2.02569&0.104475&1.60827&12.1786&3.29758&1.48648&19.5411&3.61445&44.0755&9.67635&0.759448&9.33071\\
1.050&3604.62&13616.0&469.011&2.08664&0.105664&1.81956&13.7459&3.36288&1.68210&26.6383&4.63302&50.9905&9.44236&0.654901&6.92018\\
1.000&3597.77&13935.7&499.833&2.02188&0.105664&1.81956&13.7459&3.36288&1.68210&28.2457&7.09100&53.1315&10.0121&0.694416&7.35582\\
0.950&3605.60&13962.9&515.245&1.98569&0.105664&1.81956&13.7459&3.36288&1.68210&32.4879&10.0686&52.9608&9.96443&0.691110&7.86708\\
0.900&3576.54&13656.8&541.733&2.14949&0.107629&2.18119&16.4232&3.42819&2.01696&44.0833&20.6530&58.5387&13.5437&0.783409&7.51190\\
0.850&3614.42&14058.1&557.626&2.09235&0.107629&2.18119&16.4232&3.42819&2.01696&50.6613&22.7176&58.8941&13.2656&0.767323&7.73049\\
0.800&3686.38&14204.4&546.067&2.14949&0.107629&2.18119&16.4232&3.42819&2.01696&24.9805&33.9343&53.3817&9.76594&0.888488&12.4940\\
0.750&3833.31&14248.6&497.091&2.56919&0.107629&2.18119&16.4232&3.42819&2.01696&50.2504&8.31756&55.2756&12.3380&0.713667&6.78339\\
\hline
    \end{tabular}
    \end{table}
\end{landscape}

\begin{landscape}
  \begin{table}
   \begin{flushleft}\textbf{    Table A2 continued.}\end{flushleft}
    \centering
    \begin{tabular}{cccccccccccccccc}\hline \hline
$M$ & $T_{\rm eff}$ & $L$ & $P$ & $\kappa$ & He/H & C/H & O/H & N/H & $\delta_{\rm C}$&$\alpha$&$\dot{M}$&$v_{\rm \infty}$&$\rho_{\rm d}/\rho_{\rm g}$&$f_{\rm C}$&$\langle a \rangle$\\ \hline
      \multicolumn{3}{l}{$M_{\rm ini} = 2 M_{\rm \odot}$, $Z_{\rm ini} = 0$}&&$\times10^4$&&$\times10^3$&$\times10^5$&$\times10^5$&$\times10^3$&&$\times10^7$&&$\times10^3$&&$\times10^3$ \\ \hline
      \multicolumn{2}{l}{CNO-enhanced}&&&&&&&&&&&&&&\\
      \hline
1.150&3339.49&14291.4&610.218&4.68494&0.120970&1.08508&7.56827&2.25733&1.00940&14.0328&6.88367&33.4370&6.39255&0.738854&12.389\\
1.100&3319.37&14631.5&657.922&4.66262&0.120970&1.08508&7.56827&2.25733&1.00940&17.2337&12.8259&40.2044&5.89020&0.680793&13.928\\
1.060&3319.37&14620.9&675.352&4.58452&0.120970&1.08508&7.56827&2.25733&1.00940&18.0478&11.4310&41.1719&5.82789&0.673591&13.414\\
1.000&3271.61&14509.3&737.734&5.29422&0.121855&1.25492&8.92204&2.28337&1.16570&21.5347&18.5453&46.2308&7.77300&0.777945&13.282\\
0.950&3294.24&14806.8&761.586&4.93485&0.121855&1.25492&8.92204&2.28337&1.16570&24.4100&19.2805&47.3174&7.70166&0.773968&12.913\\
0.900&3347.03&14907.8&751.495&4.45024&0.121855&1.25492&8.92204&2.28337&1.16570&25.6475&20.9610&45.7600&7.60962&0.761593&13.163\\
0.870&3392.28&14828.1&727.643&4.15621&0.121855&1.25492&8.92204&2.28337&1.16570&26.4785&26.4916&45.6186&8.07064&0.807734&13.305\\
0.820&3500.37&14158.6&649.665&3.92207&0.122519&1.35533&9.70306&2.38750&1.25830&28.2349&18.2165&44.8536&7.94338&0.736492&11.008\\
0.800&3573.27&14557.1&628.565&3.69338&0.122519&1.35533&9.70306&2.38750&1.25830&29.1972&15.9277&43.9736&7.19478&0.667084&11.448\\
0.750&3900.07&14923.7&488.206&3.75327&0.122519&1.35533&9.70306&2.38750&1.25830&19.8531&1.78389&24.2369&4.88276&0.452719&6.8978\\
 \hline
\multicolumn{2}{l}{CO-enhanced}&&&&&&&&&&&&&&\\
\hline
1.100&3718.51&15066.7&443.232&1.79454&0.122083&1.30129&9.24046&2.27491&1.20889&11.5672&1.11605&24.7375&7.14649&0.689691&11.5388\\
1.072&3718.03&15021.3&450.342&1.77842&0.122083&1.30129&9.24046&2.27491&1.20889&12.6125&1.49146&25.7992&5.97437&0.576572&9.91552\\
1.030&3697.04&14885.3&469.897&1.79859&0.122943&1.46127&10.4594&2.30974&1.35668&19.3967&2.38930&40.9011&7.95007&0.683662&9.24964\\
1.000&3695.35&15152.8&489.095&1.76245&0.122943&1.46127&10.4594&2.30974&1.35668&22.8812&4.50433&42.2656&7.48530&0.643695&9.09147\\
0.950&3703.56&15270.8&507.228&1.74662&0.122943&1.46127&10.4594&2.30974&1.35668&25.0306&7.73868&44.0875&7.76989&0.668168&9.54643\\
0.900&3718.75&14540.6&496.562&1.83962&0.123551&1.56243&11.1908&2.41422&1.45052&27.9964&8.94191&46.9987&8.22067&0.661195&9.03381\\
0.850&3751.32&15220.9&522.871&1.87171&0.123551&1.56243&11.1908&2.41422&1.45052&31.1536&9.01878&45.8575&7.83613&0.630266&8.86316\\
0.800&3829.09&15379.6&511.494&2.10146&0.123551&1.56243&11.1908&2.41422&1.45052&31.9521&6.18426&42.7639&8.46901&0.681169&8.66284\\
\hline
      \multicolumn{3}{l}{$M_{\rm ini} = 3 M_{\rm \odot}$, $Z_{\rm ini} = 10^{-4}$}&&$\times10^4$&&$\times10^3$&$\times10^4$&$\times10^4$&$\times10^4$&&$\times10^7$&&$\times10^3$&&$\times10^3$ \\ \hline
      \multicolumn{2}{l}{CNO-enhanced}&&&&&&&&&&&&&&\\
      \hline
2.000&3015.63&22232.3&893.878&19.3012&0.101197&1.45730&1.28232&1.52832&13.2907&12.9080&1.33334&39.9491&8.32059&0.730388&6.5853\\
1.950&2980.03&21988.8&944.510&23.2151&0.101736&1.55750&1.34679&1.52873&14.2282&15.8151&1.30993&49.0924&8.74770&0.717283&6.8049\\
1.900&2961.70&22044.1&988.392&24.5965&0.101736&1.55750&1.34679&1.52873&14.2282&16.8991&1.93883&51.7274&8.63146&0.711262&7.1828\\
1.850&2909.93&22038.6&1072.78&32.1942&0.102569&1.70563&1.45269&1.54045&15.6036&18.4978&2.79164&53.7915&10.5294&0.787270&6.4778\\
1.800&2893.75&21994.3&1116.66&34.0360&0.102569&1.70563&1.45269&1.54045&15.6036&19.6371&6.21234&56.3819&10.9576&0.825301&7.2759\\
1.765&2884.04&21916.9&1143.66&35.1872&0.102569&1.70563&1.45269&1.54045&15.6036&20.0889&5.50874&56.6878&10.6411&0.795623&7.3496\\
1.740&2840.90&21861.5&1217.93&45.5476&0.103697&1.88861&1.60004&1.54620&17.2861&22.7527&12.1768&58.3655&11.4786&0.774708&6.9923\\
1.700&2827.96&21889.2&1265.18&47.3895&0.103697&1.88861&1.60004&1.54620&17.2861&23.8294&14.0703&60.0060&12.7515&0.860621&7.6718\\
1.650&2813.94&21861.5&1312.44&49.6918&0.103697&1.88861&1.60004&1.54620&17.2861&27.1918&24.3158&62.2434&12.7081&0.857690&9.3932\\
1.610&2804.23&21822.8&1352.95&51.0732&0.103697&1.88861&1.60004&1.54620&17.2861&28.3828&17.3372&62.5936&11.7366&0.800490&8.4547\\
1.550&2753.54&21789.6&1484.59&67.8801&0.105069&2.11080&1.78883&1.55134&19.3192&35.1209&26.7236&68.8731&12.5486&0.757796&8.0252\\
1.500&2742.75&21817.3&1545.35&69.7220&0.105069&2.11080&1.78883&1.55134&19.3192&38.7608&44.0972&70.9562&14.4273&0.871254&8.3376\\
1.450&2735.20&21822.8&1602.73&70.6429&0.105069&2.11080&1.78883&1.55134&19.3192&40.4044&53.3852&71.4569&14.1214&0.852780&8.8723\\
1.400&2730.89&21784.1&1649.99&70.6429&0.105069&2.11080&1.78883&1.55134&19.3192&42.1429&41.2658&71.1171&13.8080&0.833854&8.6252\\
1.350&2693.14&21684.4&1778.26&89.5219&0.106638&2.36348&2.00525&1.55257&21.6296&50.6735&53.2345&77.4225&14.8148&0.799091&7.9518\\
1.300&2695.30&21800.7&1828.89&86.5289&0.106638&2.36348&2.00525&1.55257&21.6296&53.6790&129.891&78.3726&17.0171&0.917879&8.7478\\
1.250&2706.08&21850.5&1862.65&80.7731&0.106638&2.36348&2.00525&1.55257&21.6296&57.1269&78.5636&78.2777&16.1485&0.871027&7.9453\\
1.200&2725.50&21883.7&1869.40&72.9452&0.106638&2.36348&2.00525&1.55257&21.6296&60.6020&91.1719&76.7541&15.5582&0.839189&7.7713\\
1.150&2755.70&21889.2&1852.52&62.5848&0.106638&2.36348&2.00525&1.55257&21.6296&63.8397&91.6873&78.1608&16.2348&0.875684&8.2444\\
1.120&2777.27&21856.0&1825.52&56.5988&0.106638&2.36348&2.00525&1.55257&21.6296&67.0162&84.7668&77.3281&15.7920&0.851798&8.1264\\
\hline
    \end{tabular}
    \end{table}
\end{landscape}

\begin{landscape}
  \begin{table}
   \begin{flushleft}\textbf{    Table A2 continued.}\end{flushleft}
    \centering
    \begin{tabular}{cccccccccccccccc}\hline \hline
$M$ & $T_{\rm eff}$ & $L$ & $P$ & $\kappa$ & He/H & C/H & O/H & N/H & $\delta_{\rm C}$&$\alpha$&$\dot{M}$&$v_{\rm \infty}$&$\rho_{\rm d}/\rho_{\rm g}$&$f_{\rm C}$&$\langle a \rangle$\\ \hline
      \multicolumn{3}{l}{$M_{\rm ini} = 3 M_{\rm \odot}$, $Z_{\rm ini} = 10^{-4}$}&&$\times10^4$&&$\times10^3$&$\times10^4$&$\times10^4$&$\times10^4$&&$\times10^7$&&$\times10^3$&&$\times10^3$ \\ \hline
\multicolumn{2}{l}{CO-enhanced}&&&&&&&&&&&&&&\\
\hline
1.250&3565.28&22295.2&680.163&2.09803&0.112988&1.01316&2.74936&1.89958&7.38224&7.28083&1.24420&16.5888&2.69006&0.425129&14.4195\\
1.200&3572.26&22260.3&694.587&2.06211&0.112988&1.01316&2.74936&1.89958&7.38224&9.08139&1.66900&16.7601&2.19562&0.346989&14.5219\\
1.150&3521.90&21873.2&742.530&2.29131&0.114533&1.28629&2.98222&1.89975&9.88068&22.4317&14.1567&39.7778&5.76723&0.680969&15.5770\\
1.100&3542.82&22155.4&759.829&2.22274&0.114533&1.28629&2.98222&1.89975&9.88068&24.2924&31.8863&37.8784&3.37954&0.399041&15.3359\\
1.050&3579.19&22208.8&758.473&2.16397&0.114533&1.28629&2.98222&1.89975&9.88068&24.5468&31.3496&37.9850&6.70059&0.791176&14.5386\\
1.015&3613.74&22168.8&747.957&2.13703&0.114533&1.28629&2.98222&1.89975&9.88068&25.2638&21.5587&37.0235&6.57241&0.776041&15.4065\\
0.980&3591.01&21826.5&774.415&2.39906&0.116819&1.70538&3.33594&1.90026&13.7179&40.3560&37.3278&45.2459&8.58464&0.730101&13.0975\\
0.950&3648.29&22162.1&758.133&2.35743&0.116819&1.70538&3.33594&1.90026&13.7179&40.2803&35.6810&43.6517&8.59415&0.730909&13.7282\\
0.900&3786.50&22308.8&690.970&2.45538&0.116819&1.70538&3.33594&1.90026&13.7179&41.3448&14.4090&40.4245&5.86793&0.499051&10.5988\\
\hline
      \multicolumn{3}{l}{$M_{\rm ini} = 3 M_{\rm \odot}$, $Z_{\rm ini} = 10^{-5}$}&&$\times10^4$&&$\times10^3$&$\times10^4$&$\times10^4$&$\times10^4$&&$\times10^7$&&$\times10^3$&&$\times10^3$ \\ \hline
      \multicolumn{2}{l}{CNO-enhanced}&&&&&&&&&&&&&&\\
      \hline
1.950&3009.24&21538.1&892.832&21.0329&0.101222&1.56216&1.24712&1.00473&14.3745&14.1400&1.24103&46.1341&8.71975&0.707716&6.6426\\
1.900&2987.14&21668.6&939.831&22.4509&0.101222&1.56216&1.24712&1.00473&14.3745&16.6370&1.85042&52.4538&8.30533&0.674080&7.0388\\
1.850&2941.73&21567.1&1013.69&28.4151&0.101878&1.68577&1.33080&1.00506&15.5269&18.3658&2.11308&50.8081&10.7392&0.806927&6.1932\\
1.800&2923.32&21588.8&1057.33&30.1668&0.101878&1.68577&1.33080&1.00506&15.5269&19.1460&3.44584&55.3983&10.6713&0.801822&6.6586\\
1.750&2866.87&21574.4&1158.04&40.9473&0.102970&1.86931&1.45433&1.01967&17.2388&22.5456&6.48865&58.1546&11.7476&0.795044&6.8126\\
1.700&2850.91&21559.9&1208.39&43.1126&0.102970&1.86931&1.45433&1.01967&17.2388&23.7606&8.13202&60.3070&11.4520&0.775039&7.0684\\
1.650&2836.18&21523.6&1258.75&45.1232&0.102970&1.86931&1.45433&1.01967&17.2388&25.8751&13.8307&62.6051&11.4845&0.777234&7.8384\\
1.600&2782.18&21443.9&1376.25&61.6722&0.104412&2.10154&1.64560&1.03195&19.3698&32.0997&17.3409&67.1045&12.6609&0.762579&7.1745\\
1.550&2768.68&21501.9&1440.03&64.3015&0.104412&2.10154&1.64560&1.03195&19.3698&35.0375&24.6764&69.0913&12.5090&0.753435&8.1265\\
1.500&2757.64&21501.9&1493.74&66.0028&0.104412&2.10154&1.64560&1.03195&19.3698&37.3028&32.7386&70.3525&13.9301&0.839029&7.9968\\
1.450&2749.05&21494.6&1550.81&67.0855&0.104412&2.10154&1.64560&1.03195&19.3698&39.4166&41.1780&70.9359&13.1075&0.789482&8.6812\\
1.410&2745.36&21451.2&1584.38&67.3948&0.104412&2.10154&1.64560&1.03195&19.3698&41.0530&45.7026&71.6043&13.5512&0.816203&8.1277\\
1.380&2701.18&21291.7&1698.52&89.6664&0.106248&2.38621&1.90461&1.03311&21.9575&49.0258&46.5672&77.6232&16.1737&0.859359&7.7259\\
1.350&2697.50&21429.4&1748.88&89.3571&0.106248&2.38621&1.90461&1.03311&21.9575&50.8176&55.0252&77.6706&16.2000&0.860756&7.8651\\
1.300&2699.95&21501.9&1795.88&86.5732&0.106248&2.38621&1.90461&1.03311&21.9575&54.1770&66.7586&79.0794&16.4012&0.871443&7.9068\\
1.250&2708.54&21545.4&1829.45&81.6239&0.106248&2.38621&1.90461&1.03311&21.9575&59.7619&84.2515&79.0835&15.9341&0.846626&7.9505\\
1.200&2725.73&21559.9&1842.88&73.8907&0.106248&2.38621&1.90461&1.03311&21.9575&60.9272&121.704&79.0037&16.8505&0.895315&8.7991\\
1.150&2753.96&21567.1&1829.45&64.3015&0.106248&2.38621&1.90461&1.03311&21.9575&67.2020&131.499&78.4620&16.8787&0.896815&8.3430\\
1.125&2841.09&21538.1&1809.31&58.5790&0.106248&2.38621&1.90461&1.03311&21.9575&64.3482&74.2307&76.8827&16.2465&0.863225&7.7135\\
 \hline
\multicolumn{2}{l}{CO-enhanced}&&&&&&&&&&&&&&\\
\hline
1.150&3669.76&21933.9&639.024&1.64758&0.114683&1.28961&3.15448&19.8499&9.74162&15.3286&2.75553&25.0336&4.77237&0.571544&11.6534\\
1.100&3682.16&21965.3&652.737&1.64051&0.114683&1.28961&3.15448&19.8499&9.74162&17.7682&4.70819&28.8469&4.85846&0.581854&13.3947\\
1.056&3699.28&21896.7&658.859&1.65112&0.114683&1.28961&3.15448&19.8499&9.74162&19.7741&4.66134&28.1421&5.01614&0.600738&12.5878\\
1.000&3681.57&21785.3&694.857&1.79968&0.116778&1.65531&3.47668&19.8518&13.0764&34.8799&24.5790&42.4493&7.43256&0.663126&12.6145\\
0.950&3738.84&21956.7&686.776&1.87396&0.116778&1.65531&3.47668&19.8518&13.0764&36.4313&19.2438&41.3920&7.16523&0.639275&11.9728\\
0.900&3833.22&22011.0&652.492&2.08973&0.116778&1.65531&3.47668&19.8518&13.0764&37.5735&30.9386&40.7391&7.95782&0.709990&10.0633\\
\hline
    \end{tabular}
    \end{table}
\end{landscape}

\begin{landscape}
  \begin{table}
   \begin{flushleft}\textbf{    Table A2 continued.}\end{flushleft}
    \centering
    \begin{tabular}{cccccccccccccccc}\hline \hline
$M$ & $T_{\rm eff}$ & $L$ & $P$ & $\kappa$ & He/H & C/H & O/H & N/H & $\delta_{\rm C}$&$\alpha$&$\dot{M}$&$v_{\rm \infty}$&$\rho_{\rm d}/\rho_{\rm g}$&$f_{\rm C}$&$\langle a \rangle$\\ \hline
      \multicolumn{3}{l}{$M_{\rm ini} = 3 M_{\rm \odot}$, $Z_{\rm ini} = 10^{-6}$}&&$\times10^4$&&$\times10^3$&$\times10^4$&$\times10^5$&$\times10^4$&&$\times10^7$&&$\times10^3$&&$\times10^3$ \\ \hline
      \multicolumn{2}{l}{CNO-enhanced}&&&&&&&&&&&&&&\\
      \hline
1.900&3090.39&21640.1&826.329&14.0478&0.107099&1.53665&1.07087&5.73803&14.2956&14.0679&1.31479&43.5430&8.73270&0.712675&6.5167\\
1.840&3052.12&21404.2&880.362&16.7795&0.107600&1.62608&1.12975&5.74357&15.1311&16.2432&2.52741&49.1540&9.54671&0.736091&6.8871\\
1.800&3032.07&21557.2&922.388&17.7101&0.107600&1.62608&1.12975&5.74357&15.1311&18.4198&5.25820&51.4604&8.49704&0.655157&7.3182\\
1.750&2982.87&21493.5&998.435&22.7366&0.108351&1.76695&1.21963&5.78645&16.4499&20.3402&4.67849&55.5300&10.5377&0.747359&6.6427\\
1.700&2963.74&21544.5&1046.46&24.1113&0.108351&1.76695&1.21963&5.78645&16.4499&21.9032&8.33625&56.8996&10.2316&0.725653&7.0933\\
1.640&2909.06&21397.8&1142.52&32.4794&0.109424&1.95476&1.34670&5.85009&18.2009&25.6788&25.7539&61.6647&11.3057&0.724690&7.3059\\
1.600&2894.49&21.4935&1190.55&33.9737&0.109424&1.95476&1.34670&5.85009&18.2009&27.5721&12.3439&61.6702&11.6330&0.745668&7.1334\\
1.550&2881.73&21487.1&1240.58&35.4082&0.109424&1.95476&1.34670&5.85009&18.2009&30.7266&18.5587&63.2823&12.2792&0.787088&8.0537\\
1.510&2872.62&21423.3&1272.60&36.2450&0.109424&1.95476&1.34670&5.85009&18.2009&31.8780&26.2352&64.0918&12.3664&0.792677&8.4409\\
\hline
\multicolumn{2}{l}{CO-enhanced}&&&&&&&&&&&&&&\\
\hline
1.200&3712.30&21914.9&593.181&1.54595&0.119582&1.229240&2.78029&197.206&9.51211&9.77604&1.02568&17.5339&3.02928&0.371543&9.72747\\
1.150&3716.93&21941.6&610.051&1.53126&0.119582&1.229240&2.78029&197.206&9.51211&13.6675&2.31529&25.9417&4.97348&0.610000&14.3414\\
1.090&3684.49&21415.9&640.079&1.62921&0.121382&1.552080&3.05415&197.206&12.4667&27.4791&30.9771&39.9155&8.89785&0.832687&10.2609\\
1.050&3697.01&21815.9&661.335&1.61942&0.121382&1.552080&3.05415&197.206&12.4667&29.1651&19.1169&40.7562&7.77809&0.727897&11.3047\\
1.000&3725.74&21903.5&668.082&1.65860&0.121382&1.552080&3.05415&197.206&12.4667&31.1523&37.0494&41.2446&7.37330&0.690016&14.1229\\
0.950&3772.55&21876.8&660.660&1.75655&0.121382&1.552080&3.05415&197.206&12.4667&32.8810&17.4306&38.5650&6.84473&0.640550&11.7188\\
0.900&3806.85&21739.7&661.672&2.07980&0.124399&2.097770&3.52555&197.206&17.4522&51.9964&17.7039&48.5809&8.88470&0.593937&8.70068\\
0.850&3994.43&22013.9&582.385&2.85362&0.124399&2.097770&3.52555&197.206&17.4522&49.1353&4.38012&42.9501&9.04255&0.604489&7.19564\\
\hline
      \multicolumn{3}{l}{$M_{\rm ini} = 3 M_{\rm \odot}$, $Z_{\rm ini} = 0$}&&$\times10^4$&&$\times10^3$&$\times10^4$&$\times10^5$&$\times10^3$&&$\times10^7$&&$\times10^3$&&$\times10^3$ \\ \hline
      \multicolumn{2}{l}{CNO-enhanced}&&&&&&&&&&&&&&\\
      \hline
1.705&3193.12&20915.0&769.728&9.40080&0.162251&1.83376&1.32750&4.41197&1.70101&16.0610&1.10253&43.0910&9.72756&0.667181&5.7245\\
1.650&3158.45&20944.1&821.882&10.4698&0.162556&1.89158&1.36560&4.42104&1.75502&16.5100&2.00024&46.2026&8.83906&0.587585&5.4973\\
1.600&3133.16&20766.3&858.738&11.5145&0.162828&1.94259&1.40051&4.43918&1.80254&18.8256&2.19967&50.3418&8.40362&0.543912&5.8097\\
1.557&3115.36&20900.5&900.462&12.0854&0.162828&1.94259&1.40051&4.43918&1.80254&21.3379&2.64154&51.3439&8.70258&0.563261&6.0833\\
1.500&3085.38&20911.4&960.245&13.5405&0.163100&1.99361&1.43543&4.45731&1.85007&23.9164&3.95056&54.9920&10.0183&0.614844&6.1652\\
1.440&3065.70&20577.6&999.120&14.7368&0.163383&2.04689&1.45924&4.47545&1.90097&26.4285&7.25363&57.8056&10.8110&0.663498&6.6509\\
1.400&3047.90&20958.6&1059.34&15.3826&0.163383&2.04689&1.45924&4.47545&1.90097&28.0893&14.2387&59.0958&10.4794&0.643146&7.2935\\
1.370&3043.22&20933.2&1077.63&15.4779&0.163383&2.04689&1.45924&4.47545&1.90097&30.0380&13.1574&58.3090&9.90343&0.607796&7.3926\\
1.325&3029.16&20788.1&1120.32&16.5578&0.163688&2.10585&1.51162&4.50266&1.95469&33.2075&20.4616&60.1866&11.1212&0.663778&8.3752\\
1.300&3024.48&20969.5&1151.57&16.6319&0.163688&2.10585&1.51162&4.50266&1.95469&35.4404&20.3661&60.5248&11.0766&0.661111&7.9328\\
1.252&3024.48&20984.0&1182.06&16.3037&0.163688&2.10585&1.51162&4.50266&1.95469&37.9885&29.2257&63.4203&10.1911&0.608261&8.1991\\
1.200&2999.27&21084.0&1265.51&18.9499&0.164645&2.28597&1.63134&4.55163&2.12284&51.7773&72.8822&69.8954&13.5973&0.747278&8.2983\\
1.150&3014.83&21238.9&1289.34&17.4436&0.164645&2.28597&1.63134&4.55163&2.12284&51.3141&55.9495&69.3102&13.1216&0.721137&7.9170\\
1.100&3044.16&21182.0&1281.16&15.3349&0.164645&2.28597&1.63134&4.55163&2.12284&54.3185&59.0855&70.0126&12.8698&0.707299&8.0798\\
 \hline
    \end{tabular}
    \end{table}
\end{landscape}

\begin{landscape}
  \begin{table}
   \begin{flushleft}\textbf{    Table A2 continued.}\end{flushleft}
    \centering
    \begin{tabular}{cccccccccccccccc}\hline \hline
$M$ & $T_{\rm eff}$ & $L$ & $P$ & $\kappa$ & He/H & C/H & O/H & N/H & $\delta_{\rm C}$&$\alpha$&$\dot{M}$&$v_{\rm \infty}$&$\rho_{\rm d}/\rho_{\rm g}$&$f_{\rm C}$&$\langle a \rangle$\\ \hline
      \multicolumn{3}{l}{$M_{\rm ini} = 3 M_{\rm \odot}$, $Z_{\rm ini} = 0$}&&$\times10^4$&&$\times10^3$&$\times10^4$&$\times10^5$&$\times10^4$&&$\times10^7$&&$\times10^3$&&$\times10^3$ \\ \hline
\multicolumn{2}{l}{CO-enhanced}&&&&&&&&&&&&&&\\
\hline
1.400&3582.34&21949.8&605.728&2.03051&0.167051&2.63632&1.84644&9.18902&2.45168&29.1694&4.47676&55.1713&9.75756&0.464328&5.90261\\
1.350&3564.13&21874.4&632.271&2.04792&0.167488&2.71915&1.90008&9.19465&2.52914&38.7477&3.81161&59.5465&12.0635&0.556476&4.87773\\
1.300&3558.59&21259.2&635.079&2.08050&0.167894&2.79141&1.94633&9.2409&2.59678&40.5105&4.71607&59.9084&10.9348&0.491273&4.36087\\
1.250&3549.09&21826.1&678.135&2.02788&0.167894&2.79141&1.94633&9.2409&2.59678&46.9114&17.4596&63.7578&14.6628&0.658764&5.07966\\
1.200&3543.46&21512.5&693.579&2.05043&0.168329&2.87500&2.00647&9.28716&2.67435&53.4059&15.0103&68.0937&14.4337&0.629658&5.23448\\
1.150&3549.09&21814.1&720.255&1.99280&0.168329&2.87500&2.00647&9.28716&2.67435&58.7739&21.7333&69.0556&14.0522&0.613019&5.28812\\
1.100&3567.30&21247.1&712.299&1.98027&0.168504&2.89764&2.02034&9.33341&2.69561&62.8817&31.5352&66.8338&15.1342&0.655012&6.06577\\
1.050&3588.68&21759.8&736.635&1.91943&0.168504&2.89764&2.02034&9.33341&2.69561&70.2173&28.4094&65.9217&13.2492&0.573430&6.06063\\
0.995&3636.98&20481.2&688.899&1.96022&0.168504&2.89764&2.02034&9.37967&2.69561&67.6089&26.3891&66.6502&13.6526&0.590891&5.79633\\
0.950&3680.53&21693.5&719.319&1.94519&0.168504&2.89764&2.02034&9.37967&2.69561&77.4801&41.0327&65.0408&13.5589&0.586833&6.19480\\
0.900&3763.66&21802.1&693.111&2.11308&0.168504&2.89764&2.02034&9.37967&2.69561&80.7719&57.2826&65.1831&16.7247&0.723849&5.37783\\
0.850&3919.57&21289.4&605.458&2.69192&0.168504&2.89764&2.02034&9.42593&2.69561&72.2165&7.49052&60.0441&12.2836&0.531639&4.84606\\
\hline
      \multicolumn{3}{l}{$M_{\rm ini} = 4 M_{\rm \odot}$, $Z_{\rm ini} = 10^{-4}$}&&$\times10^4$&&$\times10^4$&$\times10^4$&$\times10^3$&$\times10^4$&&$\times10^7$&&$\times10^3$&&$\times10^3$ \\ \hline
      \multicolumn{2}{l}{CNO-enhanced}&&&&&&&&&&&&&&\\
      \hline
1.600&3021.27&26263.1&1223.82&9.03603&0.143781&6.03060&1.63390&1.62851&4.39670&4.21939&4.02808&11.6332&1.17635&0.312145&25.399\\
1.550&3015.89&26292.5&1261.45&9.03603&0.143781&6.03060&1.63390&1.62851&4.39670&5.90510&6.72881&17.0646&1.70121&0.451416&34.152\\
1.500&2934.34&26174.9&1425.25&13.5478&0.144560&7.52633&1.73001&1.62851&5.79632&11.0935&40.1708&35.7293&3.85647&0.776218&26.790\\
1.450&2935.24&26272.9&1465.09&13.3096&0.144560&7.52633&1.73001&1.62851&5.79632&11.4519&29.5855&31.1308&3.55209&0.714955&28.299\\
1.405&2940.62&26243.5&1487.22&12.9448&0.144560&7.52633&1.73001&1.62851&5.79632&11.8894&40.9672&29.4709&3.60388&0.725379&27.289\\
1.350&2874.30&26272.9&1666.51&19.1990&0.145623&9.47664&1.85625&1.62851&7.62039&16.4957&41.6472&37.2322&4.74226&0.726031&24.590\\
1.300&2892.23&26351.3&1679.79&17.8133&0.145623&9.47664&1.85625&1.62851&7.62039&18.4112&133.004&34.3871&5.16348&0.790519&24.637\\
1.250&2919.21&26351.3&1673.15&16.1677&0.145623&9.47664&1.85625&1.62851&7.62039&20.5669&106.194&34.2135&3.74819&0.573840&22.780\\
 \hline
\multicolumn{2}{l}{CO-enhanced}&&&&&&&&&&&&&&\\
\hline
1.150&3675.76&26520.3&757.937&1.85780&0.151401&10.3069&2.28134&1.98162&8.02556&11.6577&1.99288&17.4756&2.30061&0.334437&13.9821\\
1.100&3711.32&26554.5&756.304&1.84896&0.151401&10.3069&2.28134&1.98162&8.02556&15.1591&2.69638&20.8993&3.20787&0.466324&14.2606\\
1.050&3721.98&26505.6&772.902&1.98897&0.152675&12.8513&2.47235&1.98181&10.3790&24.0850&6.72319&29.8800&4.90957&0.551870&11.5767\\
1.000&3802.33&26647.6&743.244&2.05529&0.152675&12.8513&2.47235&1.98181&10.3790&21.4958&3.77084&22.3935&3.57476&0.401828&9.57462\\
0.972&3864.91&26579.2&713.314&2.16730&0.152675&12.8513&2.47235&1.98181&10.3790&18.5270&2.86993&19.5625&2.62650&0.295236&8.05221\\
\hline
      \multicolumn{3}{l}{$M_{\rm ini} = 4 M_{\rm \odot}$, $Z_{\rm ini} = 10^{-5}$}&&$\times10^4$&&$\times10^4$&$\times10^4$&$\times10^3$&$\times10^4$&&$\times10^7$&&$\times10^4$&&$\times10^3$ \\ \hline
      \multicolumn{2}{l}{CNO-enhanced}&&&&&&&&&&&&&&\\
      \hline
1.600&3093.68&26124.4&1114.66&6.83974&0.142862&5.37275&1.52002&1.64125&3.85273&1.24775&1.30532&1.87644&4.92134&0.149026&19.300\\
1.550&3088.12&26054.8&1146.62&6.85780&0.142862&5.37275&1.52002&1.64125&3.85273&1.79558&2.67970&6.61737&5.17000&0.156556&20.701\\
1.500&2991.22&25915.7&1315.53&10.5429&0.143618&6.78398&1.61044&1.64136&5.17354&8.35267&13.6928&23.8596&24.4276&0.550859&33.056\\
1.450&2990.42&25972.6&1352.05&10.4165&0.143618&6.78398&1.61044&1.64136&5.17354&8.52058&12.1903&21.5878&27.0495&0.609985&32.755\\
1.400&2914.67&25764.1&1515.03&15.6998&0.144499&8.46971&1.71993&1.64136&6.74978&13.2999&26.8137&34.7882&40.9655&0.708069&27.571\\
1.350&2922.61&25980.0&1552.19&15.0336&0.144499&8.46971&1.71993&1.64136&6.74978&14.2953&38.9306&34.8483&41.7788&0.722126&28.868\\
1.300&2938.67&26027.0&1566.42&14.0956&0.144499&8.46971&1.71993&1.64136&6.74978&15.5810&68.1279&32.5261&42.9516&0.742397&28.512\\
1.272&2950.97&25966.1&1563.65&13.4819&0.144499&8.46971&1.71993&1.64136&6.74978&16.0643&78.6956&32.6764&40.2158&0.695110&29.3247\\
 \hline
    \end{tabular}
    \end{table}
\end{landscape}

\begin{landscape}
  \begin{table}
   \begin{flushleft}\textbf{    Table A2 continued.}\end{flushleft}
    \centering
    \begin{tabular}{cccccccccccccccc}\hline \hline
$M$ & $T_{\rm eff}$ & $L$ & $P$ & $\kappa$ & He/H & C/H & O/H & N/H & $\delta_{\rm C}$&$\alpha$&$\dot{M}$&$v_{\rm \infty}$&$\rho_{\rm d}/\rho_{\rm g}$&$f_{\rm C}$&$\langle a \rangle$\\ \hline
      \multicolumn{3}{l}{$M_{\rm ini} = 4 M_{\rm \odot}$, $Z_{\rm ini} = 10^{-5}$}&&$\times10^4$&&$\times10^4$&$\times10^4$&$\times10^3$&$\times10^4$&&$\times10^7$&&$\times10^4$&&$\times10^3$ \\ \hline
\multicolumn{2}{l}{CO-enhanced}&&&&&&&&&&&&&&\\
\hline
1.100&3785.17&26285.2&695.912&1.57039&0.150927&11.0177&2.15468&1.95640&8.86302&13.1504&1.69817&15.9886&22.5177&0.296408&10.3599\\
1.050&3820.36&26303.3&696.293&1.63229&0.150927&11.0177&2.15468&1.95640&8.86302&12.2253&1.71194&14.7973&25.3087&0.333146&9.88877\\
1.000&3837.96&26310.6&709.626&1.83125&0.152537&14.2860&2.40283&1.95654&11.8832&24.6081&4.09965&27.2018&49.5283&0.486260&8.87579\\
0.950&3940.20&26444.8&670.771&2.13465&0.152537&14.2860&2.40283&1.95654&11.8832&22.7546&2.26200&21.4764&31.8215&0.312417&7.05192\\
\hline
      \multicolumn{3}{l}{$M_{\rm ini} = 4 M_{\rm \odot}$, $Z_{\rm ini} = 10^{-6}$}&&$\times10^4$&&$\times 10^4$&$\times10^4$&$\times10^3$&$\times10^4$&&$\times10^7$&&$\times10^4$&&$\times10^3$ \\ \hline
      \multicolumn{2}{l}{CNO-enhanced}&&&&&&&&&&&&&&\\
      \hline
1.600&3056.28&25099.8&1123.72&8.22070&0.142591&6.00088&1.62127&1.76576&4.37961&1.75155&1.78657&5.65076&5.42277&0.144455&16.220\\
1.550&3048.34&25065.2&1160.62&8.26799&0.142591&6.00088&1.62127&1.76576&4.37961&2.95001&3.10306&9.93794&8.93528&0.238023&22.341\\
1.500&2955.41&24869.2&1324.78&12.8547&0.143407&7.48672&1.72073&1.76576&5.76599&10.2694&11.6918&29.8459&31.3296&0.633909&29.344\\
1.450&2951.44&24973.0&1369.05&12.8075&0.143407&7.48672&1.72073&1.76576&5.76599&10.9875&1.91926&32.6459&32.4373&0.656324&29.304\\
1.407&2953.02&24938.4&1394.87&12.5710&0.143407&7.48672&1.72073&1.76576&5.76599&11.4217&21.8926&32.8370&36.3615&0.735725&28.996\\
1.350&2884.39&24920.7&1568.48&18.5363&0.144385&9.25377&1.83611&1.76576&7.41766&15.1808&44.8265&37.5033&46.8478&0.736833&25.805\\
1.300&2895.89&25012.3&1594.29&17.5425&0.144385&9.25377&1.83611&1.76576&7.41766&16.0896&66.8600&36.3088&47.7197&0.750547&26.304\\
1.250&2915.58&25029.5&1600.36&16.1603&0.144385&9.25377&1.83611&1.76576&7.41766&17.9144&82.3470&33.9071&50.0389&0.787024&26.256\\
 \hline
\multicolumn{2}{l}{CO-enhanced}&&&&&&&&&&&&&&\\
\hline
1.090&3775.79&24994.8&674.877&1.54866&0.152228&13.0274&2.2125&2.05395&10.8149&17.7968&2.88524&23.7196&50.2826&0.542428&10.3451\\
1.050&3797.12&25354.7&688.286&1.58295&0.152228&13.0274&2.2125&2.05395&10.8149&18.1129&2.87756&22.1834&36.5738&0.394543&9.18635\\
1.000&3841.09&25413.5&685.238&1.68090&0.152228&13.0274&2.2125&2.05395&10.8149&18.2486&2.86271&20.9475&30.8867&0.333193&8.35958\\
0.950&3871.56&25318.0&687.676&1.98455&0.154504&17.5785&2.5712&2.05449&15.0073&40.8401&6.83862&39.7812&55.7683&0.433542&8.26986\\
0.900&4037.32&25549.4&616.976&2.67512&0.154504&17.5785&2.5712&2.05449&15.0073&24.3846&1.47272&24.3480&55.3056&0.429946&5.73953\\
\hline
      \multicolumn{3}{l}{$M_{\rm ini} = 4 M_{\rm \odot}$, $Z_{\rm ini} = 0$}&&$\times10^4$&&$\times10^4$&$\times10^4$&$\times10^3$&$\times10^4$&&$\times10^7$&&$\times10^3$&&$\times10^2$ \\ \hline
      \multicolumn{2}{l}{CNO-enhanced}&&&&&&&&&&&&&&\\
      \hline
1.550&3004.09&24654.4&1206.50&10.9489&0.191974&7.31515&1.98422&2.35083&5.33093&3.94484&2.09875&11.6642&1.26130&0.276033&1.9355\\
1.500&2934.46&24635.1&1347.82&15.1860&0.192583&8.60712&2.07809&2.36755&6.52903&10.1968&11.8647&34.4380&2.66649&0.476473&2.7915\\
1.450&2928.71&24731.0&1396.54&15.2088&0.192583&8.60712&2.07809&2.36755&6.52903&10.7991&17.2217&37.2506&3.10405&0.554661&3.0968\\
1.400&2898.26&24588.8&1482.08&17.8334&0.192984&9.40438&2.13842&2.36755&7.26596&11.8515&15.2341&35.6180&3.80695&0.611268&2.4041\\
1.350&2900.30&24734.9&1526.72&17.3887&0.192984&9.40438&2.13842&2.36755&7.26596&12.2301&15.7419&32.5745&3.85461&0.618919&2.3183\\
1.300&2867.66&24418.4&1617.01&21.3985&0.193604&10.6662&2.23583&2.36755&8.43037&14.3101&23.2360&35.9948&4.30871&0.596276&2.1278\\
1.250&2879.90&24763.3&1658.61&20.0718&0.193604&10.6662&2.23583&2.36755&8.43037&15.6117&43.5326&35.9164&4.72500&0.653886&2.1286\\
1.200&2902.34&24799.9&1662.66&18.3691&0.193604&10.6662&2.23583&2.36755&8.43037&18.3262&58.1988&36.1284&4.79241&0.663214&2.2690\\
 \hline
    \end{tabular}
    \end{table}
\end{landscape}

\begin{landscape}
  \begin{table}
   \begin{flushleft}\textbf{    Table A2 continued.}\end{flushleft}
    \centering
    \begin{tabular}{cccccccccccccccc}\hline \hline
$M$ & $T_{\rm eff}$ & $L$ & $P$ & $\kappa$ & He/H & C/H & O/H & N/H & $\delta_{\rm C}$&$\alpha$&$\dot{M}$&$v_{\rm \infty}$&$\rho_{\rm d}/\rho_{\rm g}$&$f_{\rm C}$&$\langle a \rangle$\\ \hline
      \multicolumn{3}{l}{$M_{\rm ini} = 5 M_{\rm \odot}$, $Z_{\rm ini} = 10^{-4}$}&&$\times10^4$&&$\times10^4$&$\times10^4$&$\times10^3$&$\times10^4$&&$\times10^7$&&$\times10^4$&&$\times10^2$ \\
\hline
      \multicolumn{2}{l}{CNO-enhanced}&&&&&&&&&&&&&&\\
      \hline
1.450&3156.83&31888.7&1338.92&5.06238&0.165217&5.78915&2.42856&1.73913&3.36059&1.58642&4.09535&5.31577&3.71425&0.128944&2.4288\\
1.400&3100.55&31820.7&1465.27&6.71101&0.165788&6.98985&2.50124&1.73913&4.48861&7.63765&19.3213&19.1998&18.7954&0.488523&3.6688\\
1.350&3126.53&31937.4&1462.93&6.31041&0.165788&6.98985&2.50124&1.73913&4.48861&8.14753&20.0081&15.9822&16.2788&0.423115&3.2800\\
1.330&3137.71&31888.7&1458.25&6.17174&0.165788&6.98985&2.50124&1.73913&4.48861&8.12276&21.9824&14.6053&14.1822&0.368621&3.0160\\
\hline
      \multicolumn{2}{l}{CO-enhanced}&&&&&&&&&&&&&&\\
      \hline
1.100&3870.29&32226.1&774.846&1.91793&0.169942&10.7233&2.84210&1.88837&7.88120&10.7892&1.31327&12.3453&15.7767&0.233545&0.9440\\
      \hline
      \multicolumn{3}{l}{$M_{\rm ini} = 5 M_{\rm \odot}$, $Z_{\rm ini} = 10^{-5}$}&&$\times10^4$&&$\times10^4$&$\times10^4$&$\times10^3$&$\times10^4$&&$\times10^7$&&$\times10^4$&&$\times10^2$ \\
\hline
      \multicolumn{2}{l}{CNO-enhanced}&&&&&&&&&&&&&&\\
      \hline
1.450&3189.53&30675.0&1239.85&4.53491&0.166764&5.68728&2.41216&1.83212&3.27512&1.06547&1.25136&1.54306&3.80944&0.135700&2.3862\\
1.400&3115.15&30664.3&1389.61&6.23243&0.167337&6.93248&2.49292&1.83221&4.43956&8.44758&19.9802&22.3217&21.8578&0.574398&4.3401\\
1.350&3136.66&30664.3&1391.63&5.89292&0.167337&6.93248&2.49292&1.83221&4.43956&8.30818&22.7109&20.8376&20.7950&0.546470&4.2711\\
1.300&3096.33&30707.3&1502.94&7.59044&0.168007&8.20920&2.56376&1.83230&5.64544&12.7565&50.7684&24.7468&25.9658&0.536599&3.2370\\
1.250&3138.45&30739.5&1472.58&6.85485&0.168007&8.20920&2.56376&1.83230&5.64544&13.8330&48.4231&24.5847&25.9371&0.536008&3.4423\\
\hline
      \multicolumn{3}{l}{$M_{\rm ini} = 5 M_{\rm \odot}$, $Z_{\rm ini} = 10^{-6}$}&&$\times10^4$&&$\times10^4$&$\times10^4$&$\times10^3$&$\times10^4$&&$\times10^7$&&$\times10^4$&&$\times10^2$ \\
\hline
      \multicolumn{2}{l}{CNO-enhanced}&&&&&&&&&&&&&&\\
      \hline
1.450&3142.78&30215.0&1294.98&5.36104&0.163058&6.11749&2.27678&1.78645&3.84071&2.11325&4.72581&7.09111&4.76467&0.144733&2.2432\\
1.417&3150.60&30190.3&1301.72&5.24999&0.163058&6.11749&2.27678&1.78645&3.84071&2.42082&5.82691&7.44480&5.26745&0.160006&2.4163\\
\hline
      \multicolumn{2}{l}{CO-enhanced}&&&&&&&&&&&&&&\\
      \hline
1.050&3917.13&30586.5&728.383&1.84213&0.170688&13.8482&2.88997&1.96628&10.9582&10.7141&1.50135&15.5986&15.9509&0.169821&0.5885\\
      \hline
    \end{tabular}
    \end{table}
\end{landscape}



\bsp
\label{lastpage}
\end{document}